\UseRawInputEncoding
\documentclass[aps,prb,showpacs,notitlepage,floatfix,
 twocolumn]{revtex4-1}
\usepackage{bbm}
\usepackage{mathrsfs}
\usepackage{epsfig}
\usepackage{graphicx}
\usepackage{amsfonts}
\usepackage[figuresright]{rotating}
\usepackage{amssymb}
\usepackage{amsmath}
\usepackage{dcolumn}
\usepackage{bm}
\usepackage{braket}
\usepackage[colorlinks,linkcolor=blue,anchorcolor=blue,citecolor=blue,urlcolor=blue]{hyperref}
\usepackage{multirow}
\usepackage{longtable}
\usepackage{pbox}
\usepackage{float}
\usepackage[normalem]{ulem}

\def\avg#1{\langle#1\rangle}
\def\Re {\mbox{Re}}
\def\Im {\mbox{Im}}
\def\tr {\mbox{tr}}
\def\be{\begin{equation}}       \def\ee{\end{equation}}
\def\bea{\begin{eqnarray}}      \def\eea{\end{eqnarray}}
\def\bp{\begin{pmatrix}} \def\ep{\end{pmatrix}}
\def\beaa{\begin{equation}\begin{aligned}}
\def\eeaa{\end{aligned}\end{equation}}

\def\nn{\nonumber}

\begin{document}
\title{The Symmetry Principle in Condensed Matter Physics (I)}
\author{Congjun Wu}
\affiliation{Department of Physics, School of Science,
Westlake University, Hangzhou 310024, Zhejiang, China
}
\affiliation{
Institute for Theoretical Sciences, Westlake University,
Hangzhou 310024, Zhejiang, China
}
\affiliation{
Key Laboratory for Quantum Materials of Zhejiang Province,
School of Science, Westlake University, Hangzhou 310024, China
}
\affiliation{
Institute of Natural Sciences, Westlake Institute for Advanced Study,
Hangzhou 310024, Zhejiang, China }

\begin{abstract}
Symmetry distills the simplicity of natural laws from the complexity
of physical phenomena.
The symmetry principle is of vital importance in various aspects of
modern physics, including analyzing atomic spectra, determining
fundamental interactions in the Standard Model, and unifying
physics at different energy scales.
In this chapter, novel applications of this principle are
reviewed in condensed matter physics and cold atom physics for
exploring new states of matter.

First, the concept of {\it space-time group} generalizes
crystalline space group symmetries to their dynamic counterparts,
including nonsymmorphic space-time symmetries (e.g. {\it time-screw}
rotation, {\it time-glide} reflection, and {\it time-shift} rotary reflection).
It includes and goes beyond the Floquet theory framework,
and applies to a large class of dynamic systems such as laser-driven
solid crystals, dynamic photonic crystals, and optical lattices, {\it etc}.
Second, the perspective of high symmetries (e.g. SU($N$) and Sp($N$))
bridges large-spin cold fermion systems with high energy physics.
For example, a generic SO(5), or, isomorphically Sp(4) symmetry is
proved in spin-$\frac{3}{2}$ systems.
Moreover, an exact SO(7) symmetry is identified as possessing an
extraordinary unifying power: Its $\chi$-pairing operator extends
Yang's $\eta$-pairing to a high-rank Lie algebra, integrating 21 orders
in both particle-hole and particle-particle channels into a unified framework.
Such systems also exhibit multi-fermion orderings, including quartetting superfluidity (charge $4e$) and quartet density wave, which are $\alpha$-particle-like, or, baryon-like orderings.
The resonant quantum plaquette states of SU(4) antiferromagnetism
are described by a high-order gauge theory.
A quantum phase transition occurs from the
Slater region to the Mott region in the SU(6) Hubbard model.
A tendency of convergence of itineracy and
locality is revealed in 1D SU($N$) systems as $N$ goes large.
Third, a new mechanism is presented to generate spin-orbit coupling based on ``spin-from-isospin" via many-body Fermi surface instabilities
of the Pomeranchuk type.
In contrast, the conventional wisdom views spin-orbit coupling
as a single-body relativistic effect.
This mechanism generalizes itinerant ferromagnetism to the unconventional
symmetry versions (e.g. $p$-wave), which can also be viewed
as magnetic multipolar orderings in momentum space.
\end{abstract}
\maketitle
\tableofcontents

\section{Introduction}
\label{sect:intro}
I feel honored to contribute to this Festschrift for {\it the Yang Centenary}.
Professor C. N. Yang is the role model for Chinese physicists of my generation.
Throughout our careers, we have been inspired by his milestone contributions
to theoretical physics, including parity violation in the weak interaction
\cite{lee1956}, Yang-Mills gauge theory \cite{yang1954},
Yang-Baxter equation\cite{yang1967}, and monopole gauge theories
\cite{wu1975,wu1976}, {\it etc}.
Among these masterpieces, the symmetry principle is a threading theme,
which is also a distinct style of his research.

I learned to appreciate the symmetry principle under the
guidance of my Ph. D. advisor Professor Shoucheng Zhang, who himself was
deeply influenced by Professor Yang.
Symmetries and their applications in condensed matter
physics and cold atom physics are my major research directions.
Hence, I shall review progresses along this line for this Festschrift.

\subsection{General backgrounds}
The appreciation of symmetry at a fundamental level has a long
history.
The ancient Greeks proved the existence of only five types
of convex regular polyhedra (the Platonic solids): tetrahedron, cube, octahedron, dodecahedron, and icosahedron.
They hypothesized that these regular polyhedra correspond to
the classic elements of water, earth, fire, air, and ether, respectively
\cite{weyl}.
Galileo's relativity principle implies the homogeneity of
space and time (translational symmetry), the isotropy of space (rotational symmetry), and the equivalence of all the inertial
reference frames \cite{gross1996}.
Einstein's relativity is a profound victory of the symmetry principle:
The Lorentz symmetry is viewed as a fundamental symmetry of space-time,
which is not only a property of Maxwell's equations but also
the primary constraint to all physical laws \cite{gross1996}.
In high energy physics,
Yang stated, {\it ``Symmetry dictates interaction"}, {\it i.e.},
interactions among fundamental particles in the Standard Model
are determined by their fundamental gauge symmetries
\cite{yang1996}.

The first application of the symmetry principle in physics
actually started in the field of condensed matter.
Soon after the establishment of group theory by Galois and
Cauchy in the 1830s-1840s, it was applied to analyze crystalline symmetries.
In the 1890s, Sch\"onflies and Fedorov completed the construction of
the 230 space groups \cite{lax2012}.
Each space group corresponds to one type of crystalline structure
in three dimensions (3D), which is a subgroup symmetry of
3D flat space containing a discrete translational group
as its normal subgroup.

In the 1880s, the concept of group was generalized to continuous groups,
i.e., Lie groups, by Sophus Lie, and then calculus and differential
equations entered the study of symmetry \cite{georgi1999}.
Lie group and its generators Lie algebra became the main tools to analyze
symmetries.
Noether proved that each continuous symmetry
gives rise to a local conservation law: Momentum
conservation arises from the translational symmetry; angular momentum conservation arises from the rotational symmetry \cite{noether1918}, {\it etc}.

The application of group theory in quantum physics was pioneered by
Wigner \cite{wigner1959} and Weyl \cite{weyl1950}.
Because of the linear nature of quantum mechanics, the eigenstates of a time-independent Hamiltonian form irreducible
representations of its symmetry group $G$.
Its generators commute with the Hamiltonian, and thus are conserved
quantities.
This principle is extremely successful in classifying the atomic and
molecular optical spectra and explaining selection rules for
optical transitions.

Two remarkable examples of hidden
symmetries of simple systems are the hydrogen
atom \cite{fock1935} and the harmonic oscillator \cite{sakurai2010}.
The $N$-dimensional hydrogen atom possesses the SO($N+1$) symmetry
due to the conserved Runge-Lentz vectors.
Classically, the Runge-Lentz vector specifies the orientation of the
elliptical orbit.
The $N$-dimensional harmonic oscillator possesses the SU($N$) symmetry
which transforms among the complex space spanned by the complex
combination of coordinate and momentum
$a_i=\frac{1}{\sqrt 2} (x_i +i p_i)$.

One central theme in modern physics is the unification by the symmetry
principle.
Electricity and magnetism are unified by the Lorentz group.
The interaction between matter and the electromagnetic field is
described by the U(1) gauge theory.
In particle physics, the electromagnetic and weak interactions are unified
by the SU$_L$(2) $\otimes$ U(1) gauge theory as
the electroweak interaction,
where $L$ refers to left-handed leptons and quarks \cite{Glashow1959,Salam1959,Weinberg1967}.
The quantum chromodynamics is described by the SU(3) color
gauge theory, and quarks of three colors (R, G, B)
form the fundamental representation of the SU(3) group.
Mesons are quark-antiquark bound states and bayrons
are three-quark bound states, both of which are color singlets.
In addition, bayrons and mesons can be classified as multiplets of the
approximate SU(3) flavor symmetry \cite{peskin1995}.

Spontaneous symmetry breaking is a crucially important concept, which was first proposed by L. Landau for constructing a general framework of
phase transitions\cite{landau1937,ginzburg1950,landau1980}.
Most second-order phase transitions are related to certain kinds of
symmetry breaking of order parameters (the matter fields).
For instance, the magnetic phase transition breaks time-reversal
and rotational symmetries; the charge-density-wave breaks
translational symmetry; superfluidity breaks the U(1) symmetry.
If a continuous global symmetry $G$ is spontaneously broken, the
transverse fluctuations of order parameters are gapless,
which are the Goldstone modes as reminiscences of the original
symmetry before its breaking \cite{goldstone1962}.
The Goldstone manifold is represented as the coset of $G/H$, where
$H$ represents the residual subgroup symmetry after symmetry breaking.

Even more profound physics occurs when a gauge symmetry is spontaneously
broken.
For example, superconductivity is a consequence of the U(1) gauge
symmetry breaking \cite{anderson1963,higgs1964}.
The electromagnetic properties of superconductors are characterized
by the London equation $\mathbf{j}=-\rho_s \mathbf{A}$,
where $\rho_s$ is the superfluid density, giving rise to the
celebrated Meissner effect.
This is is due to the Anderson-Higgs mechanism that the gauge boson
(photon) becomes massive and acquires its longitudinal component
by absorbing the Goldstone mode of phase fluctuations.
Consequently, the electromagnetic field can only enter the
superconductor surface at the penetration depth $\lambda$
with the relation of $\rho_s=c/(4\pi\lambda^2)$.

The Anderson-Higgs mechanism is essential in high energy physics
\cite{peskin1995}.
The gauge bosons become massive, once the corresponding gauge symmetries
are broken.
This cures the apparent discrepancy between the short-range weak
and strong interactions and the
massless Yang-Mills gauge fields \cite{yang1954}.
This was the major obstacle to applying the Yang-Mills theory
as the paradigm for formulating fundamental interactions.
Furthermore, the Higgs field generates masses for fermions of
quarks and leptons as shown in the Glashow-Weinberg-Salam theory,
which unifies the weak and electromagnetic interactions
into the electroweak interaction
\cite{Glashow1959,Salam1959,Weinberg1967}.

In the context of condensed matter physics, the symmetry principle
is employed to unify seemingly unrelated phenomena.
For example, Yang's pseudo-spin SU(2) symmetry based on the
$\eta$-pairing unifies the charge-density-wave ordering
and superconductivity \cite{yang1989,yang1990,zhang1991}.
Its extension to the SO(5) theory of high T$_c$ superconductivity
by Zhang views antiferromagnetism and $d$-wave superconductivity
on the equal footing as different components of a 5-vector
\cite{zhang1997,demler2004}.
The sharp resonance modes of neutron scattering spectroscopy could
be interpreted as the pseudo-Goldstone excitations in the superconducting
ground state towards the direction of antiferromagnetism
\cite{demler1995,zhang1997}.

Some new applications of the symmetry principle in
condensed matter and ultra-cold atom physics will be reviewed below
focusing on exploring novel states of matter.
The motivation and outline of the main results for each topic are briefly
explained below.

\subsection{Space-time group for dynamic systems}

\begin{figure} \centering
\includegraphics[width=\linewidth]{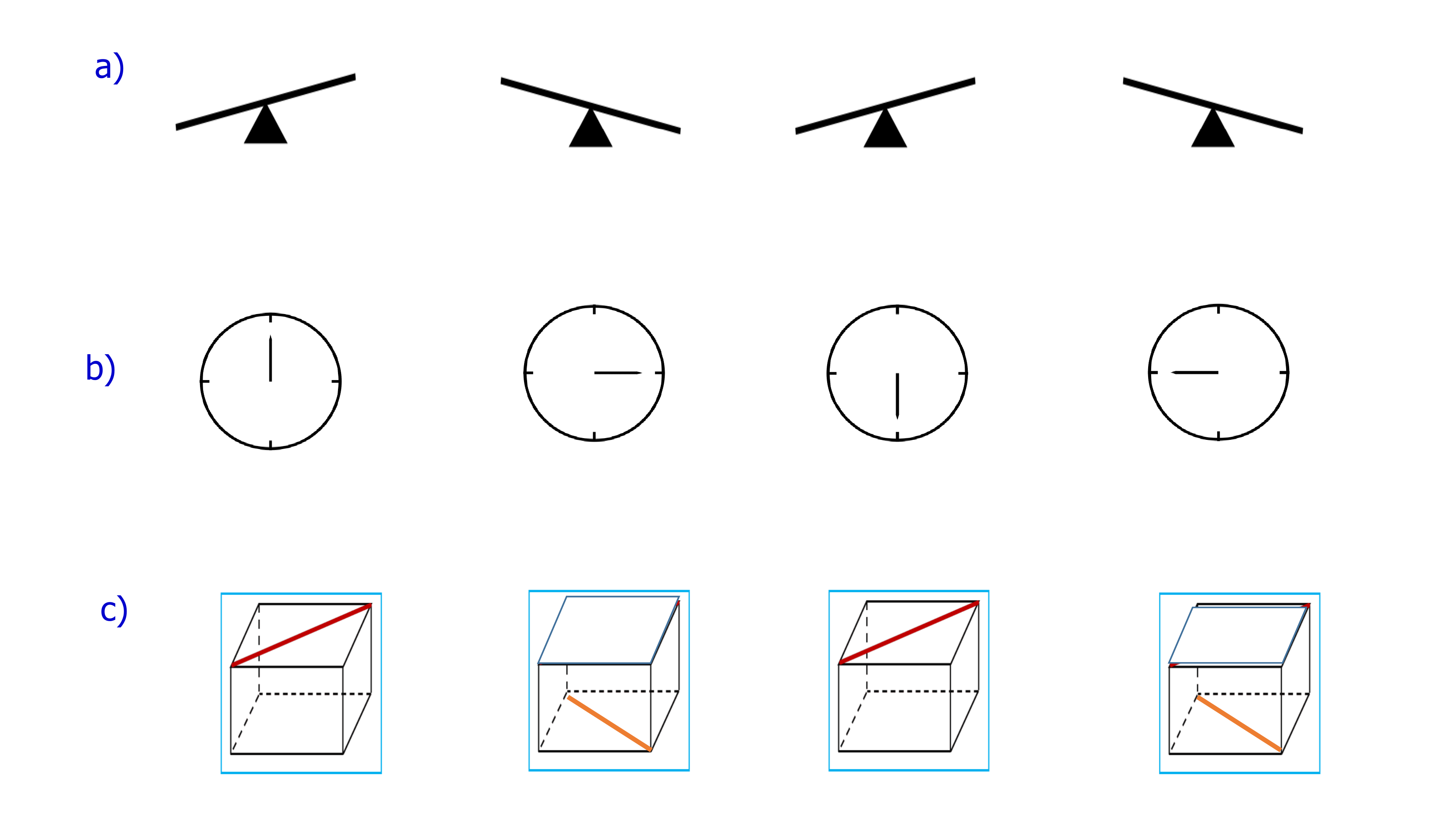}
\caption{Time sequence configurations for three representative
space-time nonsymmorphic symmetries.
$a$) {\it Time-glide} reflection symmetry.
A see-saw is invariant by a reflection followed by a
time-shift of half a period.
$b$) {\it Time-screw} rotation symmetry.
A clock is invariant by a rotation followed by a
fractional time translation.
$c$)  {\it Time-shift} rotary reflection symmetry, i.e.,
rotary reflection followed by a fractional time translation.
Time-glide reflection and time-screw rotation are analogies of
glide reflection and screw rotation of space group symmetries,
respectively,
while time-shift rotary reflection has no counterpart in 3D
space group operations.
}
\label{fig:nonsym}
\end{figure}

A solid state textbook typically starts with crystalline symmetries,
which are classified according to the 230 space groups,
and then proceeds with the Bloch theorem setting up the framework of
electron's quantum behavior in solids \cite{kittel1987}.
Space group symmetries include the discrete translational symmetry of
the underlying Bravais lattice, and point group symmetries
(e.g. {\it rotation, reflection}, and {\it rotary reflection}).
Space group possesses non-symmphoric symmetries, which means that
under such operations there are no fixed points,
including {\it screw rotation} and {\it glide reflection}.
Screw rotation is the symmetry of a screw: A rotation is insufficient
to maintain a screw invariant which needs to be followed by
a certain translation along the rotation axis.
Glide reflection is a symmetry of a row of footprints, i.e., a
reflection followed by a translation of half a period
\cite{lax2012}.

Symmetry literally means ``balanced proportions'',
and thus is commonly viewed as a static concept.
However, time dynamics is an important topic in various subjects of
physics.
The recent experimental progresses, such as the
pump-prob measurements \cite{gedik2013,zhangxiang2018}
and shaken cold-atom optical lattice
experiments\cite{parker2013,anderson2017}
, have stimulated the study of dynamically driven systems.

A natural question is how to analyze symmetries of dynamic systems?
Systems under periodical driving are often denoted as the Floquet ones.
In such systems, time translational symmetry is violated
while a discrete version still exists, which
is the counterpart of the discrete spacial translational symmetry in crystals.
However, within the Floquet framework, temporal symmetry is decoupled from
the spacial one
\cite{rechtsman2013a,rudner2013,thakurathi2013a,vonKeyserlingk2016b,
gu2011,lindner2011,else2016,vonKeyserlingk2016a,potter2016a,roy2016,
nathan2015,flaschner2016}.

Just like that a 3D crystal is typically not the direct-product
between a 2D crystal in the $ab$-plane with a 1D crystal along
the $c$-axis, a dynamic crystalline system is {\it not} just
the direct-product between a static crystal with a Floquet periodicity.
We construct the symmetry group of dynamic systems and
dub it {\it space-time group}, which is a dynamic
extension of the crystalline space group \cite{xu2018}.
The Bloch theorem is also generalized accordingly.
This concept applies to a large class of dynamic systems beyond the Floquet
framework, including laser-driven solid crystals, dynamic photonic
crystals, and optical lattices, etc.

There exist nonsymmorphic versions of space-time symmetries
as depicted in Fig. \ref{fig:nonsym} \cite{xu2018}.
(Please do not confuse them with Lorentz symmetries).
For example, a see-saw is invariant by a reflection followed
by a time-shift of half a period, and this symmetry is dubbed
{\it time-glide reflection} (Fig. \ref{fig:nonsym}($a$)).
A clock does not exhibit the rotation symmetry but
a rotation combined with a suitable time-translation leaves it
invariant, and this symmetry is dubbed {\it time-screw rotation}
(Fig. \ref{fig:nonsym}($b$)).
These are actually symmetries of their world lines in analogy to
screw rotation and glide reflection of space group.
Another space-time nonsymmorphic symmetry,  3D rotary-reflection
followed by a time-translation, does not
have a space group counterpart (Fig. \ref{fig:nonsym}($c$)).

A complete classification in 1+1D gives rise to
13 space-time groups in contrast to the 17 wallpaper space
groups for the 2D static crystals, and in 2+1D we have found 275
space-time groups \cite{xu2018}.
Space-time group symmetries also protect spectral degeneracies.

Time-screw rotation and time-glide reflection symmetries were also
proposed by Morimote {\it et. al.} independently for studying novel
topological band structures in driven systems
\cite{morimoto2017}, but the concept of ``space-time group"
was not proposed there.

\subsection{High symmetry perspective to large-spin cold fermion systems}
\begin{figure} \centering
\includegraphics[width=0.8\linewidth]{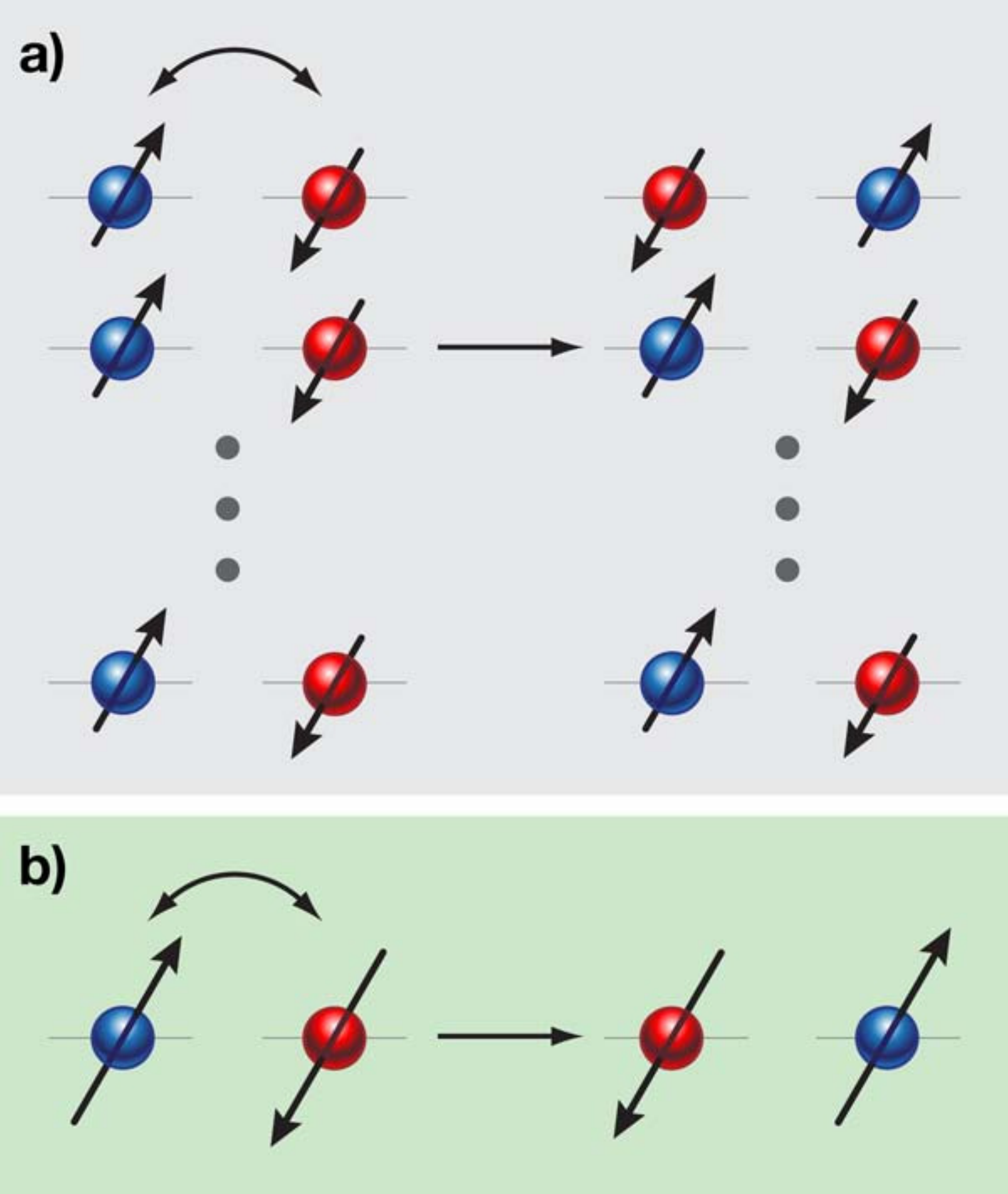}
\caption{Superexchange processes in ($a$) large-spin solid state
systems and ($b$) large hyperfinespin cold fermion systems.
In solids, quantum magnetic fluctuations are suppressed
by the large-$S$ effect; while quantum fluctuations are enhanced
by the large number of spin components $N = 2S + 1$.
Hence, the appropriate viewpoint for large-spin fermions
is the large-$N$ physics of a high symmetry group rather than
the large-$S$ physics of the SU(2) group.
This feature bridges high energy physics and ultracold atom physics
in spite of hugely different energy scales.
From Ref. [\onlinecite{wu2010}].
}
\label{fig:largespin}
\end{figure}

High symmetries (e.g. SU($N$) and Sp($N$)) are essential in high
energy physics, nevertheless, their applications in condensed matter
physics are often to provide \textcolor{red}{a} mathematical
tool of the large-$N$ expansion to handle strong correlations
\cite{affleck1985,arovas1988,affleck1988,read1990}.
On the other hand, cold atom physics has become a new frontier of
condensed matter physics for creating novel quantum states
of matter, particularly those uneasy to access in solids.

Many fermionic atoms possess large-hyperfine-spins.
We have been working on exploring new states of large-spin fermions
from the new perspective of high symmetries of SU($N$) and Sp($N$)
since 2003\cite{wu2003,chen2005,wu2005,wu2006,xu2008,wu2010a,hung2011,
gao2020}.
It works as a guiding principle to explore beautiful many-body physics,
providing a natural connection between cold atom physics and high energy physics.
It is amazing to see that physics at dramatically different energy scales
is deeply related.
Systematic studies have been performed in exploring high symmetry
effects, including the unification of competing orders\cite{wu2003,wu2005}
, novel quantum magnetism \cite{xu2008,hung2011}, and non-Abelian topological defects \cite{wu2010a}.

High-symmetry cold fermions have attracted considerable attentions from
various research groups in the cold atom community
\cite{gorshkov2010,hermele2009,xu2010,cazalilla2009,
controzzi2006,lecheminant2005,hattori2005,tu2006,tu2007,
ostlund2005,bartenstein2005,rapp2007,rapp2008}.
This direction has also become an active experiment focus:
Takahashi's group realized the SU(6) symmetric alkaline-earth
fermions of $^{173}$Yb \cite{taie2010,taie2012,sugawa2011,hara2011}.
Fallani's group studied the 1D systems of $^{173}$Yb
with tunable component numbers\cite{fallani2014}.
The 10-component $^{87}$Sr systems ($F=I=\frac{9}{2}$) have been
studied by Killian's group \cite{desalvo2010,mickelson2010},
Sengstock's group \cite{heinze2013,krauser2012},
and Ye's group \cite{bishof2011,bishof2011a,martin2012}, etc.
For non-technical introductions to the experimental progress,
please refer to Refs. \cite{wu2010,wu2012}.

A fundamental difference exists between large-spin cold fermion
systems and large-spin solid state systems
as shown in Fig. \ref{fig:largespin} \cite{wu2010}.
In solids, quantum magnetic fluctuations are suppressed in
the large-$S$ limit:
Hund's rule coupling aligns spins of several electrons into a large
spin, however, the intersite coupling is dominated by the exchange
of a single pair of electrons, hence, spin fluctuations scale as $1/S$ as
$S$ goes large.
In contrast, this restriction does not occur in cold atom systems
because each large-hyperfine-spin fermion moves as an entire object.
The exchange of a single pair of atoms completely flips the spin configuration.
The large number of spin components actually enhanced quantum
fluctuations.
and they are actually even stronger than the spin-$\frac{1}{2}$ case.
Hence, the large-spin physics of ultracold atoms is governed
by the large-$N$ physics of a high symmetry group
where $N=2S+1$.

An exact and generic hidden Sp(4), isomorphically SO(5), symmetry
is proved for hyperfine-spin-$\frac{3}{2}$ alkali and alkaline fermions
without fine-tuning \cite{wu2003,wu2005,wu2006,wu2010a}.
The candidate atoms for realizations include
$^{132}$Cs, $^9$Be,$^{135}$Ba, $^{137}$Ba, and $^{201}$Hg
\cite{}.
Yang's $\eta$-pairing pseudospin SU(2) symmetry can be generalized
to the spin-3/2 Hubbard model defined on a bipartite lattice
\cite{wu2003,wu2006}.
Such a system could exhibit an SO(7) symmetry which unifies
the singlet superconductivity and the spin-quadruple density-wave order
with the 7-vector representation.
The adjoint representation of SO(7) can unify the quintet
superconductivity, spin and spin-octupole density-wave
order, and charge-density-wave, which are in total 21-dimensional.

The large-spin fermions also exhibit similar physics to that
in quantum chromodynamics -- the multi-particle clustering
orderings.
With attractive interactions, Pauli's exclusion principle allows
$N$-fermions to form an SU(N) singlet state, a ``baryon-like''
multiple-fermion instability \cite{schlottmann1994,stepanenko1999,wu2005,lecheminant2005}.
For the super-exchange physics in the Mott-insulating states, if each
site is in the fundamental representation, it also needs $N$ sites to
form an SU(N) singlet \cite{chen2005,xu2008}.

How interaction effects scale with the component number $N$ is also
an interesting question.
For the SU($N$) Hubbard models, systematic quantum Monte Carlo (QMC)
simulations free of the sign problem have been performed for the 2D
square lattice \cite{wang2014,wang2019},
the square lattice with flux
\cite{zhou2016}, and the honeycomb lattice
\cite{xu2019}, and also in 1D \cite{xu2018a}.

\subsection{Unconventional magnetism and spontaneous spin-orbit ordering}
Spin-orbit coupling plays an important role in the research focus of
topological states of matter.
Conventionally, it is viewed as a single-particle property inherited from
the relativistic Dirac equation, not directly related to many-body physics
\cite{kittel1987}.
We have explored another possibility -- the spontaneous generation of spin-orbit coupling as a many-body effect based on Fermi surface instabilities of
the Pomeranchuk type \cite{pomeranchuk1959}.
This mechanism is essentially itinerant magnetic phase
transitions with unconventional symmetries (e.g. $p$-wave),
which is also magnetic multipolar orderings in
momentum space \cite{wu2004a,wu2007}.

\begin{figure} \centering
\includegraphics[width=0.3\linewidth]{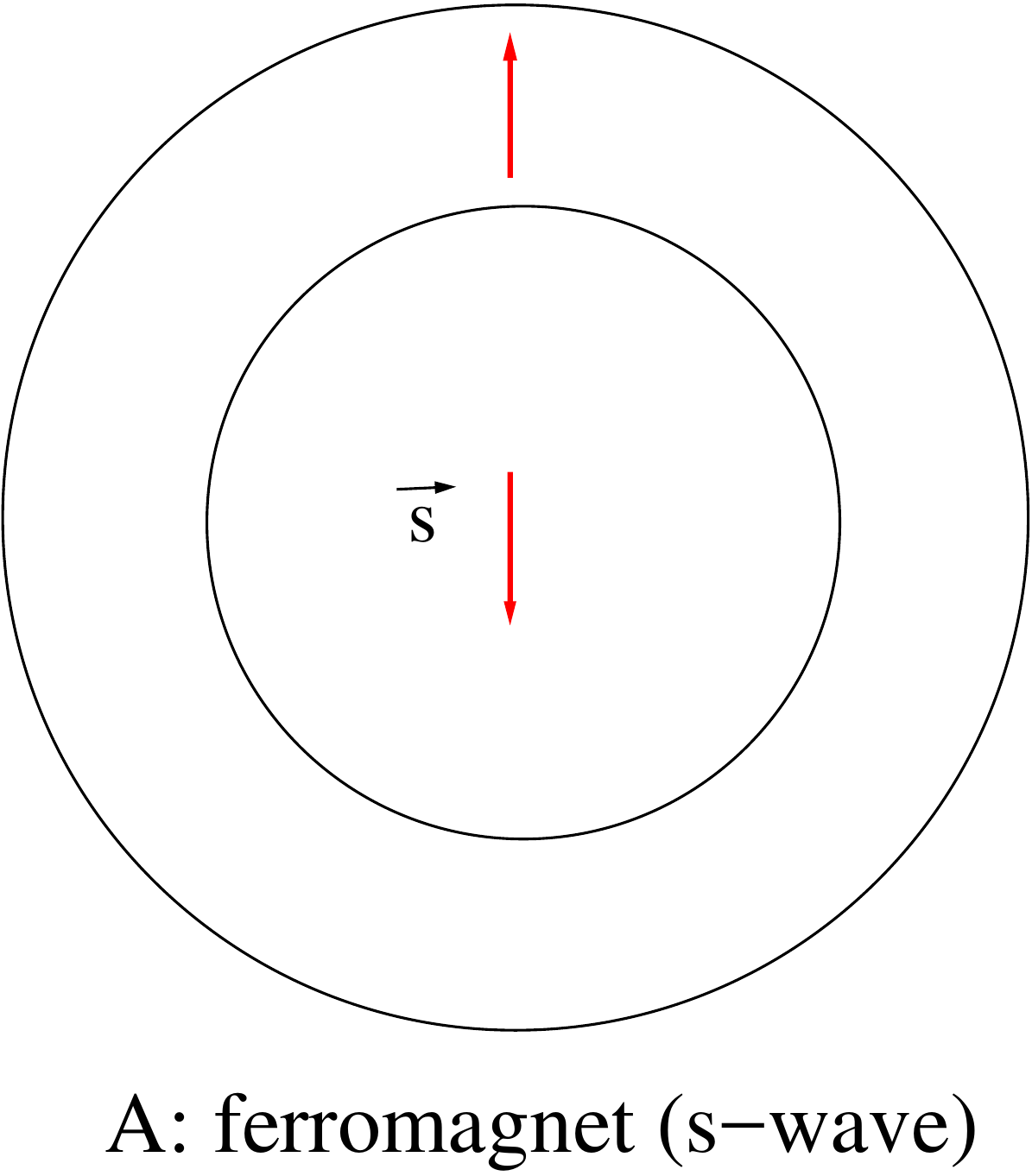}
\hspace{3mm}
\includegraphics[width=0.6\linewidth]{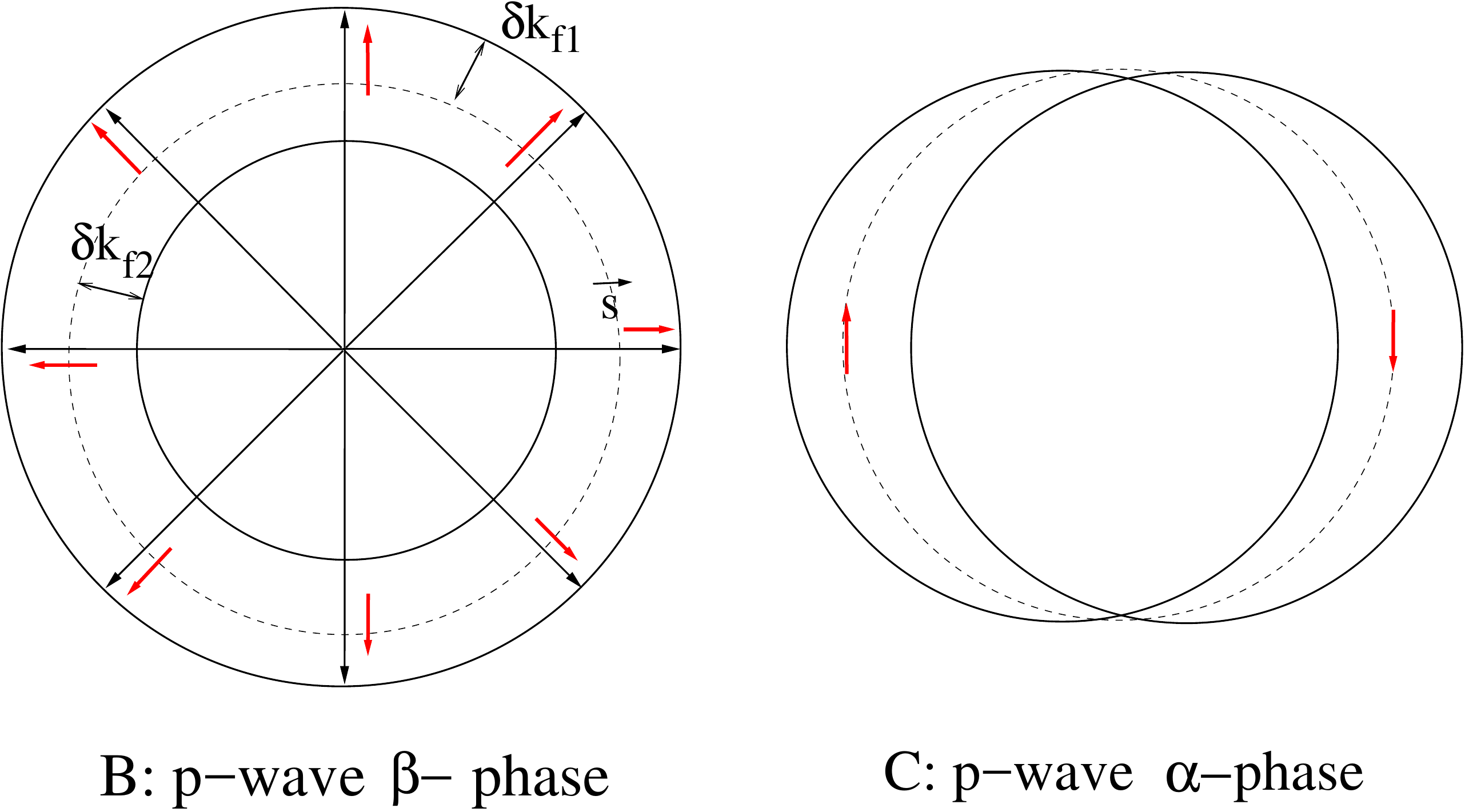}
\caption{
Fermi surface configurations of the ferromagnetic
phase (A) and the unconventional magnetic phases
in the $p$-wave channel ( the isotropic $\beta$-phase (B) and the
anisotropic $\alpha$-phase (C)).
The ferromagnetic state can be viewed as an $s$-wave type magnetism
since it does not break the orbital rotational symmetry.
The $p$-wave itinerant magnetism exhibits dipolar magnetic ordering
over the Fermi surface.
The $\beta$-phase breaks the relative spin-orbit symmetry spontaneously,
which is a particle-hole analogy to the superfluid $^3$He-B phase.
The anisotropic $\alpha$-phase is the analogy of the superfluid
$^3$He-A phase.
From Ref. \cite{wu2007}.
}
\label{fig:pwave_mag}
\end{figure}

In ferromagnetic metals, the rotational symmetry is broken in the spin channel.
However,  spin polarizes along the same direction around Fermi surfaces
independent of the direction of momentum, hence, the orbital rotational symmetry is unbroken as shown in Fig. \ref{fig:pwave_mag} ($a$).
This is similar to conventional $s$-wave superconductors whose gap
function phase keeps constant over the Fermi surface.
Therefore ferromagnetism can be viewed as the ``$s$-wave" magnetism.

As for superconductivity (fermion pairing superfluidity),
there exist unconventional pairing structures, including the $d$-wave
high $T_c$ cuprates \cite{xiang2022}
and the $p$-wave superfluid $^3$He
\cite{leggett1975}.
In analogy to unconventional superconductivity, we have
generalize ferromagnetism to the cases of unconventional
symmetries, in which spin no longer polarizes along a unique
direction but varies with momentum.
These unconventional magnetic states have close connections
to many directions in condensed matter physics, including unconventional superconductivity \cite{sigrist1991}, spin-orbit coupling and spintronics,
and electron liquid crystal states in strongly correlated systems
\cite{fradkin2010}.

The unconventional magnetism includes both isotropic and anisotropic cases,
as shown in Fig. \ref{fig:pwave_mag} ($b$) and ($c$), respectively.
They are dubbed the $\beta$ and $\alpha$-phases analogues to the
superfluid $^3$He B and A-phases, respectively.
The isotropic $\beta$-phases still exhibit circular, or, spherical
Fermi surfaces developing nontrivial spin-texture configurations in
momentum space, providing a mechanism for dynamic generation of
spin-orbit coupling independent of relativity.
The anisotropic $\alpha$-phases are electron liquid crystal states
with spin degree of freedom, exhibiting anisotropic Fermi surface distortions.
Both types of phases arise from the Pomeranchuk instability
of Fermi surfaces in the spin channel,
which include ferromagnetism as a special example.

The symmetry breaking pattern of the isotropic $\beta$-phase is subtle,
which breaks the relative spin-orbit symmetry \cite{leggett1975}.
In non-relativistic physics, spin is an internal degree of freedom,
i.e., the spin rotational symmetry $SO_S(3)$ is independent of
the orbital $SO_L(3)$.
The $\beta$-phase is invariant only if rotations in the two channels
are performed exactly in the same way.
In contrast, if there exists a difference between two rotations,
i.e., the relative spin-orbit rotation, the system indeed changes.
This symmetry breaking pattern is denoted as $[SO_L(3)\otimes SO_S(3)]/SO_{L+S}(3)$.
In other words, the total angular momentum $J=L+S$ in the $\beta$-phase
is conserved, but $L-S$ is not.

The concept of relative spin-orbit symmetry breaking was first
introduced by Leggett \cite{leggett1975} in the context of
superfluid $^3$He-B phase, whose Cooper pairing has a $p$-wave
and spin-triplet like structure, i.e. $L=S=1$.
The pair wavefunction in the B-phase is
\bea
\Psi_{pair}(\mathbf{r}_{12})= \sum_{i=x,y,z}f_{p_i}(\mathbf{r}_{12})
\chi_i,
\eea
where $f_{p_i}(\mathbf{r}_{12})$ describes the radial wavefunction
with the orbital symmetry of $p_i (i=x,y,z)$,
and $\chi_x=\frac{1}{\sqrt 2}\left(|\uparrow \uparrow\rangle +|\downarrow \downarrow\rangle\right)$,
$\chi_y=\frac{1}{\sqrt 2i}\left(|\uparrow \uparrow\rangle -|\downarrow \downarrow\rangle\right)$,
and
$\chi_z=\frac{1}{\sqrt 2}\left(|\uparrow \downarrow\rangle +|\downarrow \uparrow\rangle\right)
$.
The total angular momentum $J=L+S$ of Cooper pairs is zero,
and thus the pairing is isotropic.
Hence, the $\beta$-phase is the particle-hole channel analogy to the
$^3$He-B phase.

In Sect. \ref{sect:pomeranchuk}, we shall review  how spin-orbit
coupling can be dynamically generated without relativity but from
phase transitions, in a similar way to ferromagnetism.
We have also extended the Fermi-liquid theory to systems with spin-orbit coupling.

\subsection{Outline}
The remaining part of this article is organized as follows:
The concept of space-time group for dynamic systems \cite{}
is reviewed in Sec. \ref{sect:spacetime};
the high symmetry perspective of ultracold fermion physics
is reviewed in Sect. \ref{sect:spin32atom};
unconventional magnetism and spontaneous spin-orbit symmetry
breaking
is reviewed in Sect. \ref{sect:pomeranchuk}.
Conclusions are presented in Sect. \ref{sect:conclude}.

\section{Space-time group for dynamic systems}
\label{sect:spacetime}
The fundamental concept of crystal and band theory based on the Bloch
theorem lay the foundation of condensed matter physics \cite{kittel1987}.
In recent years, the study of dynamic systems such as the ``pump-prob"
systems becomes a new focus direction \cite{zhangxiang2018,gedik2013}.
The simplest dynamic systems exhibit space-time periodicity, and a
natural question is how to classify their symmetries by extending
the static crystalline symmetries.
There existed previously the framework of Floquet systems, i.e.,
systems under periodical driving.
However, in such a framework, the spacial and temporal
symmetries are decoupled, hence, it cannot be the generic case
\cite{rechtsman2013a,rudner2013,thakurathi2013a,vonKeyserlingk2016b,
gu2011,lindner2011,else2016,vonKeyserlingk2016a,potter2016a,roy2016,
nathan2015,flaschner2016}.

We construct a new framework, dubbed {\it space-time group},
to describe the general intertwined
space-time periodicities in $D+1$ dimensions, which include both the
static crystal and the Floquet crystal as special cases
\cite{xu2018}.
Compared to previously known space- and magnetic groups, space-time group
is augmented by ``time-screw'' rotation, ``time-glide'' reflection,
and ``time-shift" rotary-reflection, involving
fractional translations along the time direction.
We have classified that there are 13 space-time groups in 1+1D
and 275 space-time groups in 2+1D.

\subsection{Space-time unit cell and momentum-frequency Brillouin zone}

\begin{figure} \centering
\includegraphics[width=\linewidth]{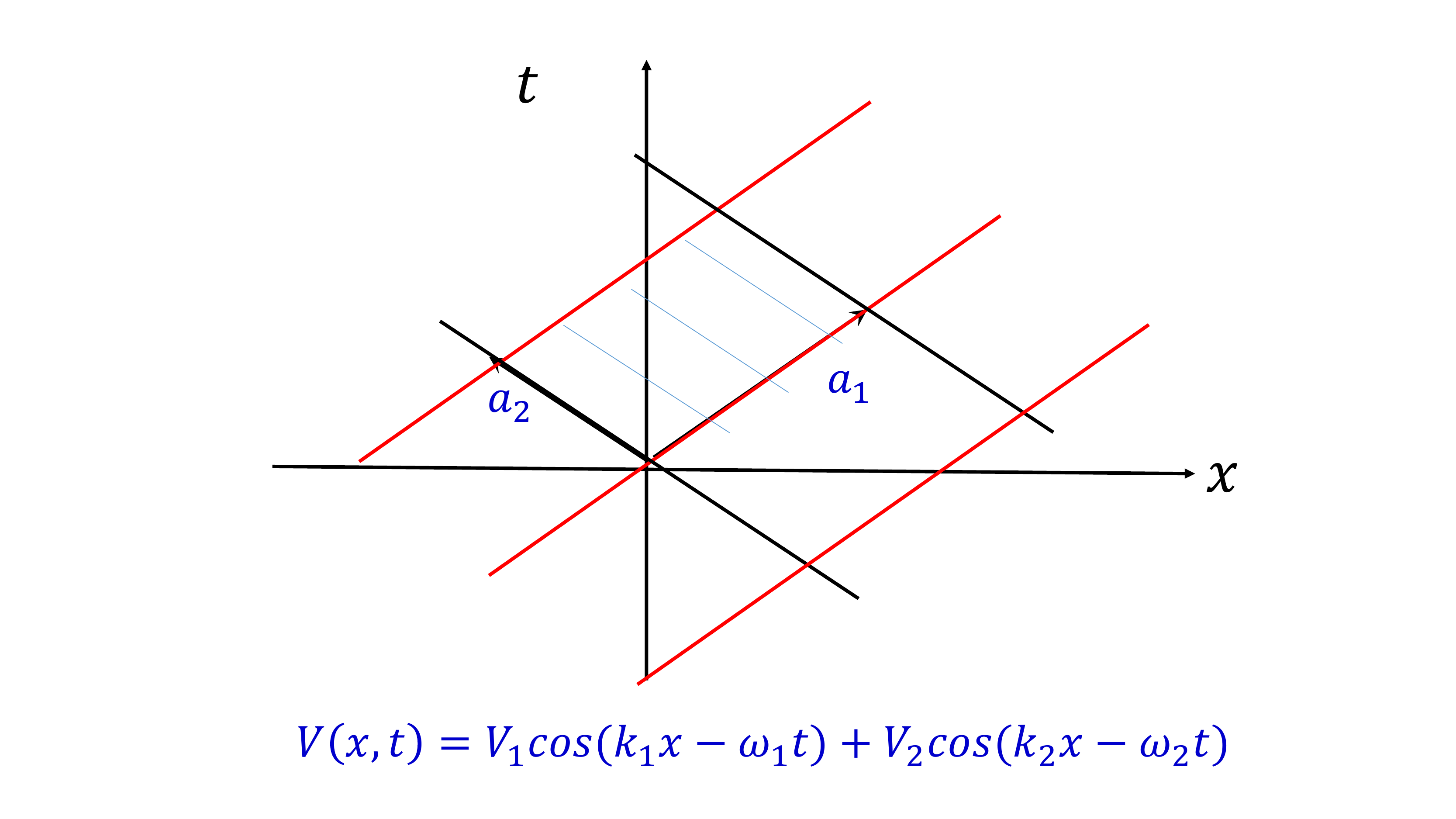}
\caption{The simplest space-time crystal in 1+1 D.
In the general case, the space-time unit cell is a parallelogram
which cannot be decomposed into a direct product between space and
time domains.
It exhibits neither translational nor time-translational symmetries,
but does possess the combined space-time translation symmetries.
}
\label{fig:1plus1D}
\end{figure}

Let us begin with a simplest example of space-time crystalline symmetry.
Consider a $1+1$ D system, whose time-dependent potential is the
superposition of two plane waves as plotted in Fig. \ref{fig:1plus1D},
\bea
V(x,t)=V_1\cos (k_1 x-\omega_1 t)+V_2 \cos (k_2 x-\omega_2 t).
\label{eq:1dspacetime}
\eea
The wavevectors $k_{1,2}$ and frequencies $\omega_{1,2}$
are supposed to be incommensurate.
If we fix a spacial position, say $x=0$, and look at the time-dependence of
$V(0,t)$, there is no temporal periodicity.
For Floquet problems, the time-evolution operator $U(T)$ of one period
is often used to map them into time-independent problems.
Clearly, here this method generally does not apply.
Similarly, if we take a snap shot at a fixed time, say $t=0$, $V(x,0)$
has no spatial periodicity either.
Hence, the ordinary Bloch theorem cannot straightforwardly be applied
here.

The periodicity only appears when we extend to space-time.
The unit cell is a space-time parallelogram, {\it not} a direct
product between space and time domains.
The unit vectors $\mathbf{a}_1$, $\mathbf{a}_2$ are space-time coupled,
\bea
\mathbf{a}_1=\left( \begin{array}{c}
\frac{2\pi\omega_2}{k_1 \omega_2 -k_2 \omega_1}\\
\frac{2\pi k_2}{k_1 \omega_2 -k_2 \omega_1}
\end{array}
\right),
\ \ \,
\mathbf{a}_2=\left( \begin{array}{c}
\frac{2\pi\omega_1}{k_1 \omega_2 -k_2 \omega_1}\\
\frac{2\pi k_1}{k_1 \omega_2 -k_2 \omega_1}
\end{array}
\right),
\eea
which define space-time coupled translation symmetries.
For the general case, a potential $V(\mathbf{r},t)$ exhibiting
the intertwined discrete $D+1$ dimensional space-time translational
symmetry satisfies
\bea
V(\mathbf{r}, t)=V(\mathbf{r}+\mathbf{u}^i, t+\tau^i),
\ \ \ i=1,2,...,D+1,
\label{eq:Ht}
\eea
where $(\mathbf {u}^i, \tau^i)=a^i$ is the primitive basis vector of
the space-time lattice.

We move to the reciprocal space and define reciprocal
lattice vectors, which can be done in a similar way to solid state physics.
The reciprocal lattice is spanned by the momentum-energy basis vectors
$b^i=(\mathbf{G^i}, \Omega^i)$ defined through
\bea
b^i\cdot a^j= \sum_{m=1}^D G^i_m u^j_m -\Omega^i \tau^j =2\pi \delta^{ij}.
\eea
This minus sign is due to quantum mechanics phase convention.
The $D+1$ dimensional momentum-energy Brillouin zone  may not
be a direct product between a momentum volume and frequency domain either.
The reciprocal lattice vectors contain both momentum and frequency components.

We emphasize that the above framework is already beyond that of Floquet.
Floquet systems only have one fundamental frequency, while,
in our case each reciprocal lattice vector has an independent frequency.
The $D+1$ dimensional space-time crystals can exhibit at most $D+1$
incommensurate frequencies.
Hence, they are related to certain types of quasi-crystals.

\subsection{The generalized Bloch-Floquet theorem}
For dynamic crystal systems with space-time periodicity, the Bloch
and Floquet theorems should be treated at equal footing.
Below they are combined and generalized.

Consider the time-dependent Schr\"odinger equation
$i\hbar\partial_t \psi(\mathbf{r}, t)=
H(\mathbf{r},t)\psi(\mathbf{r}, t)$.
Its solutions are denoted by the good quantum number of the (lattice)
momentum-energy vector $\kappa=(\mathbf{k},\omega)$, which is defined
modulate the reciprocal lattice vectors.
The Floquet-Bloch state labeled by $\kappa$ takes the form of
\bea
\psi_{\kappa,m}(\mathbf{r}, t)&=&e^{i (\mathbf{k}\cdot \mathbf{r}-
\omega t)}
u_m(\mathbf{r}, t),
\label{eq:gen_sol}
\eea
where $m$ marks different states sharing the common $\kappa$.
$u_m(\mathbf{r},t)$ is periodical in the space-time unit cell,
which is expanded as Fourier series only involving momentum-energy
reciprocal lattice vector as
\bea
u_m(\mathbf{r}, t)=\sum_{b} c_{m,b}
e^{i(\mathbf{G}\cdot \mathbf{r}-\Omega t)}
\eea
with $b=(\mathbf{G}, \Omega)$ taking all the momentum-energy reciprocal
lattice vectors.
The spectra $\omega_m$ can be solved through the secular equation,
\bea
\sum_{b^\prime}
\{ [\varepsilon_0(\mathbf{k+\mathbf{G}})-\Omega]
\delta_{b,b'}+V_{b-b'} \} c_{m,b^\prime} 
=\omega_m c_{m,b}, \ \ \, \ \ \,
\label{eq:eigen_nonint}
\eea
where $\varepsilon_0(\mathbf{k})$ is the free dispersion, and
$V_b$ is the momentum-energy Fourier component of the
space-time lattice potential $V(\mathbf{r},t)$.

The above procedure is very similar to the plane-wave expansion
method of the band theory in $D$-dimensions, in which the static
lattice potential only has Fourier components in momentum space.
The difference is that the effective dimensions become $D$+1,
since the reciprocal lattice vectors lie in the momentum-energy space
for space-time crystals.
Nevertheless, the Hilbert space of physical states remains the same
regardless of whether the potential is time-independent or not.
To reconcile this discrepancy, we notice the gauge-like redundancy
in the formalism based on and Eq. (\ref{eq:gen_sol}) and
Eq. (\ref{eq:eigen_nonint}).
The solutions in the sector labeled by $\kappa$ and those by
$\kappa+b$ are redundant since the same state in Eq. \ref{eq:gen_sol}
can also be expressed as
$\psi_{\kappa,m}(\mathbf{r}, t)=
e^{i [(\mathbf{k+G})\cdot \mathbf{r}-(\omega+\Omega) t]}
u_{m^\prime}(\mathbf{r}, t)$
with $u_{m^\prime}(\mathbf{r}, t)
=u_{m}(\mathbf{r}, t) e^{-i (\mathbf{G}\cdot \mathbf{r}-\Omega t)}$.

The dispersion based on Eq. (\ref{eq:eigen_nonint}) is generally
multiple-valued, represented by a $D$-dimensional surface in the
momentum-energy Brillouin zone which is a $D$+1 dimensional torus.
In the static case, the band dispersion only winds around the momentum
direction.
In space-time crystals, the winding patterns are richer.

Let us take the 1+1D case as a simple example.
The dispersion relation $\omega(k)$ forms closed loops in
the 2D toroidal momentum-energy Brillouin zone, each of which
is characterized by a pair of winding numbers $\mathbf{w}=(w_1,w_2)$
with $w_{1,2}$ integers.
In general, nearly all patterns $\mathbf{w}=(w_1,w_2)$ are possible
except for one constraint explained as follows.
Consider a weak lattice potential such that it can be treated as a
perturbation.
The free dispersion curve $\varepsilon(k)$ is folded into the
momentum-energy Brillouin zone with crossings.
Two states at a crossing point are connected by a reciprocal vector $b$
before folding.
The crossing is lifted if the momentum-energy Fourier component of
$V_b$ is nonzero.
The total number of states at each $k$ is independent of the
potential strength, hence crossing can only be split along the
$\omega$-direction by opening a gap of $2|V_b|$, and
$d\omega/dk$ is always finite.
Hence, the contractible loops with the winding numbers $(0,0)$
are unallowed.

Nevertheless, the winding number pattern could be constrained by spectral
degeneracies protected by symmetries.
For example, consider a magnetic group transformation applied to a 1+1 D
space-time crystal, whose unit cell is a direct product between spatial and
temporal periods $a$ and $T$, respectively.
Define the glide time-reversal operation $g_t(x,t) =(x+\frac{a}{2}, -t)$.
It operates on the Hamiltonian as
\bea
g_t^{-1} H(x,t) g_t = H^*(x+\frac{a}{2}, -t).
\eea
The corresponding transformation $M_{g_t}$ on the Bloch-Floquet
wavefunction $\psi_{\kappa}(x,t)$ of Eq. \ref{eq:gen_sol} is
anti-unitary defined as
\bea
M_{g_t}\psi_{\kappa}=
\psi^*_\kappa(g_t^{-1}(x,t)).
\eea
Consider two special lines of the momentum-energy Brillouin zone
with $\kappa_x=0$ and $\kappa_x=\pi/a$.
$M_{g_t}^2=1$ for states with $\kappa_x=0$, but it becomes a Kramers symmetry
$M_{g_t}^2=-1$ for those of $\kappa_x=\pi/a$,
\bea
M_{g_t}^2 \psi_{\kappa}= \psi_{\kappa} (x-a,t)=
e^{-i\kappa_x a}\psi_{\kappa}=-\psi_{\kappa}.
\eea
Then the crossing at $\kappa_x=\pi/a$ cannot be avoided.
Hence, the dispersion curve must wind along the momentum direction
even times, while its windings along the energy direction
cancel.
The winding number is constrained to $\mathbf{w}=(2n, 0)$.

\subsection{Definition of space-time group}

Now we are ready to formally define {\it space-time} group in analogous
to space group describing the static crystalline symmetry.
It is the discrete subgroup between the direct product of the Euclidean
group of $D$ spatial dimensions and that along the time-direction
$E_D \otimes E_1$.
In general, space-time group {\it cannot} be factorized as a direct product
between space and temporal subgroup groups.

In terms of coordinates, a space-time group operation $\Gamma$ is defined
as
\bea
\Gamma(\mathbf{r},t)=(R\mathbf{r} +\mathbf{u}, st+\tau),
\eea
where $R$ here is a point-group operation, including rotation, reflection,
rotary reflection.
$\mathbf{u}$ is a translation.
If $\mathbf{u}$ is not a symmetry by itself, then it is non-symmorphic.
Combining $R$ and $\mathbf{u}$, they span space-groups.
Further including $s=-1$, they span magnetic space-groups,
which are used to describe the symmetry properties of magnetic systems.
The last term of $\tau$ is time translation.
Combining all the point group operations, time-reversal, spacial
and temporal translations, the algebra is closed.
This new symmetry group is dubbed {\it space-time} group.

If the time translation $\tau$ itself is not a symmetry, it should be
combined with spatial transformations to form space-time
non-symmorphic symmetries as shown in Fig. \ref{fig:nonsym}.
In 1+1 D, the only available operation to combine is spatial reflection.
This is the dynamic symmetry of a see-saw [Fig. \ref{fig:nonsym} ($a$)].
A see-saw does not possess a static reflection symmetry, but it is
invariant by performing reflection and time translation at half a
period.
This symmetry is the analogy of the glide-reflection symmetry of
space group, dubbed {\it time-glide} reflection symmetry.
In 2+1D, a new possibility is to combine $\tau$ with spatial
rotation to form {\it time-screw} rotation, which can be intuitively
understood as the dynamic symmetry of a clock
[Fig. \ref{fig:nonsym} ($b$)].
Consider a simplified clock with only one pointer rotating.
It does not exhibit the rotational symmetry due to the pointer,
but a rotation combined with time translation can leave the clock
invariant.
This is the analogy of screw rotation of space group, dubbed
``time-screw" rotation.

There also exist new possibilities that nonsymmorphic space-time
symmetries have no analogies in static space groups.
In 3+1D, a fractional time translation $\tau$ can be combined with
the rotary reflection operation $R$, dubbed {\it time-shift rotary
reflection} with an example depicted in Fig. \ref{fig:nonsym} ($c$).
(Rotary reflection $R$ is a rotation followed by a reflection whose
$\det R=-1$ with eigenvalues $\{-1, e^{\pm i\theta}\}$
and $\theta\neq 0$.
Another possibility is a space-time translation $(\mathbf{u},\tau)$
followed by a point group operation $R$.
In other words, it is non-symmorphic space group operation
followed by a fractional time translation $\tau$.

Naturally, quantum mechanical wavefunctions can be employed to span
representations of space-time group.
A spacial care needs to be taken is that the representation is
anti-unitary when $s=-1$, i.e., time-reversal is involved.
The operation of $\Gamma$ on the Hamiltonian
is defined as
\bea
\Gamma^{-1} H(\mathbf{r},t) \Gamma =
\Big\{
\begin{array}{c}
H(\Gamma(\mathbf{r},t))  \mbox{~~ for~~}  s=1,\\
H^*(\Gamma (\mathbf{r},t)) \mbox{~~for ~~} s=-1.
\end{array}
\eea
Correspondingly, the transformation $M_\Gamma$ on the Bloch-Floquet
wavefunctions $\psi_\kappa(\mathbf{r},t)$ is
\bea
M_\Gamma \psi_\kappa=
\Big\{
\begin{array}{c}
\psi_\kappa(\Gamma^{-1}(\mathbf{r},t)) \mbox{~~ for~~}  s=1, \\
\psi_\kappa^*(\Gamma^{-1}(\mathbf{r},t))  \mbox{~~for ~~} s=-1.
\end{array}
\eea

\subsection{Classifications of space-time group}
\begin{figure} \centering
\includegraphics[width=\linewidth]{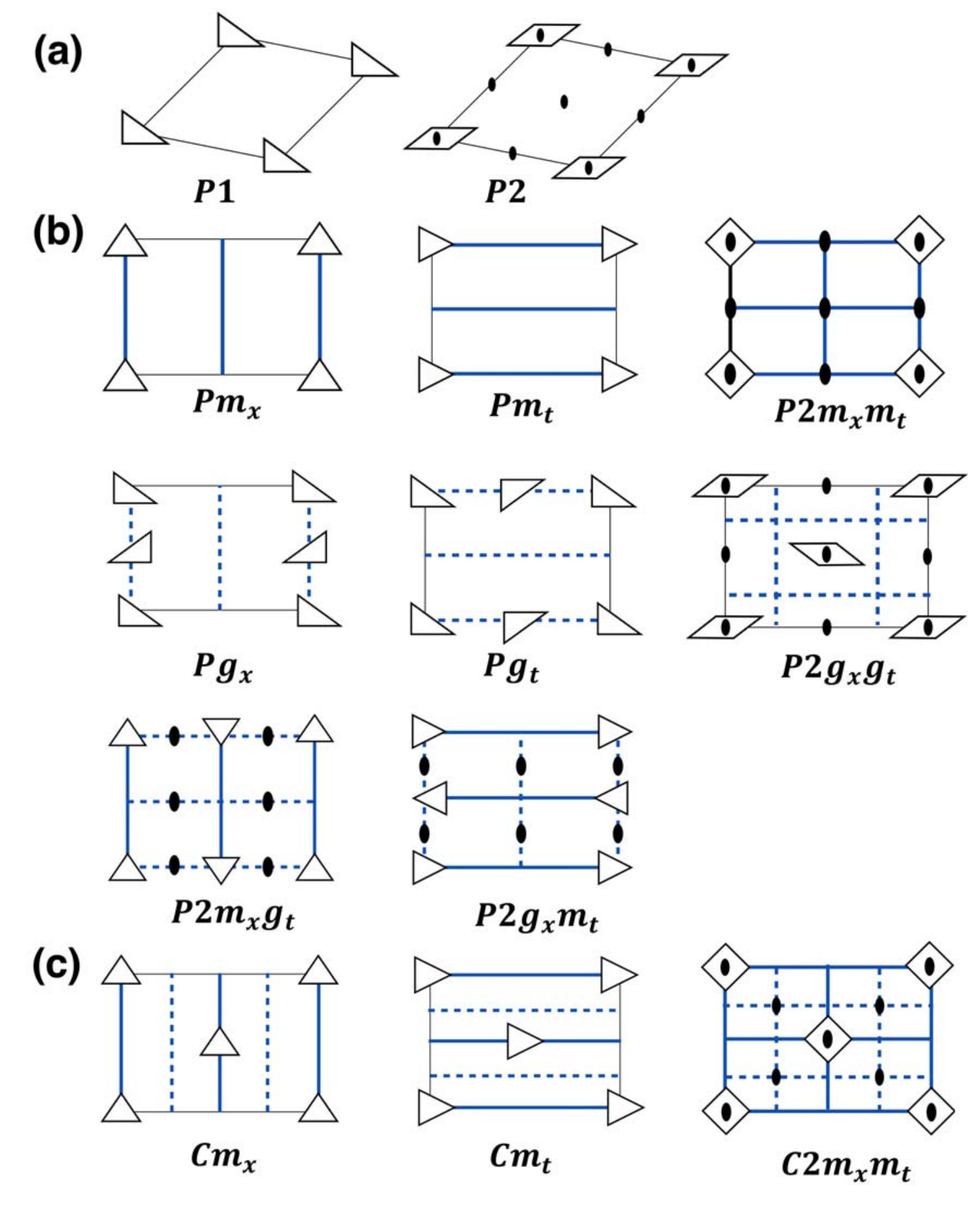}
\caption{Crystal configurations of 13 space-time groups in 1+1D.
The solid oval marks the 2-fold space-time axis, and the parallelogram
means the 2-fold axis without reflection symmetries.
The thick solid and dashed lines represent reflection and
glide-reflection axes, respectively.
Configurations of triangles and the diamond denote the local
symmetries under reflections.
($a$) The oblique Bravais lattice. 2 space-time crystals with (P1)
and without (P2) 2-fold axes in this crystal system.
They generally do not possess a Floquet period, but exhibit
space-time mixed translation symmetries.
($b$) The primitive orthorhombic Bravais lattice with 8 space-time
crystals. They are denoted as $Pm_x$, $Pm_t$, $P2m_xm_t$ $Pg_x$, $Pg_t$, $P_2g_xg_t$,
$P2m_xg_t$, and $P2g_xm_t$ according to their reflection and
glide reflection symmetries.
($c$) The centered orthorhombic Bravais lattice with 2 space-time
crystals denoted as $Cm_x$, $Cm_t$, $C2m_xm_t$.
Two unit cells are plotted to show crystalline symmetries in this
class.
Among 13 space-time crystals in 1+1D, 5 of them
($Pg_x$, $Pg_t$, $P_2g_xg_t$,
$P2m_xg_t$, and $P2g_xm_t$) are non-symmorphic, and the
other 8 are symmorphic.
From Ref. [\onlinecite{xu2018}].}
\label{fig:sptm_group}
\end{figure}

The 2D space groups are particularly intuitive with a popular
name of wallpaper groups.
There exist 17 wallpaper groups corresponding to different types
of planar patterns.
Actually, all these patterns have been already used for ornaments
since ancient times \cite{weyl}.

We classify space-time crystals based on their space-time group
symmetry structures.
A natural starting point is to classify 1+1D space-time groups,
which is an analogous problem to the 2D wallpaper groups.
Due to the non-equivalence between spatial and temporal directions
of the non-relativistic Sch\"odinger equations, we cannot really
rotate space and time into each other.
Hence, only the 2-fold space-time rotation is allowed, i.e.
$(x\to -x, t\to -t)$ and 3,4,6-fold rotations are not,
which eliminates quite a few possibilities.
On the other hand, the non-equivalence between space and time
also brings richness.
Spatial reflection $m_x$ and time-reversal $m_t$
are of a different nature.
The former is a unitary operation, and the latter is anti-unitary.
Similarly, as for two glide operations, a time-glide with a spacial
reflection $g_x$ is different from a space-glide with a
time-reversal $g_t$.

Taking the above considerations into account, in total there are 13 types
of space-time crystals as shown in Fig. \ref{fig:sptm_group}.
It is obvious that only two space-time crystal systems are allowed
in 1+1D -- oblique and orthorhombic.
No square and hexagonal space-time crystals exist.
Considering the Bravais lattices, the oblique case is monoclinic,
and the orthorhombic case has two possibilities:
the primitive one and the centered one.
The oblique Bravais lattice generates 2 types of space-time crystals,
the orthorhombic one generate 8, and the centered orthorhombic one
generates 3, as shown in Fig. \ref{fig:sptm_group}($a$, $b$, $c$),
respectively.
For the centered orthorhombic lattices, actually their primitive
cells are space-time rhombohedral.
To explicitly demonstrate the full symmetries,
two unit cells are plotted Fig. \ref{fig:sptm_group}($c$).
There are 5 space-time groups that are non-symmorphic, and
all of them belong to the orthorhombic Bravais lattice.
And the rest 8 are symmorphic.

As a concrete example, look at a crystal configuration depicted
in Fig. \ref{fig:sptm_group}($b$), the $Pg_x$ one.
This is the symmetry group of an array of see-saws,
which is actually non-symporphic.
Such a system does have a Floquet period, but it is insufficient
to show its complete space-time symmetries.
In contrast, the space-time group goes inside the Floquet period
and extracts all the space-time symmetries.
In the case of $Pg_x$, it shows the symmetry between the first and
second halves of the Floquet period.

The classifications of the space-time groups in higher dimensions
are generally complicated.
The simple method of enumeration is cumbersome.
We have classified 2+1D space-time groups based on the method
of group cohomology, and the details will be presented elsewhere.
This is an analogous problem to the 3D space groups.
There exist 275 space-time groups with 72 of them
symmorphic and the rest 203 non-symmorphic.

There are still 7 crystalline systems and 14 Bravais lattices for
2+1D space-time groups, whose numbers are the same as in the 3D case,
but the situation is different.
The cubic space-time crystal does not exist in 2+1D since we cannot compare
the length along the time direction with that in the $ab$-plane, i.e.,
there does not exist a universal velocity like the light velocity
in non-relativistic physics.
Instead, there exist two different types of monoclinic space-time
crystals.
``Monoclinic" here means that the $c$-axis is perpendicular to
the $ab$-plane,
but the $a$ and $b$ axes are non-perpendicular to each other.
The $c$-direction could be chosen as time, or, one of the spatial
direction corresponding to the $T$- and $R$- monoclinic space-time
crystals.
Other crystal systems such as tetragonal, orthorhombic, trigonal,
hexagonal, and triclinic ones can be similarly constructed.

\subsection{Spectral degeneracy protected by nonsymmorphic space-time
symmetry}
For static crystals, it has been extensively studied that non-symmorphic
space-group symmetries can protect spectral degeneracies and enrich
topological phases \cite{fu2011,parameswaran2013, young2015,
wang2016a, watanabe2016}.
In this subsection, we show that the intertwined space-time
nonsymmorhic symmetries also protect non-trivial spectral
degeneracies of the driven system.

We express a general space-time group element as
\bea
g=T_\mathbf{r}(\mathbf{u})T_t(\tau) R m_t^s,
\eea
where $T_\mathbf{r}(\mathbf{u})$ and $T_t(\tau)$ are
spatial and temporal translations, respectively.
$R$ is a point group operator; $m_t$ is the time-reversal
operation with $s=1$ or $0$ determining whether $g$
is anti-unitary or not, respectively.
If two operations $g_1$ and $g_2$ belong to the little group of a
high symmetry point $\kappa=(\mathbf{k},\omega)$, whose
point group operations commute, then
\bea
g_1g_2&=& T g_2g_1
\eea
with $T$ a translation of integer unit vectors.
$T$ is decomposed into spatial and temporal parts as
$T=T_\mathbf{r}(\tilde{\mathbf{u}})T_t(\tilde{t})$,
where
\bea
\mathbf{\tilde{u}}&=&(\mathbf{I}-R_2)\mathbf{u_1}
-(\mathbf{I}-R_1)\mathbf{u_2},   \ \ \,
\tilde{t}=2 s_2 t_1-2 s_1 t_2.
\label{SM:degeneracy}
\eea
Below we review degeneracies protected by this symmetry.

The representation matrices $M_{g_{1,2}}$ acting on
Floquet-Bloch wavefunctions with $\kappa$ satisfy
\bea
M_{g_1}M_{g_2}=
e^{i \mathbf{k}\cdot\tilde{\mathbf{u}}-i
\omega \tilde{t}}M_{g_2} M_{g_1}.
\label{eq:commute}
\eea
The following three cases need to be examined depending on whether
$g_{1,2}$ are unitary or anti-unitary.

First, if neither $g_1$ nor $g_2$ reverses the direction of time.
In this case, $\tilde{t}=0$, and then the phase factor
in Eq. (\ref{eq:commute}) does not involve time.
If $\mathbf{k} \cdot \mathbf{u}=2\pi p/q$ with $p$ and $q$ coprime,
then the Bloch-Floquet wavefunctions exhibit a $q$-fold
degeneracy at $\kappa=(\mathbf{k},\omega)$
proved as follows.
Since $g_1$ belongs to the little group, a Bloch-Floquet eigenstate
$\psi_\kappa$ of quantum number $\kappa$ can be chosen as an eigenstate
of $M_{g_1}$ satisfying $M_{g_1} \psi_{\kappa}=\mu \psi_{\kappa}$.
Then consider the sequence of
\bea
\psi_{\kappa}, ~M_{g_2}\psi_{\kappa},
~M_{g_2}^2\psi_{\kappa}, ~...., ~M_{g_2}^{q-1}\psi_{\kappa},
\eea
all of which share the same $\kappa$ since $g_2$ belongs to the
little group of $\kappa$.
Moreover, they are also $g_1$'s eigenstates exhibiting a set of different
eigenvalues as
\bea
\eta, \mu \eta, \mu\eta^2, ..., \mu \eta^{q-1}
\eea
with $\eta=e^{i2\pi p/q}$.
Then they are orthogonal to each other spanning a $q$-fold degeneracy.

Second, we consider the case that only one of $g_{1,2}$ involves
time-reversal.
Without loss of generality,
$M_{g_1}$ is assumed to be unitary while $M_{g_2}$ is anti-unitary.
Then the prefactor in Eq. (\ref{eq:commute})
exhibits frequency dependence.
Again since $g_1$ is in the little group, the Floquet-Bloch eigen
state $\psi_\kappa$ can be chosen as an $M_{g_1}$ eigenstate,
expressed as
\bea
M_{g_1} \psi_\kappa=e^{i \mathbf{k}\cdot \mathbf{u_1}-i \omega t_1}
e^{i\theta} \psi_\kappa,
\label{eq:phase1}
\eea
in which $\theta$ is extracted to be \textit{only} dependent on the
point group operation $R_1$.
Based on Eq. (\ref{eq:commute}), $M_{g_2} \psi_\kappa$
is also an eigenstate of $g_1$ as
\bea
M_{g_1} \left( M_{g_2} \psi_\kappa\right)=e^{i\mathbf{k}\cdot (-\mathbf{u_1}+\tilde{\mathbf{u}})+i \omega (-\tilde t +t_1)}e^{-i\theta}
\left( M_{g_2}  \psi_\kappa \right).
\label{eq:phase2}
\eea
Plugging in $\tilde{t}=2t_1$, then the frequency dependence in the phase factor
in Eq. (\ref{eq:phase2}) disappears.
We conclude that if $e^{i \mathbf{k}\cdot (2\mathbf{u}_1-\tilde{\mathbf{u}})+2i\theta}\neq1$,
then two phases in Eqs. (\ref{eq:phase1}) and (\ref{eq:phase2})
do not equal.
Hence, the degeneracy is protected.
Nevertheless, further applying high powers of $M_{g2}$ does
not bring new phases.

The last case is when both $g_1$ and $g_2$ flip the time-direction,
i.e., both $M_{g_1}$ and $M_{g_2}$ are anti-unitary.
By defining $g=g_1g_2$ which is unitary again and combining $g$
and $g_2$, we have
\bea
g g_2 =T_{\mathbf r} (\mathbf{u}) T_t (\tau) g_2 g ,
\eea
which is reduced to case 2.

We emphasize that in none of the above three cases, the degeneracy condition
depends on the frequency component of $\kappa$.
This is expected since one can always shift the frequency of the spectrum
by adding a constant to the Hamiltonian.

\subsection{Discussions}
So far the concept of {\it space-time group} has received considerable
attentions \cite{milton2019,rudner2020,harper2020,guo2021,yu2021,giergiel2019,
peng2019,kleiner2021,gao2021,muenchinger2022}.
We expect it would serve as a
guiding principle for quantum dynamic studies, in analogy to the
role of space group to static crystalline symmetries \cite{lax2012}.
The classification of space-time group in 3+1D would be of fundamental
importance if it is completed successfully, which is currently
under investigation.

Actually the lattices in solids are dynamic, and the quantized
lattice vibrations are phonons.
However, phonons are typically thermally driven and incoherent.
If a certain type of phonon mode is coherently excited, say,
optically, or, by other pumping methods, it cannot be treated
perturbatively \cite{zhangxiang2018,zhangniu2015}.
Instead, the time-dependent motions of lattice ions should be treated at
the zeroth order, i.e., we should include them in the time-dependent
lattice potential of the Schr\"odinger equation.
In artificial lattices, such as the phononic, photonic crystals, and
optical lattice for ultracold atoms, lattice potentials could be
manipulated on purpose \cite{anderson2017,
kleiner2021,milton2019},
In these cases, space-time group should replace space group
as the symmetry guidance of quantum dynamics.

Certainly, the semi-classic transport in dynamic crystals is
of importance.
When the periodicity of lattice potential is weakly broken by
slowly varying external fields both spatially and temporally,
semi-classic equations of motion for a quantum particle
could be developed \cite{kittel1987}.
We should distinguish two different types of dynamics:
the fast changing periodical lattice potential which should be taken by the
band structure calculation, and the slowly changing external
field which should be treated in an adiabatic way.
A challenging problem is how to generalize the Berry curvature
formalism to the dynamic version and incorporate it into
equations of motion \cite{xiao2010}.
The work in this direction would provide a general framework
for further studying topological properties in dynamic systems
\cite{chiu2016}.

Another direction to explore is the connection to the research
of time crystal
\cite{Kyp2021,shapere2012,wilczek2012,bruno2013,watanabe2015,
sacha2015,else2016,yao2017,khemani2016,google2021,randall2021}.
The current study of time crystal is concentrated on the
spontaneous breaking of the discrete time-translational symmetries,
which is a profound interaction effect.
Nevertheless, the symmetry breaking pattern typically is just
the Floquet type.
It would be interesting to combine these two directions together,
for example, to consider how to spontaneously generate dynamic crystals
with non-symmorphic space-time symmetries.
More philosophically, we could ask the problem of discrete subgroups
of different types of space-time symmetries, including Galilean,
Poincar\'e, anti-de Sitter symmetries, etc.


\section{High symmetry perspective to large-spin cold fermion systems}
\label{sect:spin32atom}
The study of ultracold atom systems has become a new frontier for
condensed matter physics as a way of creating novel quantum states
of matter.
We have proposed a new perspective of high symmetries (e.g. SU($N$)
and Sp($N$)) since 2003 to study the alkali and alkaline-earth fermion
systems \cite{wu2003,chen2005,wu2005,wu2006,xu2008,wu2010a,hung2011,
gao2020}, where $N$ is the fermion component number and
typically even.
It is exciting to explore, in atomic systems, complex and
beautiful many-body physics difficult to realize in usual solids
\cite{wu2010,wu2012}.
It also significantly enriches the physics of large-$N$ quantum
magnetism by providing a realistic system.


\subsection{The generic SO(5) symmetry of spin-$\frac{3}{2}$ cold
fermions}
In this subsection, we review the proof of an exact and generic hidden
symmetry of Sp(4), or, isomorphically SO(5) symmetry in spin-$\frac{3}{2}$
fermion systems (e.g.,$^{132}$Cs, $^9$Be, $^{135}$Ba, $^{137}$Ba, $^{201}$Hg)
\cite{wu2003,wu2006}.
It plays the role of SU(2) in electron systems since its
exactness is independent of dimensionality, lattice geometry, and
particle filling.
Such a high symmetry without fine-tuning is rare, which can be used
as a guiding principle for exploring novel quantum phases.

Sp(4) and SO(5) share the same Lie algebra.
Rigorously speaking, Sp(4) has spinor representations while SO(5) has not.
Sp(4) is the double covering group of SO(5),
and the relation between them is the same as that between SU(2) and SO(3).
For simplicity, we will use Sp(4) and SO(5) interchangeably
neglecting their minor difference.

Let us begin with the standard $s$-wave scattering interactions
of spin-$\frac{3}{2}$ fermionic atoms \cite{ho1999,yip1999}.
The total spin of atom is often called ``hyperfine spin"
including contributions from the nuclear spin, electron spin and
electron orbital angular momentum.
Below we follow the convention of atomic physics to use $F$
to denote atom's hyperfine spin.
For simplicity, spin and
hyperfine spin are used interchangeably.

Since the orbital wavefunction is symmetric in the $s$-wave channel,
the total spin wavefunction of two fermions is constrained by
Pauli's exclusion principle to be antisymmetric, which must be
either singlet or quintet.
The corresponding interaction parameters are denoted
$g_0$ and $g_2$, respectively.
The Hamiltonian reads,
\bea
H&=& \int d^d{\vec r} ~ \Big\{ \sum_{\alpha=\pm 3/2, \pm 1/2}
 \psi^\dagger_\alpha({\vec r})
\big ({-\hbar^2\over 2m}\nabla^2-\mu\big) \psi_\alpha({\vec r})
\nonumber \\
&+& g_0 P_{0,0}^\dagger({\vec r}) P_{0,0}({\vec r})
+
g_2 \sum_{m=\pm2,\pm1,0} P_{2,m}^\dagger({\vec r})
P_{2,m}({\vec r})\Big \}, \nn
\\
\label{eq:swave32}
\eea
with $d$ the space dimension, $\mu$ the chemical potential,
and $P^\dagger_{0,0}, P^\dagger_{2,m}$ the singlet and quintet
pairing operators defined through the Clebsh-Gordan
coefficient for two indistinguishable particles as
\bea
P_{F,m}^\dagger(\vec r)=\sum_{\alpha\beta} \avg{\frac{3}{2}
\frac{3}{2};F,m|\frac{3}{2}\frac{3}{2}\alpha\beta}
\psi^\dagger_\alpha(\vec r) \psi^\dagger_\beta(\vec r),
\eea
where $F=0,2$ and $m=-F,-F+1, ...,  F$.
Its lattice version is the spin-$\frac{3}{2}$ Hubbard model,
\bea
H&=&-t \sum_{\langle ij \rangle, \sigma} (\psi^{\dag}_{i \sigma}
\psi_{j \sigma}+h.c.)-\mu \sum_{i \sigma}\psi^{\dag}_{i \sigma}
\psi_{i \sigma} \nn
\\
&+&U_0 \sum_i P^{\dag}_{0}(i)P_{0}(i)
+U_2 \sum_{i,-2\le m\le 2}P^{\dag}_{2m}(i)P_{2m}(i),
\nn \\
\label{eq:hubbard32}
\eea
where $t$ is the hopping integral, $U_{0,2}$ are the onsite
Hubbard interaction parameters in the singlet and quintet
channels, respectively.

So far, the perspective in Eqs. (\ref{eq:swave32}) and
(\ref{eq:hubbard32}) is the usual spin SU(2) symmetry.
The 4-component spinor, singlet and quintet channels correspond to
the spin quantum numbers $\frac{3}{2}$, 0, and 2, respectively.
Below we will show that this degeneracy pattern equally well fits
in a high symmetry group of Sp(4), or, isomorphically, SO(5),
which provides a whole new perspective in spin-$\frac{3}{2}$
fermion systems.

For this purpose, we construct the Sp(4) algebra by extending
the typical charge and spin sectors.
For spin-$\frac{1}{2}$ systems, charge and spin form a complete
set for the particle-hole (p-h) channel observables.
However, they are incomplete
since there are $4^2=16$ bilinears in spin-$\frac{3}{2}$ systems,
\bea
M^I=\psi^\dagger_{i,\alpha} M^I_{\alpha\beta} \psi_{i,\beta}
\ \ \, (I=1\sim 16).
\eea
To systematically decompose the 16 matrix kernels of $M^I_{\alpha\beta}$,
high rank spin tensors are employed,
\bea \label{s32_algebra}
&\mbox{particle number:}& ~~I; \nonumber \\
&\mbox{spin:} & ~~ F^i, \ \ \hspace{20mm} i=1,2,3; \nonumber \\
&\mbox{spin quadrupole:} & ~~\xi^a_{ij} F_i F_j, \hspace{15mm} a=1,..,5; \nonumber \\
&\mbox{spin octupole:} & ~~ \xi^L_{ijk} F_i F_j F_k, \hspace{9mm} L=1,..,7,
\eea
where $\xi$'s are the typical fully symmetric, traceless tensors
converting 3-vector into spherical tensors.

The five spin quadrupole matrices are denoted $\Gamma^a=\xi^a_{ij} F_i F_j$,
which remarkably anticommute with each other forming a basis of the
Dirac $\Gamma$ matrices satisfying
\bea
\{\Gamma_a, \Gamma_b\}=2\delta_{ab}.
\eea
Explicitly, they are
\bea
\Gamma^1&=&\left (
\begin{array} {cc}
0 & -i I\\
i I& 0
\end{array} \right) , \ \ \
\Gamma^{2,3,4}=\left ( \begin{array}{cc}
{\vec \sigma}& 0\\
0& {-\vec \sigma} \end{array}\right), \nn \\
\Gamma^5&=&\left( \begin{array} {cc}
0& I \\
I & 0 \end{array} \right ),
\label{eq:gamma}
\eea
where $I$ and
$\vec{\sigma}$ are the 2$\times$ 2 unit and  Pauli matrices,
respectively.
They are explicitly expressed by the spin matrices as
\bea
\Gamma_1&=& \frac{1}{\sqrt 3} ( F_x F_y +F_y F_x ), \ \ \
\Gamma_2= \frac{1}{\sqrt 3} ( F_z F_x +F_x F_z ),
\nn \\
\Gamma_3&=& \frac{1}{\sqrt 3} ( F_z F_y +F_y F_z ), \ \ \
\Gamma_4= F_z^2-\frac{5}{4},
\nn \\
\Gamma_5&=& \frac{1}{\sqrt 3} (F_x^2 -F_y^2 ).
\eea
Moreover, the 3 spin and 7 spin octupole matrices together can be
organized into 10 commutators of $\Gamma$-matrices defined as
\bea
\Gamma^{ab}=
-{i\over 2} [ \Gamma^a, \Gamma^b] \ \ \ (1\le a,b\le5).
\eea

Consequently, the 16 particle-hole channel bilinear operators
are classified according to their properties under
the Sp(4) transformations as scalar, vector, and
anti-symmetric tensors (generators) as
\bea
n({\vec  r})&=& \psi^\dagger_\alpha({\vec r})
\psi_\alpha({\vec r}), \ \ \
n_a({\vec r} )= \frac{1}{2}
\psi^\dagger_\alpha({\vec r}) \Gamma^a_{\alpha\beta}
\psi_\beta({\vec r}), \nonumber \\
L_{ab}({\vec  r})&=&  -\frac{1}{2}\psi^\dagger_\alpha({\vec r})
\Gamma^{ab}_{\alpha\beta} \psi_\beta({\vec r}).
\label{ch3:phbilinear}
\eea
The SO(5) generators $L_{ab}$ and its vectors $n_a$ together
span the SU(4) algebra.
They satisfy the commutation relations as
\bea
&&[L_{ab}, L_{cd}]=-i(\delta_{ac}L_{bd}+\delta_{bd}L_{ac}-\delta_{ad}L_{bc}
-\delta_{bc}L_{ad}), \nonumber \\
&&\left[L_{ab}, n_c\right]= -i( \delta_{ac} n_b -\delta_{bc} n_a), \nn \\
&&\left[n_a, n_b \right]=-i L_{ab}.
\eea
It is well known that the SU(4) algebra is isomorphically to SO(6),
and SU(4) is the double covering group of SO(6).

In order to study pairing operators in the particle-particle
(p-p) channel and time-reversal transformation, we introduce
the charge conjugation matrix $R$:
The combination of $R$ and the creation operators
$R_{\alpha\beta}\psi^\dagger_\beta$ transforms the same as
the annihilation operator $\psi_\alpha$ under the Sp(4) transformation.
The existence of $R$ is based on the pseudoreality of Sp(4) spinor
representation, satisfying
\bea
R^2&=&-1,\ \ \ R^\dagger=R^{-1}=~^t R=-R, \nn \\
R \Gamma^a R&=&-^t\Gamma^a, \ \ \ R \Gamma^{ab} R=~ ^t\Gamma^{ab},
\eea
where $^t\Gamma^{a, ab}$ are the transposed matrices of $\Gamma^{a,ab}$.
In the representation of Eq. (\ref{eq:gamma}), $R$ is expressed as
$R=\Gamma_1\Gamma_3$.

Under the assistance of $R$, the fermion pairing operators is expressed
as \cite{wu2010a}
\bea
\eta^\dagger({\vec r})&=&\Re \eta+ i~\Im \eta=
\frac{1}{2} \psi^\dagger_\alpha({\vec r})  R_{\alpha\beta}
\psi^\dagger_\beta({\vec r}),\nonumber\\
\chi^\dagger_a({\vec r})&=& \Re \chi_a + i~\Im \chi_a =
 -\frac{i}{2}
\psi^\dagger_\alpha({\vec r}) (\Gamma^a R)_{\alpha\beta}
\psi^\dagger_\beta ({\vec r}).
\nn \\
\label{eq:pairing}
\eea
Clearly, $\eta^\dagger({\vec r})$ is an Sp(4) scalar, and
$\chi^\dagger_a({\vec r})$'s are a set of Sp(4) vector.
They are related to the spin SU(2) representation via
\bea
P^\dagger_{0,0}&=&  -\frac{1}{\sqrt 2}\eta^\dagger, \nn \\
P^\dagger_{2,0}&=& -i \frac{1}{\sqrt 2}\chi^\dagger_4, \nn \\
P^\dagger_{2,\pm1}&=& \frac{1}{2}(-\chi^\dagger_3\pm i\chi^\dagger_2),
\nn \\
P^\dagger_{2,\pm2}&=& \frac{1}{2}(\mp\chi^\dagger_1+ i\chi^\dagger_5).
\eea
The anti-unitary time-reversal transformation $T^2=-1$ is expressed as
\bea
T=R~ C,
\label{ch3:Rspin32}
\eea
where $C$ denotes complex conjugation.
$L_{ab}$'s consist of spin and spin-octupole operators \cite{wu2003,wu2006a}.
Since they are odd rank spin tensors, they are time-reversal odd.
$n_a$'s and $N$ are time-reversal even.
It is also straightforward to check that they transform differently
under the $T$ transformation
\bea
T n T^{-1} &=& n,~~  T n_a T^{-1} =n_a,\nn \\
T L_{ab} T^{-1}&=& -L_{ab}.
\label{eq:evenodd}
\eea

Now we are ready to prove the generic SO(5) symmetry of Eq. (\ref{eq:swave32})
and Eq. (\ref{eq:hubbard32}).
The kinetic energy part has an explicit SU(4) symmetry which is the
unitary transformation among four spin components.
The singlet and quintet interactions are proportional to
$\eta^\dagger({\vec r}) \eta({\vec r})$ and $\chi_a^\dagger({\vec r})
\chi_a({\vec r})$, respectively, thus reducing the symmetry group
from SU(4) to SO(5).
When $g_0=g_2$, the SU(4) symmetry is restored because
$\chi^\dagger_a,\eta^\dagger$ together form the 6 dimensional
antisymmetrical tensor representation of SU(4).

For the continuum model, the odd partial wave scatterings include the
spin triplet and septet channels, whose interactions are denoted
as $g_1$ and $g_3$, respectively.
The SO(5) symmetry is broken if $g_1\neq g_3$, and restored
at $g_1=g_3$ since the triplet and septet together could form
the 10D adjoint representation of SO(5).
However, to the leading order, $p$-wave scattering is weak for neutral
atoms, and thus can be safely neglected.
For the lattice model, the onsite interactions do not allow the
triplet and septet interactions.

For later convenience, the lattice Hubbard model of Eq. \ref{eq:hubbard32}
can be rewritten in another manifestly Sp(4) invariant form as
\bea
\label{hmlattice2}
H_0&=& -t \sum_{\langle i,j \rangle}\left( \psi^\dagger(i) \psi(j)
+ h.c.\right), \nonumber \\
H_I&=& \sum_{i, 1\le a\le 5}\left[
{3 U_0+5 U_2 \over 16} \left(n(i)-2\right)^2-{U_2-U_0\over 4} n_a^2(i) \right]\nn \\
&-& (\mu -\mu_0) \sum_i n(i),
\label{eq:hubbardsp4}
\eea
where the SU(4) symmetry appears at $U_0=U_2$ as before.
At half-filling, $\mu_0=(U_0+5U_2)/2$ to ensure
the particle-hole symmetry.
Here half-filling means the average particle number per site
equals 2, half of the component number.

\subsection{The SO(7) unification and the $\chi$-pairing}
The spin-$\frac{1}{2}$ Hubbard model defined in a bipartite lattice
in any dimensions actually possesses a pseudospin SU(2) symmetry
spanned by the $\eta$-pairing operators and particle number
as discovered by Yang \cite{yang1989} and by Yang and Zhang
\cite{yang1990}.
In this subsection, we review the extension of the pseudo-spin symmetry
to the SO(7) symmetry in the spin-$\frac{3}{2}$ Hubbard model,
and define the quintet $\chi$-pairing operators.
It exhibits much richer unifying power in treating a variety
of competing orders at equal footing \cite{wu2003}.

The $\eta$-pairing operator in spin-$\frac{1}{2}$ systems
sums over the onsite singlet pairing operators with opposite signs on
two sublattices.
The pseudospin SU(2) algebra is particularly useful for unifying competing
orders in the negative-$U$ Hubbard model \cite{yang1989,yang1990,zhang1991}.
The complex order parameters of superconductivity and the charge-density-wave
are unified forming a 3D representation.
The $\eta$-pairing generator transforms superconductivity into
charge-density-wave and vice versa.
At half-filling, the pseudospin SU(2) symmetry is exact, and these
two types of orders are degenerate.
Away from half-filling, the SU(2) symmetry is explicitly broken, and
the superconducting ground state is selected.
However, when applying the $\eta$-pairing operator to the
ground state, it creates well-defined excitations, which are the
pseudo Goldstone modes rotating superconductivity into charge-density-wave.

Before moving on, let us fully explore the symmetry
structure of spin-$\frac{3}{2}$ systems.
The largest algebra formed by 4-component fermions is actually
SO(8) \cite{lin1998}, including 16 p-h channel fermion bilinears
and the other 12 in the p-p channel.
On each site, the local SO(8) generators $M_{ab}(i)~(0\le a<b\le 7)$
are organized as follows \cite{wu2003,wu2006},
\begin{widetext}
\bea
\label{so8algebra}
M_{ab}(i)=
\left( \begin{array}{cccc}
0&  \Re \chi_1(i)~ \sim~ \Re \chi_5(i) & N(i)  &   \Re \eta(i)  \\
 &                                  &\Im \chi_1(i) &n_1(i)  \\
 &        L_{ab}(i)                    & \sim      &\sim \\
 &                                  &\Im\chi_5(i)  &n_5(i) \\
 &                                  &   0       &-\Im\eta(i) \\
 &                                  &           &0 \\
\end{array} \right),
\nn \\
\eea
\end{widetext}
with $N(i) =(n(i)-2)/2$.
For $1\le a<b\le 5$, they are just $L_{ab}(i)$ forming
its SO(5) subalgebra.
The global SO(8) generators are defined as
\bea
M_{ab}&=&\sum_i M_{ab}(i),
\eea
or,
\bea
M_{ab}&=&\sum_i (-)^i M_{ab}(i),  ~~~(0\le a<b \le 7),
\eea
depending on $M_{ab}$ lying in the p-h or p-p channels,
respectively.
More explicitly, we write
\bea
L_{ab}&=&M_{ab}, \hspace{17mm} n_a=M_{a7} \nn \\
N&=&M_{06}, \hspace{17mm}
\eta^\dagger=M_{06}-i M_{67}, \nn \\
\chi_a^\dagger &=&M_{0a}+iM_{a6},~
\label{eq:glso8gen}
\eea
with $1\le a<b\le 5$.
$L_{ab}$, $n_a$ and $N$ lie in the p-h channel, and
$\eta$ and $\chi^\dagger_a$ lie in the p-p channel.
The $\eta^\dagger$ operator is the spin-$\frac{3}{2}$ generalization
of Yang's $\eta^\dagger$, both of which are spin singlet.
In contrast, the $\chi_a^\dagger$ pairing operator is a non-trivial
quintet generalization.

It is easy to check that the $H_0$ part of the Hamiltonian
Eq. (\ref{eq:hubbardsp4}) satisfies $[H_0, M_{ab}]=0$.
However, $H_{int}$ typically breaks the SO(8) symmetry
unless it vanishes within the framework of 4-fermion
interactions.

The next highest algebra is SO(7) spanned by $M_{ab}$ with $0\le a<b\le 6$,
which is the high-rank Lie algebra generalization of Yang's
pseudospin SU(2) algebra.
Explicitly, they include the SO(5) generators $L_{ab}$, the $\chi$-pairing
operators $\Re \chi_a, \Im \chi_a$, and the particle number $N$.
This SO(7) symmetry becomes exact at $U_0=-3U_2$, where the
interacting part of the Hamiltonian Eq. (\ref{eq:hubbardsp4})
is reduced to
\bea
H_I=\sum_{i,0\le a <b \le 6} \Big \{ \frac {2}{3} U_2 ~
\left[M_{ab}(i)\right]^2 -(\mu-\mu_0) n(i)\Big \}.
\nn \\
\eea
At half-filling, $\mu=\mu_0$, then the global SO(7) symmetry
becomes exact.

\begin{figure} \centering
\includegraphics[width=0.8\linewidth, angle=90]{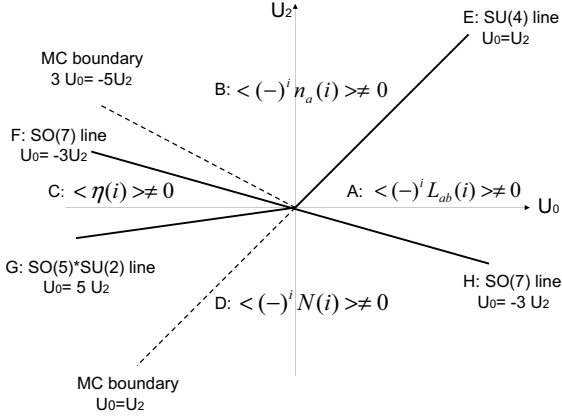}
\caption{Competing phases of spin-$\frac{3}{2}$
Hubbard model unified by high symmetries.
(A) and (B): Antiferromagnetism in
the Sp(4) adjoint and vector representations;
(C): the singlet superconductivity; (D): CDW;
(E), (F), (G), and (H): exact phase boundaries
with higher symmetries of SU(4), SO(7),
SO(5)$\otimes$ SU(2) and SO(7), respectively.
From Ref. \cite{wu2003}.}
\label{fig:phasediag}
\end{figure}

The SO(7) symmetric spin-$\frac{3}{2}$ Hubbard model can be further
divided into two cases:  $U_0=-3U_2$ with (I) $U_2>0$;
(II)$U_2<0$.
The physics of case (I) lies in SO(7)'s vector representation,
while that of case (II) lies in the adjoint representation.

In case (I), the system in the weak coupling
regime exhibits the competition between the singlet supercondutivity
and density-wave of spin quadrupole orders, whose order parameters
are organized as
\bea
V_a&=&\sum_i (-)^i M_{a6},  \mbox{~~~or,~~~~} \nn \\
V_a&=&\sum_i M_{a6}, ~~~(0\le a \le 7)
\eea
depending on $V_a$ lying in the p-h channel or the p-p channel,
respectively.
More explicitly,
superconductivity and spin quadrupole density wave
are unified as
\bea
V_0&=&\Re\Delta_s, ~~V_a=\mbox{SDW}_a ~~(1\le a\le 5),\nn \\
V_6&=&-\Im \Delta_s.
\eea
They transform according to the vector
representation of SO(7),
\bea
[M_{ab}, V_c]=i\left(\delta_{bc} V_a -\delta_{ac}V_b\right).
\eea
Hence, the Goldstone manifold is $S^6$.
Away from half-filling, SO(7) is broken and the singlet
superconductivity is selected as the ground state ordering.
The $\chi$-pairing operators remain the eigen-operators as
\bea
&&[H,\chi^\dagger_{a}]= -(\mu-\mu_{0}) \chi^\dagger_{a},
\mbox{~~and,~~ }  \nn \\
&&[H,\chi_a ]= (\mu-\mu_{0}) \chi_a.
\eea
At $\mu<\mu_0$, applying $\chi^\dagger_a$ to the superconducting
ground state $|\Omega\rangle$ creates a quintet excitation,
\bea
H \left(\chi^\dagger_a |\Omega\rangle\right) =(\mu_0-\mu) \left(\chi^\dagger_a |\Omega\rangle\right),
\label{eq:chimode}
\eea
which carries the lattice momentum $\mathbf Q=(\pi,\pi)$.
In other words, the $\chi_a$-pairing operator behaves like
a quasi-Goldstone mode, which rotates the singlet superconductivity
to the density-wave state of the $a$-th component of
spin quadrupole density-wave.

Yang's $\eta$-pairing operators were generalized to the SO(5) theory
of high $T_c$ superconductivity, which unifies the 2-component
superconductivity and 3-component antiferromagnetism into a 5-vector.
Nevertheless, the SO(5) algebra is not exact \cite{zhang1997}.
The celebrated neutron resonance modes in the superconducting states
were interpreted as the pseudo-Goldstone modes rotating from
superconductivity to antiferromagnetism, denoted as $\pi$-modes.
The $\chi$-modes here are just analogs of the $\pi$ modes
SO(5) theory \cite{zhang1997}.
However, the SO(5) algebra is not exact in high $T_c$ cuprate
systems.
In contrast, here the SO(7) symmetry is
exact.

The SO(7) unification is even more powerful in case (II) with $U_0>0$,
in which the 21D adjoint representation of SO(7) plays the role.
The order parameter manifold includes the quintet superconductivity,
the 10-fold density-wave of spin and spin octupole orders, and
charge-density-wave, which are organized as
\bea
T_{ab}=\sum_i (-)^i M_{ab},
\eea
or,
\bea
T_{ab}=\sum_i M_{ab},
~~~(0\le a<b \le 6),
\eea
depending on $T_{ab}$ lying in the p-h channel or the p-p channel,
respectively.
Explicitly, they are
\bea
\Delta_{q,a}= T_{0a}+i T_{a6}, \ \ \,
\mbox{SDW}_{ab}=T_{ab}, \ \ \,
\mbox{CDW}=T_{06}, \nn \\
\eea
where $1\le a<b\le 6$.
They transform according to the SO(7) algebra as
\bea
[M_{ab}, T_{cd}]=i\left(\delta_{ac}T_{bd}+\delta_{bd}T_{ac}
-\delta_{ad}T_{bc}-\delta_{bc}T_{ad}\right).
\nn \\
\eea

It is amazing to realize such a ``grand unification" in a
non-relativistic model.
The Goldstone manifold is SO(7)$/[$SO(5)$\times$ SO(2)$]$, which
is 10-dimensional.
When away from half-filling, the SO(7) symmetry is broken into SO(5),
and the ground state exhibit the quintet superconductivity.
Eq. (\ref{eq:chimode}) still applies.
Assuming $\avg{\Omega|\Delta_{q,b}|\Omega}\neq 0$,
$\chi^\dagger_a|\Omega\rangle$ remains a quasi-Goldenstone mode
which rotates to the spin/spin octupole density-wave state
SDW$_{ab}$ if $a\neq b$, or, the charge-density-wave state
if $a=b$.

The pseudo-spin SU(2) symmetry of the spin-$\frac{3}{2}$ version occurs
at $U_0=5~U_2$.
In this case, $H_I$ is rewritten as
\bea
H_I=\sum_{i, 1\le a,b \le 5} \left\{-U_2~L_{ab}^2(i)-(\mu-\mu_0) n(i)
\right \},
\eea
which only involve the SO(5) generators.
Then $M_{06}$, $M_{07}$, $M_{67}$ span an SU(2) algebra commuting with
all the SO(5) generators.
More explicitly, they are just the pseudo-spin SU(2) algebra
spanned by the $\eta$-pairing and particle number operators.
At $U_0=5~U_2<0$, this pseudospin SU(2) symmetry unifies the
singlet pairing and charge-density-wave order parameters
in a similar way to the spin-$\frac{1}{2}$ negative-$U$ Hubbard model.
Again, when away from half-filling, this symmetry is broken, and the
ground state is the singlet pairing state.
In this case, the $\eta$-pairing operators remain eigen-operators
\bea
&&[H, \eta^\dagger] = -(\mu-\mu_{0}) \eta^\dagger, \nn \\
&&[H, \eta] = (\mu-\mu_{0}) \eta.
\eea

We emphasize that the pseudo-spin SU(2) symmetry in the spin-$\frac{3}{2}$
system is still different from that in the spin-$\frac{1}{2}$ case.
In the latter case, the empty and doubly occupied states
form a pseudospin-$\frac{1}{2}$ representation.
In the spin-$\frac{3}{2}$ case, there are three onsite singlet
states: empty, 2-particle
singlet, and the 4-occupied states forming
a pseudo-spin-1 representation.

Based on the above analysis and assisted by mean-field calculations,
the weak-coupling phase diagram in a bipartite lattice at half-filling
in two dimensions and above is shown in Fig. \ref{fig:phasediag}.
The higher symmetries lines are as follows: The SU(4) symmetry appears
along line E with $U_0=U_2$; the SO(7) symmetry appears along lines F
and H with $U_0=-3U_2$; and the SO(5)$\otimes$ SU(2)symmetry appears
long line G with $U_0=5U_2$.
These lines are phase boundaries separating phases A, B, C, and D.
Phase A and B are regimes where repulsive interactions dominate.
Hence, they are density-wave states of spin tensors.
In phase A, the onsite singlet energy is smaller than the quintet
energy, leading to the spin quadrupole density-wave forming
the 5-vector representation of the Sp(4) group.
On the other hand, the lowest onsite states in phase B are
5-fold degenerate, leading to the spin/spin octupole density-wave
forming the 10-dimensional adjoint representation of Sp(4).
The Goldstone manifold in phase A is SO(5)/SO(4)=$S^4$, while that
in phase B is SO(5)/[SO(3)$\otimes$ SO(2)].
On line E, the SU(4) symmetry unifies the 15 dimensional
density-wave orders in all the spin channels forming the SU(4)
adjoint representation, whose Goldstone manifold
is U(4)/[U(2)$\otimes$ U(2)].

Phase C is the singlet pairing state, and phase D is the
charge-density-wave state.
Orders in phases B and C are unified along the SO(7) line F.
In contrast, the SO(7) line H unifies orders in phases A, D, and
the quintet pairing.
Orders in phases C and D are unified along the psuedospin
SU(2) symmetry line G.

At last, let us mention an interesting point
that SO(7) possesses a subgroup of G$_2$,
which is the smallest exceptional Lie group and also
the automorphism group of non-associative algebra of octonions.
A G$_2$ symmetric spin-$\frac{3}{2}$ Hubbard model is
constructed which is the common subgroup of two different
SO(7) algebras connected by the structure constant of
octonions as shown in Ref. \cite{gao2020}.
This model exhibits various interesting symmetry breaking patterns:
The G$_2$ symmetry can be spontaneously broken into SU(3), or,
SU(2)$\otimes$ U(1), both of which are essential in high energy physics.
In quantum disordered states, quantum fluctuations
generate the effective SU(3), or, SU(2)$\otimes$ U(1)
gauge theories.

\subsection{Quartet (charge-$4e$) superfluidity and quartet
density wave}
\label{sect:onedim}

Superconductivity arises from the coherent condensation of Cooper pairs,
which is the central concept of the celebrated Bardeen-Cooper-Schrieffer
(BCS) theory.
Moreover, there exist multi-fermion clustering instabilities in strong
correlation systems in various disciplines of modern physics.
The SU(3) gauge symmetry requires three quarks to form a color singlet
bound state of baryon \cite{peskin1995}; $\alpha$-particles are
4-body bound states of two protons and two neutrons, and biexcitons
are bound states of two electrons and two holes.
These states go beyond the framework of the BCS theory since they
cannot be reduced to a 2-body problem.
The competitions among the quartetting (charge $4e$) and pairing
($2e$) superfluidities, quartet and pair density wave orderings
are investigated in 1D 4-component fermion systems \cite{wu2005,
lecheminant2005}.
In recent years, charge-$4e$ superconductivity has been proposed
as a consequence of strong fluctuations of the pair-density-wave
state in high T$_c$ cuprates \cite{berg2009,agterberg2020}.
Competitions of 4-fermion orderings in the context of
superconducting phase fluctuations have recently received
attentions \cite{jian2021,fernandes2021,zeng2021}.
Excitingly experimental evidence of Little-Parks oscillations at the
periods of $hc/(4e)$ and $hc/(6e)$ have been observed in the Kagome
superconductor CsV$_3$Sb$_5$
\cite{gewang2022}.

Spin-$\frac{3}{2}$ systems allow the quaretteting order, i.e., 4 fermions
forming a clustering instability, which is also called ``charge-$4e$" in
condensed matter physics.
A quartet in the strong coupling limit is a 4-body maximally
entangled EPR state with all the spin components forming an SU(4) singlet,
whose order parameter is expressed as
\bea
Q(x)=\psi^\dagger_{\frac{3}{2}}(x)
\psi^\dagger_{\frac{1}{2}}(x)\psi^\dagger_{-\frac{1}{2}}(x)
\psi^\dagger_{-\frac{3}{2}}(x).
\eea
Furthermore, spin-$\frac{3}{2}$ systems could support 6 different types of
Cooper pairing states including an Sp(4) singlet and
a set of Sp(4) quintet states
whose order parameters are presented in Eq. (\ref{eq:pairing}).
It would be interesting to investigate their competitions.

Assisted by the strong coupling methods for 1D problems, we are able to
analyze the competition between the quartetting and pairing formations.
Quartets and pairs can undergo either superfluidity or density-wave
transitions depending on the charge channel interactions.
Only the quartetting states are SU(4) invariant, and the 6
pairing operators presented in Eq. (\ref{eq:pairing}) form
the rank-2 antisymmetric tensor representation of SU(4).
\footnote{In terms of SO(6), which equals SU(4)$/$Z$_2$,
they form the 6-vector representation.}
Due to the strong quantum fluctuations, non-Abeliean Lie group
symmetries cannot be spontaneously broken in 1D.
Hence, only quartet orderings, either superfluidity or density
wave, are allowed by the SU(4) symmetry.
Nevertheless, if the symmetry is reduced to Sp(4), the Sp(4) singlet
pairing could survive, while the quintet pairing still cannot
survive.
Naturally, there exist competitions between Sp(4) singlet (charge-$2e$)
pairing orders and quartteting (charge-$4e$) orders.
Between them it is an Ising order-disorder transition
in the spin channel.

Here we briefly outline the procedure of the bosonization and
renormalization group (RG) analysis, and the details were presented
in Ref. \cite{wu2006}.
The Sp(4) currents include scalar (charge), vector (spin quadrupole),
and tensor (spin plus spin octupole) ones,
\bea
J_{R,L}(z) &=&\psi^\dagger_{R,\alpha}(z) \psi_{R,\alpha}(z) \nn \\
J^a_{R,L}(z)&=&\frac{1}{2}\psi^\dagger_{R,\alpha}(z)\Gamma^a_{\alpha\beta}
\psi_{R,\beta}(z)~ (1\le a\le 5),  \nonumber \\
J^{ab}_{R,L}(z)&=&\frac{1}{2}
\psi^\dagger_{R,\alpha}(z)\Gamma^{ab}_{\alpha\beta} \psi_{R,\beta}(z)
~(1\le a<b\le 5),
\nn \\
\eea
where $R$ and $L$ refer to right and left-movers.
The low energy effective Hamiltonian density
$H=H_0+ H_{int}$ is  written as,
\bea
\label{ham1}
H_0&=& v_f \Big\{ \frac{\pi}{4} J_R J_R +\frac{\pi}{5}
(J^a_R J^a_R +J^{ab}_R J^{ab}_R) +(R\rightarrow L)  \Big\}, \nonumber \\
H_{int}&=&   \frac{g_c}{4} J_R J_L + g_v J^a_R J^a_L
+g_t J^{ab}_R J^{ab}_L,
\eea
where the chiral couplings are neglected at one-loop level since
it only renormalizes Fermi velocities.
At the tree level, these dimensionless coupling constants are
related by the pair interaction parameters
$g_0$, $g_2$ defined in Eq. (\ref{eq:swave32})  as
$g_c=  (g_0+ 5 g_2)/2,  \ \ \
g_v= (g_0-3g_2)/2,
\ \ \
g_t=-(g_0+g_2)/2$.
Certainly, they are renormalized significantly under the RG process.
At $g_v=g_t$, or, $g_0=g_2$, the SU(4) symmetry is restored.

\begin{figure} \centering
\includegraphics[width=\linewidth, angle=]{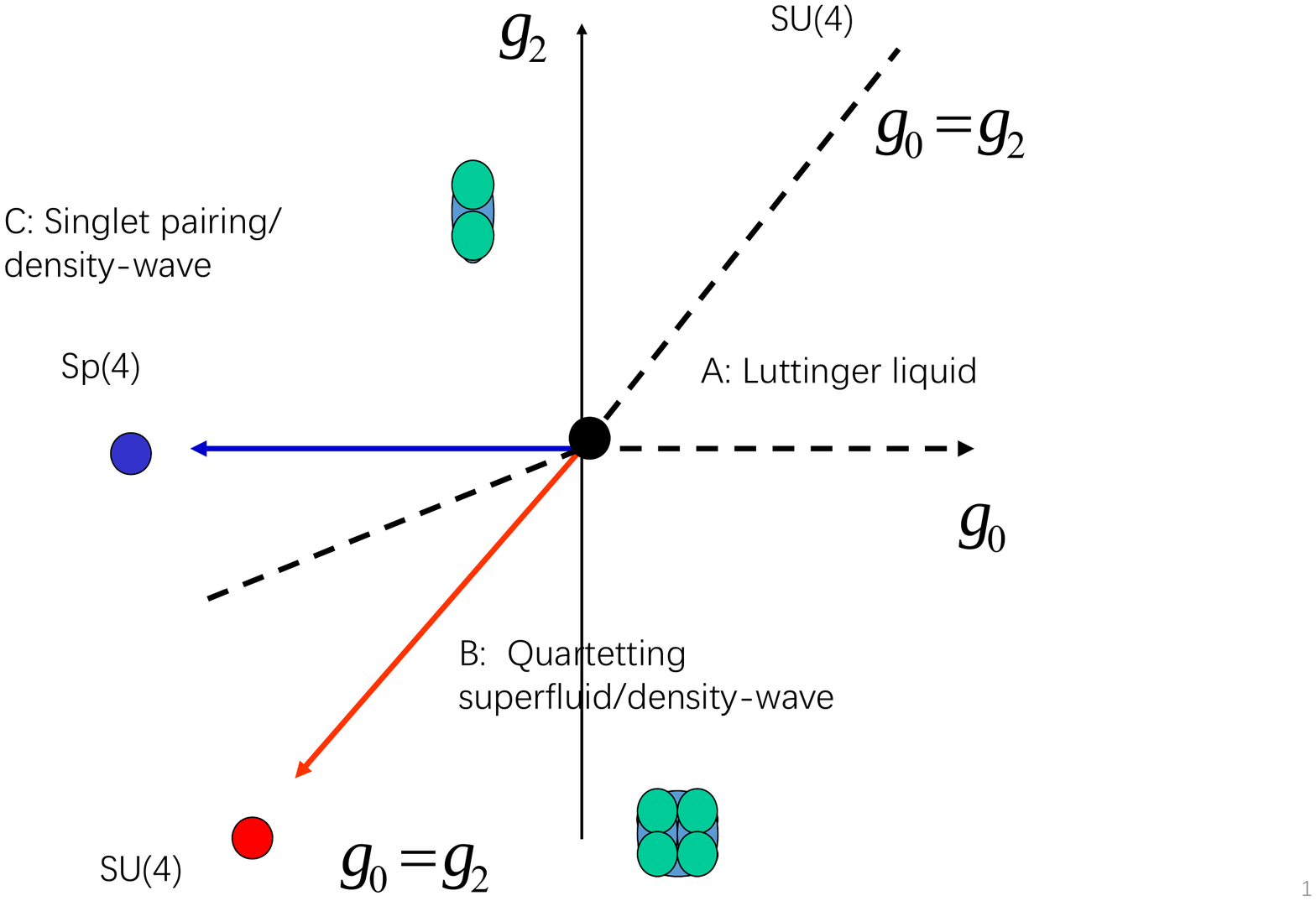}
\caption{Competition between quaretting (charge-$4e$) and singlet
pairing (charge-$2e$) phases in 1D spin-$\frac{3}{2}$ systems at
incommensurate fillings.
Combined with the charge channel Luttinger parameter $K_c$,
three phases are identified with phase boundaries marked by
dashed lines.
A) the gapless Luttinger liquid phase controlled by the
non-interacting fixed point (the black spot).
B) Quartetting superfluidity at $K_c>2$
or quartet density-wave ($2k_f$) at $K_c<2$.
They are controlled by the strong coupling fixed point along the
SU(4) line (the red spot);
C) singlet pairing superfluidity at $K_c>1/2$
or pair density wave ($4k_f$) at $K_c<1/2$.
They are controlled by the strong coupling fixed point
along $g_2=0$ (the blue spot).
Phases B and C are both gapped in spin channels, and
the transition between them is an Ising order-disorder
transition.
From Ref. [\onlinecite{wu2005}].
}
\label{fig:quartet}
\end{figure}

The phase diagram at incommensurate fillings are presented at
Fig. \ref{fig:quartet}.
The charge sector remains gapless and decouples with the spin sectors.
In the spin sector, three phases are identified:
Phase A is the gapless Luttinger liquid phase lying in the repulsive
interaction region where $0<g_2<g_0$, which is controlled by the
non-interacting fixed point.
Phase B is the quartetting phase controlled by the strong coupling
fixed point along the SU(4) line with $g_v=g_t\to +\infty$,
or, $g_0=g_2\to -\infty$.
It lies in the regime where attractive interactions dominate.
Phase C is the spin singlet pairing phase controlled by the
strong coupling point along the line of $-g_v=g_t\to +\infty$
corresponding to $g_0\to -\infty$ and $g_2\to 0$.
The pairing phase even covers the regime with a purely repulsive
interaction regime.

Within quartetting phase B, there also exist two competing orders,
the quartetting superfluidity and quartet density wave.
By checking the periodicity, the quartet density wave corresponds to
the $2k_f$ CDW.
Four fermions first form quartets, and then they either undergo superfluidity,
or, density wave ordering.
As for the charge sector, their bosonic expressions are
\bea
Q&=&\psi^\dagger_{\frac{3}{2}}\psi^\dagger_{\frac{1}{2}}
\psi^\dagger_{-\frac{1}{2}}\psi^\dagger_{-\frac{3}{2}}
\propto e^{2i\sqrt\pi \theta_c}, \nn \\
O_{qdw}&=&\psi^\dagger_{R\alpha} \psi_{L\alpha}
\propto e^{i\sqrt \pi \phi_c}.
\eea
The scaling dimensions for quartet superfluidity and density-wave
orders are $1/K_c$ and $K_c/4$, respectively.
Hence, the quartet superfluidity wins at $K_c>2$, while
the quartet density wave wins at $K_c<2$.

Similarly in phase C, Cooper pairs can either undergo superfluidity,
or, pair density wave ordering.
The pair density wave corresponds to the 4$k_f$ charge density
wave.
As for the charge sector, their bosonic expressions are
\bea
\Delta_s&=&\eta^\dagger=\psi^\dagger_{\frac{3}{2}}
\psi^\dagger_{-\frac{3}{2}}
-\psi^\dagger_{\frac{1}{2}}
\psi^\dagger_{-\frac{1}{2}}
\propto e^{i\sqrt\pi \theta_c},
\nn \\
O_{pdw}&=&\psi^\dagger_{R\alpha}\psi^\dagger_{R\beta}
\psi_{L\beta}\psi_{L\alpha}
\propto e^{2i\sqrt\pi \phi_c}.
\eea
The scaling dimensions for the singlet pairing and pair density wave
orders are $1/(4K_c)$ and $K_c$, respectively.
Hence, the pairing superfluidity dominates over the
pair-density-wave at $K_c>1/2$.
In the region of $1>K_c>1/2$, pairing superfluidity is the leading
instability in an overall repulsive interaction environment.

The boundary between phase B quartetting (charge-$4e$) and phase C
singlet pairing (charge-$2e$) is determined by the unstable fixed
point $(g_v=0, g_t\to \infty)$, which is approximately plotted
in Fig. \ref{fig:quartet}.
The competitions between these two phases can be mapped to a phase-locking
problem of two-band superconductivity.
The first component is $\Delta_1=\psi^\dagger_{\frac{3}{2}}
\psi^\dagger_{-\frac{3}{2}}$, and the second one
$\Delta_2=\psi^\dagger_{\frac{1}{2}}\psi^\dagger_{-\frac{1}{2}}$,
whose bosonic representations are
\bea
\Delta_1\propto e^{i\sqrt \pi \theta_1}&=&e^{i\sqrt \pi (\theta_c+\theta_r)},
\nn \\
\Delta_2\propto e^{i\sqrt \pi \theta_2}&=&e^{i\sqrt \pi (\theta_c-\theta_r)},
\eea
where the charge channel $\theta_c$ is the average phase
and $\theta_r$ is the relative phase.
In fact, $\theta_r$ and its vortex, or, dual field $\phi_r$, are
of the spin quadrupole channel.
The bosonic expressions of the pairing and quartteting order
parameters are expressed as
\bea
\Delta_s&=&\Delta_1-\Delta_2\propto e^{i\sqrt \pi \theta_c}
\cos \sqrt \pi \theta_r,
\nn \\
Q&=&\Delta_1\Delta_2=e^{i2\sqrt \pi\theta_c} \cos 2\sqrt\pi \phi_r.
\eea
$\theta_c$ is power-law fluctuating, and does not play a role in
the transition between quartetting and pairing.
It is the relative phase fluctuations that control the transition
as described by the sine-Gordon theory,
\begin{widetext}
\bea
H_{eff}=\frac{1}{2}\left\{ (\partial_x \theta_r)^2
+(\partial_x \phi_r)^2 \right\}
+\frac{1}{2\pi a^2} \left(\lambda_1 \cos 2\sqrt \pi \theta_r
+\lambda_2 \cos 2\sqrt \pi \phi_r \right),
\label{eq:Ising}
\eea
\end{widetext}
which contains cosine terms of both $\theta_r$ and $\phi_r$.

If $\lambda_1>\lambda_2$, the relative phase $\theta_r$
is pinned leading to the pairing order;
otherwise if $\lambda_1<\lambda_2$, the vortex (dual) field $\phi_r$
is pinned giving rise to the quartetting order.
The transition occurs at $\lambda_1=\lambda_2$.
Eq. (\ref{eq:Ising}) can be mapped to two free majorana fermions
with masses $m_{\pm}=|\lambda_1\pm \lambda_2|$.
One channel becomes massless at the transition,
which is the Ising critical point.

As a difference between the pairing and quartetting orders,
there exist quartet breaking processes of $4\to 1+3\to 1+1+2$
and $4\to 2+2$, which can be used to distinguish quartetting
and pairing.
The vortex lattice configurations are also different for quartetting
superfluidity.
In the quartetting superfluid, the flux quantization is $hc/(4e)$.
Hence, the number of vortices should be doubled compared to
those of pairing superfluidity.

\subsection{Color magnetism}

The prominent multi-particle correlations also manifest in the SU(N)
quantum antiferromagnetism in the Mott insulating states
at $1/N$-filling, i.e., one fermion per site.
The superexchange favors the tendency that every $N$ sites form an
SU($N$) singlet as dubbed ``color magnetism" due to its similarity to the SU(3)
gauge theory of quantum chromodynamics in which 3 quarks form
a color singlet.

In the one-dimensional Sp(4) Heisenberg chain in the fundamental
spinor representation, it has been found that the ground state exhibits
oscillations at the period of four sites \cite{wu2005,hung2011}.
The plaquette tendency was investigated in the SU(4) symmetric
Kugel-Khomskii model by diagonalization
up to the size of $4\times 4$.
\cite{bossche2000}
The Majumdar-Ghosh model was generalized to the SU(4) case in
a ladder system whose ground state is solvable as a direct
product state of SU(4) plaquettes \cite{chen2005}.
The 4-site SU(4) singlet plaquette wavefunction can be written as
\bea
\frac{1}{\sqrt{4!}}
\epsilon_{\alpha\beta\gamma\delta}\psi^\dagger_\alpha(1)
\psi^\dagger_\beta(2) \psi^\dagger_\gamma(3) \psi^\dagger_\delta(4)
|\Omega\rangle,
\eea
which is a 4-particle maximally entangled EPR state.

\begin{figure} \centering
\includegraphics[width=0.4\linewidth, angle=]{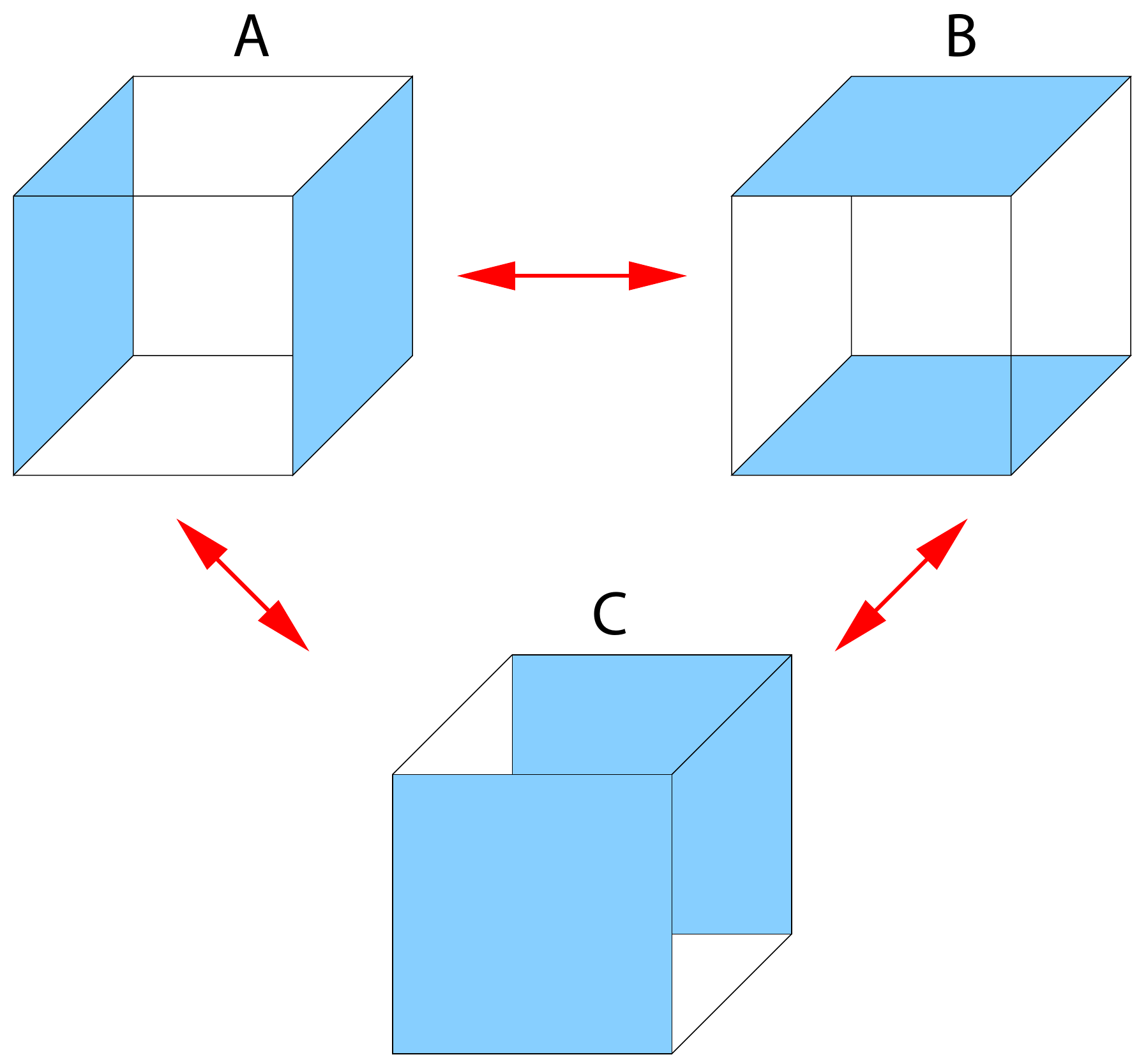}
\caption{In the SU(4) Mott-insulator at quarter-filling, i.e., one
particle per site in the fundamental representation of SU(4).
The superexchange interaction favors four sites of a plaquette form an
SU(4) singlet in analogous to the dimer formation in the SU(2)
antiferromagnetism.
For a 3D cube, there exist three flappable plaquette configurations,
based on which a quantum plaquette model can be constructed.
It can be described by an effective high-rank gauge theory
with conserved electric dipoles instead of charges.
From Ref. \cite{xu2008}.
}
\label{fig:plaquette}
\end{figure}

Consider the SU(4) antiferromagnetism with each site in the fundamental
representation in a 3D cubic lattice.
We construct the SU(4) resonating plaquette model in 3D in analogous to
the Rokhsar-Kivelson quantum dimer model in 2D square lattice
\cite{xu2008,pankov2007}.
There exist three resonant configurations: the left-right,
front-back, up-bottom plaquette coverings in a cube as shown in
Fig. \ref{fig:plaquette}.

The Rokhsar-Kivelson (RK) type Hamiltonian is constructed as
\cite{rokhsar1988}:
\bea
H&=&-t \sum_{\mbox{each cube}} \big\{ |A\rangle \langle B|
+ |B\rangle \langle C|
+|B\rangle \langle C| +h.c.\big\} \nonumber\\
&+& V\sum_{\mbox{each cube}} \big\{ |A\rangle \langle A|  +
|B\rangle \langle B| + |C\rangle \langle C| \big\},
\label{eq:qpm}
\eea
where $t$ is assumed to be positive and $V$ is the plaquette
flipping amplitude.
Similarly to the RK point of the quantum dimer model, here at $V/t=2$,
the ground state is the equal weight superposition of all plaquette
configurations connected by local flips \cite{pankov2007}.

The low energy physics of the quantum plaquette model can be mapped to
a gauge theory model, actually, it is a high order gauge theory.
We assign each face with an integer number $n$ only taking values
of 1 and 0: 1 corresponds to that the plaquette is
an SU(4) singlet, and otherwise, 0.
The ``electric field" at site $i$ is defined as a rank-2 symmetric
traceless tensor
\bea
E_{i, \mu\nu} = \eta(i)(n_{i+\frac{1}{2}\hat{\mu} +
\frac{1}{2}\hat{\nu}} - \frac{1}{2}),
\eea
where $\eta(i)=\pm 1$ marking the sublattice,
and $i+\frac{1}{2}\hat{\mu} +
\frac{1}{2}\hat{\nu}$ refers to the location of a face center.
Since each site can only join one singlet, the sum of $n$ over all the
twelve faces sharing the same site is constrained to be 1, which can
be represented as
\bea
\nabla_x \nabla_y E_{xy}+\nabla_y
\nabla_z E_{yz} + \nabla_z \nabla_x E_{zx}=5\eta(i),
\label{constraint}
\eea
where $\nabla$ is the lattice derivative.
According to the standard electrodynamics, $E$ is conjugate
to the vector potential $A_{i,\mu\nu}$, which is
also a rank-2 tensor, as
\begin{eqnarray}
[E_{i,\mu\nu}, A_{j,\rho\sigma} ] &=& i\delta_{ij}
(\delta_{\mu\rho}\delta_{\nu\sigma} +
\delta_{\mu\sigma}\delta_{\nu\rho} ).
\label{eq:EA}
\end{eqnarray}
Since $E$ is like angular momentum taking integer numbers, $A$ should
behave as an angular variable $A_{i,\mu\nu} = \eta(i) \ \theta_{i + \frac{1}{2}\hat{\mu} + \frac{1}{2}\hat{\nu}}$, which is
compact with the period of $2\pi$.
Then
\bea
[E_{i,\mu\nu}, e^{i A_{j,\nu\sigma}} ]=
(\delta_{\mu\rho}\delta_{\nu\sigma} +
\delta_{\mu\sigma}\delta_{\nu\rho} ) e^{i A_{j,\nu\sigma}}.
\eea

With these preparations,
the plaquette flipping term in Eq. \ref{eq:qpm} is represented as
\bea
H_t = &-& t\big\{\cos(\nabla_z
A_{xy} - \nabla_xA_{yz}) + \cos(\nabla_x A_{yz} - \nabla_yA_{zx})
\nn \\
&+& \cos(\nabla_y A_{zx} - \nabla_zA_{xy}) \big\}.
\nn \\
\label{eq:low1}
\eea
The associated gauge invariant transformation is,
\bea
A_{\mu\nu} \rightarrow A_{\mu\nu} + \nabla_\mu\nabla_\nu f,
\label{eq:gauge}
\eea
where $f$ an arbitrary scalar function.
The corresponding Gauss's law becomes
\bea
\partial_i\partial_j E_{ij}=\rho.
\eea
Its physical meaning has recently been revealed in the context of
the ``fracton'' physics, which is a recent focus in the condensed
matter community for exotic states of matter and has the potential
of applications for topological quantum memory
\cite{nandkishore2019}.


\subsection{Half-filled SU($N$) Hubbard models: Slater v.s.
Mott physics}
\begin{figure}
\centering
\includegraphics[width=0.8\linewidth, angle=]{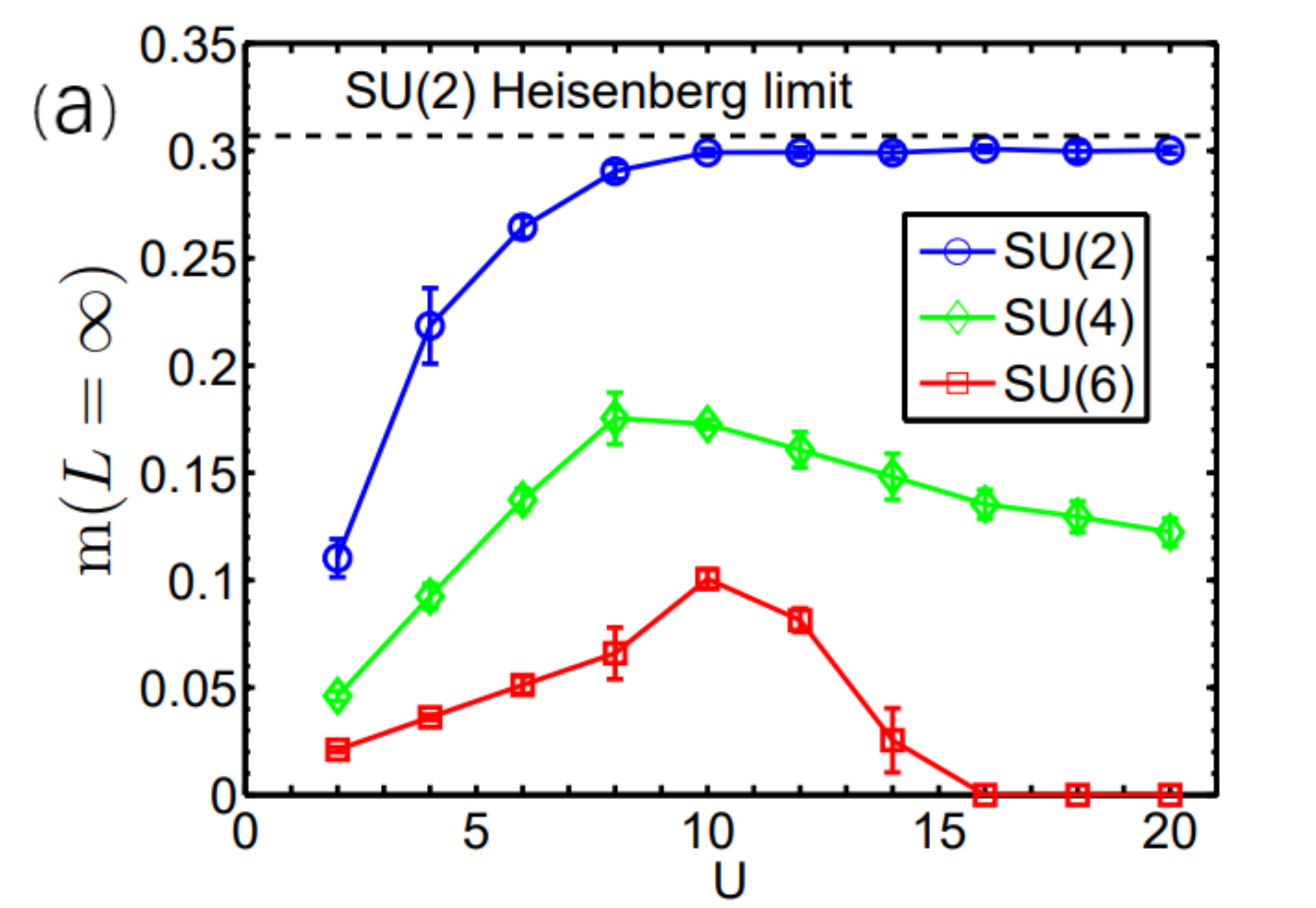}
\includegraphics[width=0.8\linewidth, angle=]{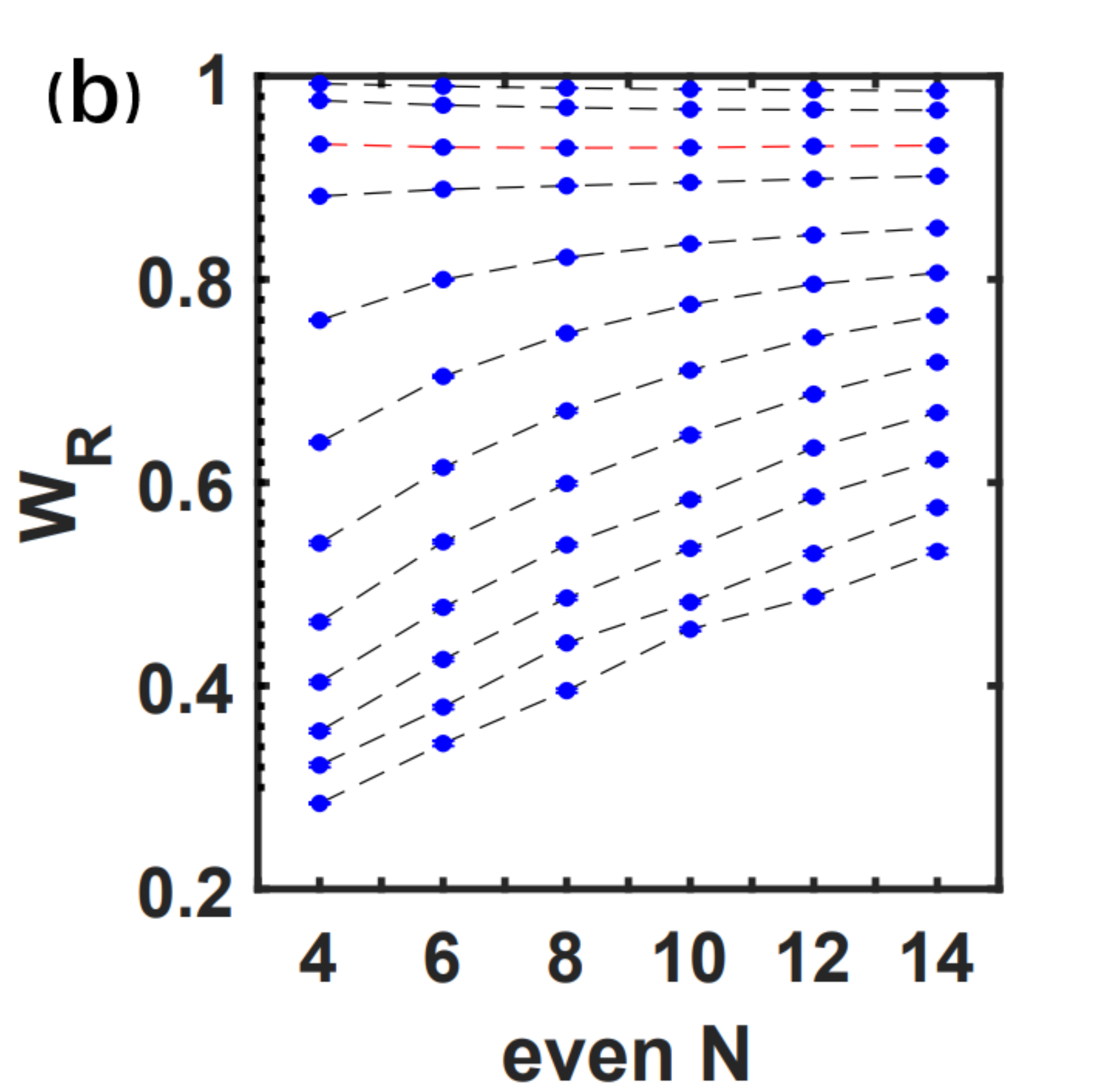}
\caption{QMC simulations on SU($N$) Hubbard models to reveal how
interaction effects scale with fermion component number $N$.
$a$) The AFM orders of the SU($N$) Hubbard models in the square lattice
as varying $U$ and $N$.
The AFM order shows a monotonically increase at $N=2$ but behaves
differently at $N=4$ and 6.
For the latter cases, the AFM orders start to grow and then drop as increasing $U$.
In particular, it is completely suppressed in the SU(6)
case at $U>U_c$ at $N=6$, after which the VBS order appears.
This shows a quantum phase transition from the Slater
physics to the Mott physics region. (From Ref. \cite{wang2014})
(b) QMC simulations of the relative bandwidth $W_R$ for the 1D
SU($N$) Hubbard model at half-filling.
The results show the convergence between itineracy and locality
as $N\to \infty$.
The dashed lines shown as a guide from top to bottom correspond
to $U/t =0.5; 1.0; 2.0; 3.0; 5.0; 7.0; 9.0; 11.0; 13.0; 15.0;
17.0; 19.0$, respectively.
The cross-over lines with $U/t=2$ (marked red) separating
the weak and strong interaction regions are nearly
$N$-independent.
(From Ref.\cite{xu2018a}.)
}\label{fig:sun_mott}
\end{figure}

How interactions drive a partially filled band into an insulating state is an outstanding problem.
There exist two basic physical pictures - the Slater physics (Fermi surface nesting)
at weak coupling, and the Mott physics at strong coupling.
For the SU(2) case, the antiferromagnetic (AFM) order increases monotonically
and smoothly.
No phase transition exhists between the Slater and Mott regions
\cite{hirsch1983,hirsch1985,hirsch1989}.

There exist qualitative differences between the Slater and Mott regimes for
the two-dimensional SU($N$) Hubbard models arising from the enhanced
spin and charge fluctuations at $N>2$.
Previous large-$N$ studies in the literature mostly focus on the
antiferromagnetic Heisenberg models \cite{read1990,read1991}.
In contrast, the interplay between charge and spin physics is even more
challenging, which could be investigated via the sign-problem free quantum
Monte-Carlo (QMC) simulations.
The following fermonic SU($N$) Hubbard model at half-filling is
employed,
\bea
H&=&-t\sum_{\avg{ij}} \left\{ c^\dagger_{i\alpha} c_{j\alpha} +h.c.
\right\} \nn \\
&+& U\sum_i \left( n(i)-\frac{N}{2}\right)^2,
\label{eq:hubbardsun}
\eea
where $N$ is an even number.
The $U$-term is written in the particle-hole symmetric form,
which pins the average particle number per site at $N/2$,
i.e., half-filling.

QMC simulations indicate the fundamental difference between
the SU($N$) case and SU(2) case as shown in Fig. \ref{fig:sun_mott} ($a$)
for a square lattice.
The AFM orders in both SU(4) and SU(6) cases start to appear at small
$U$ in agreement with the Slater physics, where the single-particle
gaps are exponentially  small.
As $U$ further increases, the AFM orders reach the maxima and then decrease.
Meanwhile, the single-particle gaps scale linearly with $U$, marking
the onsent of Mott physics.
For the SU(6) case, the AFM order is completely suppressed
at a large value of $U_c\approx 13.3$.
Fitting the simulation data shows that the critical exponents of the
AFM order with $\nu = 0.60$ and $\eta= 0.44$ \cite{wang2019}.
At $U>U_c$, the transition to the valence-bond-solid (VBS) state is found,
which can be interpreted as the transition from the Slater regime
to the Mott regime where the local-moment super-exchange dominates.

How do interaction effects scale with $N$ with fixing the filling level
and the interaction $U$?
Sign-problem free QMC simulations have been performed for the half-filled SU($N$)
Hubbard models in 1D to address this problem \cite{xu2018a}.
Based on simulation results, we conjecture the existence of a universal
interacting state as $N\to \infty$ explained as follows:
The relative bandwidth is defined to reflect the correlation strength,
\bea
W_R(U,N)=E_{k,N}(U)/E_K(0),
\eea
where $E_{K,N}(U)$ is the kinetic energy per component with the interaction
parameter $U$ and component number $N$, and $E_K(0)$ is that at $U=0$.
Hence $W_R(U=0,N)=1$ for the free system, and it becomes 0 in the strong
coupling limit at $U=\infty$.
At small values of $U/t$, say $U/t\sim 1$, fermions are nearly itinerant,
and correlations manifest through inter-component collisions.
Hence, $W_R(U,N)$ decreases monotonically as $N$ increases which
enhances the collision possibility resulting in the amplification of
correlations.
In contrast, at large values of $U/t$, say, when $U/t>10$, increasing
$N$ softens the Mott insulating background.
The kinetic energy gained from virtual hoppings scales as
$N^2 t^2/U$, hence, $W_R(U,N)$ increases linearly as increasing $N$.
In the crossover region which lies around $U/t\approx 2$,
$W_R(U,N)\approx 0.9$ nearly independent on $N$.
Although the simulation data are still inconclusive, we conjecture that
\bea
\lim_{N\to \infty}  [1-W_R(U,N)]\approx 0.1,
\eea
which means an interacting large-$N$ limit.
It means that weak and strong interacting systems are driven to
a crossover region as $N\to \infty$, but from opposite directions
exhibiting a convergence of itinerancy and Mottness.
On the other hand,
\bea
\lim_{N\to \infty} \lim_{U\to 0} [1-W_R(U,N)]=0.
\eea
Hence,
there exists a singularity at $U\to 0$ and $N\to \infty$.
Other physical quantities, including the Fermi distribution, and
the spin structure factor, also exhibit nearly $N$-independent
behavior.
More analytic and numeric works are needed to further check if there
exists a universal strongly interacting limit with vanishing charge
gaps as $N\to \infty$, and its possible connection to
non-Fermi liquid states.

\subsection{Discussions}
The perspective of high symmetries (SU($N$), Sp($N$)) brings much richness
and novelty to studying large-spin ultra-cold fermions.
The large numbers of spin components render the system in the quantum
large-$N$ regime instead of the semi-classical large-$S$ regime.
We have reviewed systematically the hidden Sp(4) symmetries in
spin-$\frac{3}{2}$ systems, the unification based on
the $\chi$-pairing which is an SO(7) generalization of Yang's
$\eta$-pairing.
Quartet superfluidity, quartet density wave state, and plaquette singlet
formation in the Mott insulating state exhibit similar features of
multi-particle clustering correlations analogous to the color
singlet in quantum chromodynamics.
Interaction effects as varying the value of $N$ are investigated,
which show a tendency of convergence of itinerancy and Mottness
as $N\to \infty$.

On the experimental side, there have been significant progresses in the
past decade.
The SU(6) symmetric $^{173}$Yb \cite{taie2010,hara2011} and SU(10) symmetric
$^{87}$Sr \cite{desalvo2010} fermion atom systems have been experimentally
realized.
The nuclear spin, as well as the electron-orbital degree of freedom,
leads to rich physics \cite{gorshkov2010,hermele2009}.
Various SU($N$) symmetric quantum degenerate gases and Mott insulators in optical lattices have been realized~\cite{gorshkov2010,hara2011,
taie2010,taie2012,desalvo2010,fallani2014}.
As for spin-$\frac{3}{2}$ systems, there are a few candidate atoms
$^{132}$Cs, $^9$Be, $^{135}$Ba, $^{137}$Ba, and $^{201}$Hg.
Considering the rapid developments in this field, we
expect that the exotic Sp(4) physics could also be
experimentally investigated.

\section{Unconventional magnetism and spontaneous spin-orbit ordering}
\label{sect:pomeranchuk}
In the non-relativistic Fermi liquid theory, spin is an internal symmetry
independent of orbital rotations, which rigorsly speaking should be
denoted as ``isospin" instead of spin from the relativistic perspective.
In the mechanism of unconventional magnetic transitions, ``isospin"
develops entanglement with momentum orientation and genuinely becomes the
physical spin, hence, it shares the same spirit of ``spin-from-isospin" in
gauge theories \cite{jackiw1976}.
The consequential states can be viewed as ``non-$s$-wave'' generalizations
of ferromagnetic metals in which spin no longer polarizes along a unique
direction but varies with momentum forming a non-trivial representation
of the rotation group \cite{wu2004,wu2007}.

In other words, effective spin-orbit coupling is generated as an order
parameter through the Pomeranchuk type of Fermi surface instabilities,
which is tunable by external parameters such as temperature and pressure.
Furthermore, similar to magnetic fluctuations in ferromagnets, this
effective spin-orbit coupling possesses its collective mode dynamics.
This gives rise to a conceptually new mechanism to generate spin-orbit
coupling dynamically without involving relativity \cite{wu2004a,wu2007}.
Due to the richness of many-body physics, unconventional
magnetism potentially provides a new way to engineer spin-orbit
couplings and to control electron spins.



\subsection{Fermi liquid theory and Pomeranchuk instabilities}
In this subsection, we briefly review the concept of the non-relativistic
Fermi liquid theory and Pomeranchuk instability \cite{pomeranchuk1959}.

A large part of our current understanding of interacting electronic systems
is based on the Landau Fermi liquid theory, which was designed originally
for the normal state $^3$He and also applies to most metals
\cite{leggett1975,baym1991}.
The central assumption is the existence of the well-defined Fermi surface and fermionic quasi-particles, which exist as long-lived states at very low
energies.
Interactions among quasi-particles, which are reflected by the forward scattering processes of quasi-particles near the Fermi surface, are described by the Landau interaction functions.
The Landau interaction function can be classified into the density
(particle-hole singlet ) and spin (particle-hole triplet) channels,
which are also traditionally denoted as symmetric ($s$) and asymmetric
($a$) channels, respectively,
\bea
f_{\alpha\beta,\gamma\delta}(\mathbf{\hat k}_1, \mathbf{\hat k}_2)
=f^s(\mathbf{\hat k}_1, \mathbf{\hat k}_2)+f^a(\mathbf{\hat k}_1, \mathbf{\hat k}_2)
\vec \sigma_{\alpha\beta}\cdot \vec\sigma_{\gamma\delta},
\nn \\
\eea
where $\mathbf{\hat k}$ is the direction of quasi-particle momentum close
to the Fermi surface.
Each of them can be further decomposed into different orbital
partial wave channels as
\bea
f^{s,a}(\mathbf{\hat k}, \mathbf{\hat k}^\prime)=\sum_l f^{s,a}_l
P_l (\mathbf{\hat k} \cdot \mathbf{\hat k}^\prime)
\label{eq:landau}
\eea
where $P_l$ is the $l$-th Legendre polynomial and
$l$ denotes the orbital angular momentum of the partial wave channel.

In the Landau Fermi liquid theory, the interactions among
quasi-particles are captured by a few dimensionless Landau
parameters
\bea
F^{s,a}_l=N_0 f^{s,a}_l,
\eea
where $N_0$ is the density of states on the Fermi surface.
Physical susceptibility in each channel acquires significant
renormalizations by the Landau interactions,
\bea
\chi^{s(a)}_{FL, l} =\chi_{0,l}
\frac{1+F^s_1/3}{1+F^{s(a)}_l/(2l+1)},
\label{eq:sus}
\eea
where$\chi_{0,l}$ is the susceptibility of free fermi gas.
Spin susceptibility lies in the $F^a_0$ channel,
and compressibility lies in the $F^s_0$ channel.

Pomeranchuk instabilities refer to a large class of
instabilities of Fermi surface distortions in both the density
and spin channels \cite{pomeranchuk1959}.
In order for the Fermi surface to be stable, Landau parameters $F_l^{s(a)}$
cannot be too negative.
Otherwise, Fermi surface distortions will occur.
The Fermi surface could be viewed as an elastic membrane in momentum space.
Let us perturb the Fermi surface and expand the energy cost in different
partial-wave channels.
We arrive at
\bea
\frac{\Delta E}{V}= \frac{2\pi}{N_0} \sum_{lm}
\Big \{(1+\frac{F^{s,a}_{l}}{2l+1})|\delta n^{s,a}_{lm}|^2\Big\},
\eea
where $\delta n^{s(a)}_{lm}$ are amplitudes of Fermi surface distortions
in the corresponding partial-wave channels, and $V$ the system volume.
The first term is the kinetic energy cost which is always
positive, while the second term is the interaction contribution,
which can be either positive or negative.
When $F_l^{s,a}<-(2l+1)$, the surface tension of the Fermi surface
goes negative, and develops instability in the corresponding
channels, which is consistent with the divergence of susceptibility in
Eq. (\ref{eq:sus}) at $F_l^{s,a}=-(2l+1)$.

The most familiar Pomeranchuk instabilities are found
in the $s$-wave channel, i.e., ferromagnetism at $F^a_0<-1$
and phase separation at $F^s_0<-1$.
Pomeranchuk instabilities in non-$s$-wave wave channels $(l\ge 1)$
have been attracting a great deal of attention
\cite{wu2004a,wu2007,hirsch1990,hirsch1990a,barci2003,
lawler2006,halboth2000, dellanna2006,
kee2003, kivelson2003, gorkov1992,
varma2005, varma2006, kee2005, honerkamp2005}.
The density channel instabilities result in uniform but
anisotropic liquid (nematic) phases \cite{oganesyan2001},
which have been investigated in the context of doped Mott insulators~\cite{kivelson1998} and high T$_c$
materials \cite{kivelson2003}.
Experimental evidence  has also been found in ultra-high mobility
two-dimensional electron gases (2DEG) in  Al$_x$Ga$_{1-x}$As-GaAs
heterostructures and quantum wells
in nearly half-filled high Landau levels at very low temperatures
\cite{lilly1999,du1999}, and near the metamagnetic
transition of the ultra-clean samples of the
bilayer ruthenate Sr$_3$Ru$_2$O$_7$
\cite{grigera2001,grigera2004,borzi2007}.

Unconventional magnetism corresponds to Pomeranchuk instabilities in the
spin channel with $l\ge 1$ \cite{hirsch1990, hirsch1990a,wu2004a,wu2007,
varma2005,varma2006,kee2005}.
In Ref. \cite{wu2004a}, these states are classified by the author and Zhang
as isotropic and anisotropic phases dubbed $\beta$ and $\alpha$-phases,
respectively.
The $\alpha$-phases was studied by many groups before: The $p$-wave phase
was first studied by Hirsch \cite{hirsch1990, hirsch1990a} under the
name of the ``spin-split'' state, and was also proposed by Varma
{\it et al.} \cite{varma2005, varma2006}
as a candidate for the hidden order phenomenon in URu$_2$Si$_2$;
the $d$-wave phase was studied by Oganesyan {\it et al.}
\cite{oganesyan2001} under the name of ``nematic-spin-nematic'' phase.
Systematic studies of ground state properties and collective
excitations in both the anisotropic $\alpha$ and isotropic
$\beta$-phases have been performed \cite{wu2004, wu2007}.
Chubukov and Maslov found that when approaching the ferromagnetic quantum
critical point, the $p$-wave channel spin Pomeranchuk instability develops
before the ferromagnetic instability \cite{chubukov2009}.

\subsection{Unconventional magnetism as multipolar orderings}
\label{sect:mulpolar}
The unconventional magnetic order parameters are defined as
multipolar parameters in momentum space but not in coordinate space \cite{wu2004a,wu2007}.
For simplicity, we first take the  2D $p$-wave case as an example.
Its order parameters are the $x$ and $y$-spatial components of
spin-dipole moments defined as
\bea
\mathbf{n}_1= \frac{|f^a_1|}{V} \sum_{\mathbf{k}} ~\mathbf{ s}(\mathbf{k})
~\hat k_x, \ \ \
\mathbf{n}_2= \frac{|f^a_1|}{V} \sum_{\mathbf{k}} ~\mathbf{s}(\mathbf{k})
~ \hat k_y,
\label{eq:spindipole}
\eea
where $f^a_1$ is the Landau interaction parameter defined in
Eq. (\ref{eq:landau});
$\hat k_{x,y}=k_{x,y}/|k|$ are the $p$-wave angular form factors;
$\mathbf{s}(\mathbf{k})=\avg{c^\dagger_{\mathbf{k}\alpha}
\vec \sigma_{\alpha\beta} c_{\mathbf{k}\beta}}$
is the spin-moment of momentum $\mathbf{k}$,
and $\avg{}$ means ground state expectation valule.
Each of $\mathbf{n}_{1,2}$ is a 3-vector in spin space.
This is a natural generalization of the ferromagnetic
moment $\mathbf{S}=\sum_\mathbf{k} \mathbf{s}( \mathbf{k})$
whose $s$-wave angular form factor is just a constant.

In the anisotropic $p$-wave $\alpha$-phase depicted in Fig.
\ref{fig:pwave_mag} (C), the order parameter configuration  is
equivalent to only one of $\mathbf{n}_1$ and $\mathbf{n}_2$ is nonzero,
or more generally, $\mathbf{n}_1 \parallel \mathbf{n}_2$.
Their orientation in spin space is arbitrary.
The order parameter configuration in the $p$-wave $\beta$ phase
depicted in Fig. \ref{fig:pwave_mag} (B) shows that
$\avg{n^x_1} = \avg{n^y_2} \neq 0$.
More generally, this is equivalent to both $\mathbf{n}_{1,2}
\neq 0$ and their orientations are perpendicular to each
other as $\mathbf{n}_1 \perp \mathbf{n}_2$.
Using an optics analogy, the spin configuration over the Fermi
surface in the $\alpha$-phase is linearly poloarized, while
that in the $\beta$-phase is circularly polarized.

This order parameter definition can be easily generalized into other
partial wave channels in 2D and 3D systems by using the corresponding
multipolar angular form factors.
For example, the 2D $d$-wave channel order parameters can be defined
as components of spin-quadrupole moments
\bea
\mathbf{n}^d_1&=&\frac{|f^a_2|}{V}\sum_{\mathbf{k}}
~\mathbf{s(\mathbf{k})} ~\cos 2 \phi_{\mathbf{k}}, \nn\\
\mathbf{n}^d_2&=&\frac{|f^a_2|}{V}\sum_{\mathbf{k}}
~\mathbf{s}(\mathbf{k})~ \sin 2  \phi_{\mathbf{k}},
\label{eq:spinquadrupole}
\eea
where $\phi_k$ is the azimuthal angle of $\mathbf{k}$.
We could also combine them as a matrix form $n^{\mu,b}$
with each column representing a 3-vector $\mathbf{n}_b$.
Below we will use the matrix and vector forms of
order parameters interchangeably.

The 3D counterpart of these expressions can be written in
terms of spherical harmonic functions.
Hence, in 3D the Latin label $b$ of the order parameter
$n^{\mu b}$ take $2l+1$ values, while the Greek index
$\mu$ still takes $x,y,z$.

We consider a 2D Fermi-liquid system focusing on a general partial-wave
channel-$l$.
Since there is no spin-orbit coupling, the symmetry is the direct product
$SO_L(2) \otimes SO_S(3)$, where $L$ and $S$ refer to the orbit
and spin channels, respectively.
The Landau interaction function $f_a^l$ could depend on the total
momentum $\mathbf{q}$ of the particle-hole excitations with the
assumption that
\bea
f(\mathbf{q}) = \frac{f^a_l}{ 1+\kappa |f^a_l| q^2},
\eea
which gives rise to an interaction range $\xi\approx \sqrt{\kappa  |f^a_l|}$.
Mean-field theory is valid when $\xi\gg d \approx 1/k_F$, where $d$
is the inter-particle distance.
After the mean-field decomposition, the mean-field Hamiltonian becomes
\begin{widetext}
\bea
H_{MF}&=&
\sum_\mathbf{k}
\psi^\dagger_\alpha (\mathbf{k}) \left[\epsilon(\mathbf{k})-\mu
-\left( \mathbf{n}_1 \cos( l\theta_\mathbf{k})
+   \mathbf{n}_2 \sin (l\theta_\mathbf{k})\right)
\cdot \vec \sigma \right] \psi_\beta(\mathbf{k})
+\frac{|n_1|^2 +|n_2|^2}{2|f^a_l|}.
\label{eq:meanfield}
\eea
\end{widetext}
The validity of mean-field theory at quantum criticality
requires an analysis of quantum fluctuations
which are not included in mean-field theory~\cite{hertz1976,millis1993}.

To determine the ground state configuration of $\vec n_{1,2}$,
the Ginzburg-Landau (GL) free energy is constructed as,
\bea
F(\mathbf{n}_1, \mathbf{n}_2)&=&\gamma_1 \partial_a \mathbf{n}_b
\cdot \partial_a
\mathbf{n}_b +
r(\mathbf{n}^2_1+\mathbf{n}^2_2) +
v_1 [\mathbf{n}^2_1+\mathbf{n}^2_2]^2\nn \\
&+&v_2 |\mathbf{n}_1 \times \mathbf{n}_2|^2.
\label{eq:GL}
\eea
The coefficients $r$, $v_{1,2}$ are calculated from mean-field
free energy in Ref. \cite{wu2007}, whose expressions are
omitted here.

When $l=1$, a new gradient term can appear which contains
the linear order spatial derivative and the cubic order of
order parameters as
\bea
\Delta F(\mathbf{n}_1, \mathbf{n}_2)&=&
\gamma_2 \epsilon_{\mu\nu\lambda} n^{\mu a}n^{\nu b} \partial_a
n^{\lambda b}\nn \\
&=&\gamma_2 \{ (\partial_x \mathbf{n}_2 -\partial_y \mathbf{n}_1)
\cdot (\mathbf{n}_1 \times \mathbf{n}_2) \}.
\label{eq:lifshitz}
\eea
Such a term is allowed because $\mathbf{n}_{1,2}$ are odd under
parity transformation and even under time-reversal transformation,
i.e., $P \mathbf{n}_{1,2} P^{-1}=-\mathbf{n}_{1,2}$, and
$T \mathbf{n}_{1,2} T^{-1}=\mathbf{n}_{1,2}$.
It does not bring much effect in the normal phase because it
is at the cubic order of the order parameter.
However, we will see in Sect. \ref{sect:goldstonebeta}, this term becomes
important in the
ordered $p$-wave $\beta$-phase, which drives a Lifshitz transition
spontaneously developing a chiral pitch.

Both $\alpha$ and $\beta$-phases are driven by the negative value of $r$,
i.e., $F_l<-2$ in 2D.
Whether the ground state takes the $\beta$ or $\alpha$-phase
depends on the sign of $v_2$.
If $v_2<0$, Eq. (\ref{eq:GL}) favors $\mathbf{n}_1 \perp \mathbf{n}_2$,
thus gives rise to the $\beta$-phase.
On the other hand, $\alpha$-phase appears at $v_2>0$,
which favors $\mathbf{n}_1 \parallel \mathbf{n}_2$.

\subsection{``Spin from isospin" in non-relativistic systems}
\label{subsect:dyn}
\begin{figure}
\centering\epsfig{file=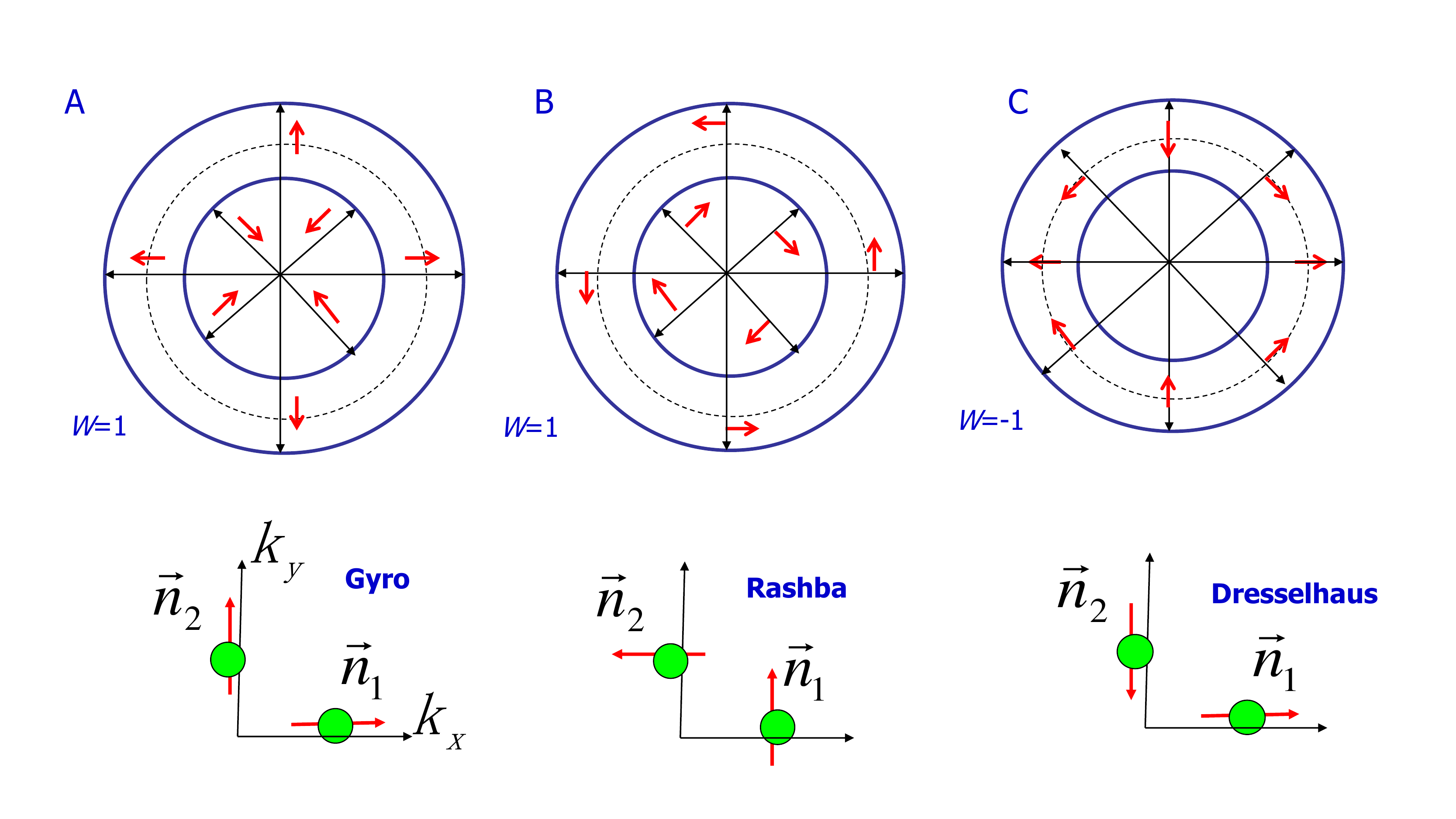,clip=1,width=\linewidth,angle=0}
\caption{Spontaneous spin-orbit orderings in the $\beta$-phases.
Order parameter configurations and the momentum space vortices
with the winding numbers $w=\pm 1 $.
(A)Gyrotropic ($w=1$), (B)Rashba ($w=1$), (C)
Dresselhaus ($w=-1$).
From Ref. \cite{wu2007}.
}\label{fig:vortex}
\end{figure}

Spin in the relativistic theory, by definition, is part of the generators
of the rotation transformation.
Hence, it is always coupled to momentum
as required by the Lorentz invariance.
While in the non-relativistic theory, it decouples from momentum,
and becomes an ``isospin" type internal degree of freedom.
The $\alpha$ and $\beta$-phases entangle spin with momentum
together via order parameters.
In this sense, spin genuinely changes from the status of ``isospin"
into spin.
As we explained before, this effective spin-orbit coupling arises
from many-body interaction instead of the single-particle
relativistic physics.

In the isotropic $\beta$-phase with $l\ge 1$, spin winds around the
Fermi surface, exhibiting a vortex-like structure in momentum space.
For the 2D $p$-wave $\beta$-phase depicted in Fig. \ref{fig:vortex} A,
its mean-field single particle Hamiltonian reads
\bea
H_{MF,\beta}&=& \sum_\mathbf{k} \psi^\dagger(\mathbf{k})
\left[ \epsilon_0(\mathbf{k})-\mu
-\bar n ( \hat k_x \sigma_x +   \hat  k_y \sigma_y ) \right]
\psi(\mathbf{k}), \nn \\
\label{eq:beta}
\eea
where $|n_1|=|n_2|=\bar n$.
It exhibits a $\vec \sigma \cdot \mathbf{k}$ type spin-orbit coupling,
which is called the gyrotropic spin-orbit coupling \cite{fu2015}.
The fermion spectrum is isotropic as $\epsilon(\mathbf{k})
=\epsilon_0(\mathbf{k})\pm \bar n$ in the $\beta$-phase.
Similarly to the ferromagnet, Fermi surfaces in the $\beta$-phase
split into large and small circles.
However, they are characterized by helicity, i.e., the spin projection
along its momentum, not by spin polarization.

The symmetry breaking in the $\beta$-phases is particularly interesting.
The normal Fermi liquid state has both spin  and orbital
rotational symmetries.
The state depicted in Fig. \ref{fig:vortex} A
is still isotropic where the total angular momentum
$\mathbf{J}=\mathbf{L} +\mathbf{S}$ remains
conserved although $\mathbf{L}$ and $\mathbf{S}$ are no longer
separately conserved.
If we fix momentum and only rotate spin, the configuration
in Fig. \ref{fig:vortex} A changes.
In other words, the relative spin-orbit symmetry is broken,
a concept first proposed by Leggett in superfluid $^3$He
systems \cite{leggett1975}.

In solid state physics, Rashba and Dresselhaus are two familiar
spin-orbit couplings whose spin configurations in momentum space
are depicted in Fig. \ref{fig:vortex} B and C, which corresponds
to order parameter configurations of
$(\mathbf{n}_1 \parallel \hat y$, $\mathbf{n}_2 \parallel -\hat x)$,
and $(\mathbf{n}_1 \parallel \hat x$, $\mathbf{n}_2 \parallel -\hat y)$,
respectively.
These two spin-orbit couplings are equivalent to the gyrotropic one
in Eq. (\ref{eq:beta}) up to a global spin rotation.
Starting from the configuration depicted Fig. \ref{fig:vortex} A,
we can arrive at the Rashba configuration by fixing $\mathbf{k}$
unchanged and rotating electron spin around the $z$-axis at 90$^\circ$
of each $\mathbf{k}$.
Similarly the Dresselhaus configuration can be obtained by
the rotation around the $x$-axis at $180^\circ$.
These ground state spin configurations exhibit, in momentum space,
the vortex structures with the winding numbers $w=\pm1$.
In principle, we can perform an abitary spin rotation to obtain
all the equivalent states, thus the ground state Goldstone
manifold is $[SO_L(2)\otimes SO_S(3)/SO(2)_{J}=SO(3)$.

This vortex picture in momentum space can be generalized into a
general $F^a_l$ channel with the winding numbers $\pm l$.
In fact, the generated spin-orbit coupling pattern
is beyond the relativity framework.
In particular, for even values of $l$, the dynamic spin-orbit orders
break time-reversal symmetry, while the relativistic spin-orbit
coupling is time-reversal invariant.
The mean-field Hamiltonian
$H_{\beta,l}$ for the $\beta$-phase in angular momentum channel $l$
can be expressed through a $d$-vector, defined by
$\mathbf{d}(\mathbf{k}) =\left(\cos (l \theta_\mathbf{k}) ,
\sin (l\theta_\mathbf{k}) , 0\right)$, as
\bea
H_{\beta,l}&=&
\sum_\mathbf{k}
\psi^\dagger(\mathbf{k}) \left[\epsilon(\mathbf{k})-\mu
-\bar n \vec d (\theta_\mathbf{k}) \cdot \vec \sigma) \right]
\psi(\mathbf{k}),
\label{eq:Hbeta}
\eea
where $\mathbf{d} (\theta_\mathbf{k})$ is the spin quantization axis for
the single particle state at $\mathbf{k}$.
Each Fermi surface is characterized by the eigenvalues $\pm 1$ of
the helicity operators  $\vec \sigma \cdot \mathbf{d}(\mathbf{\hat k})$.



The mean-field Hamiltonian in the anisotropic $\alpha$-phase
(Fig. \ref{fig:pwave_mag} C)  can be written as
\bea
H_{MF,\alpha}&=& \sum_\mathbf{k}
\psi^\dagger (\mathbf{k}) \{\epsilon_0(\mathbf{k})-\mu
-\bar n  \hat k_x \sigma_z  \} \psi(\mathbf{k}).
\label{eq:alpha}
\eea
The fermion spectra read $\epsilon(k)=\epsilon_0(k)\pm \bar n \hat k_x$,
and the spin up and down Fermi surfaces shift
to left and right, respectively.
This configuration is equivalent to the combination of
Rashba and Dresselhaus spin-orbit couplings with an equal strength.
It is an anisotropic phase in both spin and orbit channels.
Generally the Fermi surface shift can along any in-plane direction,
and the spin axis can pick any 3D direction,
thus the ground state Goldstone manifold is
$[SO_L(2) \otimes SO_S(3)]/[SO_S(2)] = SO_L(2) \otimes S_2$.

These $\beta$ and $\alpha$-phases are particle-hole channel analogies
to the triplet $p$-wave pairing superfluid $^3$He-$B$ and $A$ phases,
respectively.
The order parameters in $^3$He are defined as $x,y$ and $z$-spatial
components of the dipole-moment of the Cooper pairing amplitude
over the Fermi surfaces \cite{leggett1975,vollhardt1990}.
They are defined as
\bea
\mathbf{\Delta}_i= \sum_\mathbf{k} \mathbf{\Delta}(\mathbf{k})
~ \hat k_i \ \ \ (i=x,y,z),
\eea
where $\mathbf{\Delta}(\mathbf{k})= \avg{c^\dagger_\alpha(\mathbf{k})
(i\sigma_2\vec \sigma)_{\alpha\beta} c^\dagger_\beta(-\mathbf{k})}$.
Each one of $\mathbf{\Delta}_{x}, \mathbf{\Delta}_y$ and
$\mathbf{\Delta}_z$ is a 3-vector in spin space.
In the $B$-phase,  $\mathbf{\Delta}_{x,y,z}$ are perpendicular to
each other forming a triad.
In the $A$-phase only two of them are nonzero with a phase
difference of $\frac{\pi}{2}$, and they are parallel to each other
in spin space.
As a result, the $B$-phase is essentially isotropic with a constant
gap over the Fermi surface, while the $A$-phase is anisotropic
with nodes.

From the symmetry point of view, the unconventional magnetic
$\beta$ and $\alpha$ phases exhibit similar properties to
the $^3$He-$B$ and $A$ phases under spatial rotations.
The angular form factor of the gap functions in the
$^3$He-B and A phases are very similar to the Fermi surface
splittings in the $p$-wave magnetic $\beta$ and $\alpha$-phases,
respectively.

\subsection{Collective excitations in unconventional magnetic states}
\label{sect:collect}
As a result of spontaneous symmetry  breaking, unconventional magnetic
states exhibit low energy excitations.
In this subsection, we review the Goldstone modes in
both $\alpha$ and $\beta$-phases.
Such modes are absent in the conventional spin-orbit coupling systems.

\subsubsection{Goldstone modes in the $\alpha$-phase}
We first comment on the stability of the $p$-wave $\alpha$-phase.
The Ginzburg-Landau energy of Eq. (\ref{eq:lifshitz}) contains a cubic
term linear in spatial derivatives.
It might induce a linear derivative coupling between the massless Goldstone
modes at the quadratic level, leading
to a Lifshitz instability in the ground state.
However, as will be shown below, the Goldstone modes in the $\alpha$-phase
share the same index in either the orbital or the spin channel
as the condensed mode.
Hence, they cannot be coupled together by Eq. (\ref{eq:lifshitz}).

The $\alpha$-phases break rotational symmetries in both orbital
and spin channels, hence, the Goldstone modes can be
classified into density and spin channel modes, respectively.
Without loss of generality, we assume the ordered configuration
as shown in Fig. \ref{fig:pwave_mag} C,
\bea
\avg{n_{\mu b}}=\bar n \delta_{\mu z}\delta_{b 1},
\mbox {~~i.e.~~}, \mathbf{n}_1= \bar n \hat e_z, \mathbf{n}_2=0.
\label{eq:alpha_order}
\eea
In other words, spin configuration is along $\pm \mathbf{\hat z}$,
and Fermi surface distortion is along the $x$-axis.
Three collective modes are Goldstone modes, including one branch in the density channel, and two
branches in the spin channel.

The density channel Goldstone mode is the oscillation of the distorted
Fermi surface.
It is associated with the field
$n_{2}^z$,
\bea
n^z_2(\mathbf{q})= -\frac{f^a_1(\mathbf{q})}{V}
\sum_{\mathbf k}   \psi^\dagger_{\mathbf{k+q}}
\vec \sigma   \psi_{\mathbf k} k_y,
\eea
which describes the Fermi surface oscillation in the $y$-direction
while keeping the spin configuration unchanged.
Calculations at the random-phase approximation (RPA) level
show the effective Lagrangian,
\bea
L_{FS}^\alpha(\mathbf{q},\omega)=N_0
\left[ \frac{q\xi^2}{|F^a_l|}-i\frac{\omega}{2v_f q}
(1+2\cos 2\phi_q)
\right].
\eea
This Goldstone mode is overdamped because of the Landau damping,
and the damping is anisotropic depending on propagation directions.

The spin channel Goldstone modes $n_{sp,x\pm iy}$ describe spin
oscillations while keeping the Fermi surface unchanged,
which are spin-dipole precession modes.
In contrast, they exhibit nearly isotropic underdamped dispersion
relations at small propagating wavevectors with the dispersion relation
\bea
\omega^2_{x\pm iy}= \frac{\bar n^2}{|F^a_1|}
(q\xi)^2.
\eea
Different from spin-waves in the ferromagnets, the dispersion
relation here is linear with momentum, which is a consequence
of time-reversal symmetry.

\subsubsection{Goldstone modes in the $\beta$-phase}
\label{sect:goldstonebeta}
We further study the Goldstone modes in the $\beta$-phase.
For simplicity, we consider the 3D $\beta$-phase with the isotropic
ground state exhibiting
\bea
n^{\mu a}=\bar n \delta_{\mu a}.
\eea
In other words, $\mathbf{n_{1,2,3}}$ form an orthogonal triad.
The total angular momentum $\mathbf{J}$ remains conserved, such that
fluctuations of $\delta n^{\mu a}$ are classified into eigenstates
of $\mathbf{J}$ as $O_{jj_z}(\mathbf{q}, \omega)$.
$j=0,1,2$ mean the singlet, triplet and quintet channels respectively,
and $j_z$ is the SO(2) quantum number rotating around
the propagation direction $\mathbf{q}$.

The Goldstone modes belong to the triplet channel $(j=1)$,
which are the small relative spin-orbit rotations as
\bea
O_{1,0}(\mathbf{q}, \omega) &=&\frac{1}{\sqrt 2} \epsilon_{z \mu a}
\delta n^{\mu a}(\mathbf{q}, \omega), \ \ \
\nn \\
O_{1,\pm1}(\mathbf{q}, \omega)&=& \frac{1}{ 2} (\epsilon_{x\mu a}\pm i
\epsilon_{y\mu a}) \delta n^{\mu a}(\mathbf{q}, \omega).
\eea
The RPA approximation gives the dispersion relations,
\bea\label{eq:ch2_GS}
\omega^2= 4\bar n^2 |F^a_1|
\Big(
\frac{\kappa q^2}{N_f}+
\frac{j_z  |q| x}{18 k_f} \Big)
\ \ \ (x=\frac{\bar n}{v_f k_f}, j_z=0,\pm 1),
\nn \\
\eea
which is valid in the low energy regime
of $\omega, v_f q \ll \bar n \ll v_f k_f$.

The linear dependence on $q$ in the dispersion Eq. (\ref{eq:ch2_GS})
is due to the broken parity in the ordered $\beta$ phase.
Consequently, $\omega^2$ becomes negative for the branch with helicity
$j_z=-1$ at small $q$,
This means that the uniform ground state in the $\beta$-phase
is unstable, instead it exhibits a Lifshitz-like instability.
This behavior is a general feature in systems with broken parity
such as the spiral order in helical magnets, and the cholesteric
liquid crystals\cite{degennes}.
The true ground state configuration in the $\beta$ phase is
complicated with the preliminary analysis presented in Ref. \cite{wu2007}.

\subsubsection{Resonances in inelastic neutron scattering spectroscopy}
\label{sect:neutron}
The unconventional magnetic orders are spin-multipole moment in
momentum space and cannot couple to neutron magnetic moments
statically.
Hence, there should be no
elastic Bragg peaks.
The spin-channel Goldstone modes in the $\alpha$-phase do not couple to
neutron moments directly, either.
Nevertheless, they carry spin quantum numbers and thus couple to spin-wave modes dynamically.

Consider the following commutation relations,
\bea
[S_x,  n^y_1]= i n^z_1, \ \ \, [S_y,  n^x_1]= -i n^z_1.
\label{eq:alpha_res}
\eea
In the $p$-wave $\alpha$-phase with the configuration given in Eq. (\ref{eq:alpha_order}), $n^z_1$ can be replaced by the constants
of $\pm i \bar n$.
As a result, the Goldstone modes $n_1^x$ and $n_1^y$ become conjugate
to spin, and the coupling to spin is developed dynamically.

More formally, we can write down the following coupling Lagrangian,
\bea
L=\left( \vec n_1 \times \partial_t \vec n_1+
\vec n_2 \times \partial_t \vec n_2 \right) \cdot \vec S.
\eea
In the ordered state of Eq. (\ref{eq:alpha_order}), it is reduced to
\bea
L=\bar n \left( S_y \partial_t n_{1x} -S_x \partial_t n_{1y} \right).
\label{eq:neutron}
\eea
The RPA approximation shows that the dynamic spin-spin
correlation function behaves as
\bea
\chi_s(\mathbf{q},\omega)&=&\avg{S_+(\mathbf{q},\omega)
S_-(-\mathbf{q},-\omega)} \nn \\
&=&\frac{N_0 \frac{\omega^2}{\bar n^2}}
{\frac{\kappa q^2}{N_0}-\frac{2}{|F^a_1|}
\frac{\omega^2}{\bar n^2}-i\delta}.
\eea
Hence, it induces a resonance part in the transverse spin
wave-excitations.
The spectral functions exhibit the $\delta$-peak at
the excitation energy of the Goldstone mode,
\bea
\Im \chi_s(\mathbf{q},\omega)=\kappa \pi v_f^2
\bar n^2 q^2|F_1^a|^2 \delta(\omega^2-\omega_q^2),
\eea
which can be detected in the inelastic
neutron scattering experiments.
This is very similar to the interpretation of the SO(5) theory to
the neutron resonance mode: the $\pi$-mode lies in the particle-particle
channel which decouples from spin in the normal state,
but becomes conjugate to spin in the superconducting state
giving rise to spin resonances \cite{demler1995,demler2004,
zhang1997}.

Such a resonance peak only exhibits in the ordered phase,
and vanishes  in the disordered phase.
As shown in Eq. (\ref{eq:neutron}), in the anisotropic $\alpha$-phases
this resonance only occurs in spin-flip channels.
Similar analysis can also be performed in the isotropic $\beta$-phases,
in which the resonances occur in both spin-flip and
non-flip channels \cite{wu2007}.

\subsection{Spin-orbit coupled Fermi liquid theory}
So far we have considered the dynamic generation of spin-orbit
coupling in non-relativistic Fermi
liquid theory.
Nevertheless in materials with heavy elements, there does exist the
relativistic spin-orbit coupling.

If a system does not exhibit inversion symmetry, the relativistic
spin-orbit coupling leads to Fermi surface splitting, say, the Rashba
type.
In this case, the relativistic spin-orbit coupling behaves like
an external field which would round off the unconventional
phase transition and pin down a particular spin-orbit ordering configuration.
This situation is similar to cooling a magnet below the
transition temperature in an external magnetic field.

On the other hand, if a system Hamiltonian still preserves both parity and
time-reversal symmetries, the Fermi surface should remain doubly
degenerate.
Spin-orbit coupling does not manifest itself in the Fermi surface splitting
but should exhibit in the Landau Fermi liquid theory.
Such a situation also occurs in the presence of prominent
magnetic dipolar interactions,
which is invariant under simultaneous rotations in both orbital and
spin channels, but not under a rotation in either channel.
Landau-Fermi liquid theory has been extended to this situation
\cite{liwu2012,fu2015}.

In the inversion invariant spin-orbit coupled systems, the fermion
distribution function is reorganized in the spin-orbit coupled
bases as
\bea
\delta n_{\alpha\alpha^\prime}(\hat k)
&=&\sum_{JJ_z;LS} \delta n_{JJ_z;LS} ~
Y_{JJ_z;LS}(\hat k, \alpha\alpha^\prime),
\eea
where
$Y_{JJz;LS}(\hat k, \alpha\alpha^\prime)$
is the spin-orbit coupled spherical harmonic functions
\bea
Y_{JJz;LS}(\hat k, \alpha\alpha^\prime)
&=&\sum_{ms_z}\avg{LmSs_z|JJ_z} Y_{Lm}(\hat k)
\chi_{Ss_z,\alpha\alpha^\prime},
\nn \\
\label{eq:socouple}
\eea
and
$\chi_{Ss_z,\alpha\alpha^\prime}$ is the bases for the particle-hole singlet
(density) channel and triplet (spin) channel, respectively.
The Landau interaction function is generalized to the
interaction matrix,
\bea
\frac{N_0}{4\pi } f_{\alpha\alpha^\prime;\beta\beta^\prime}
(\hat k, \hat k^\prime)&=&
\sum_{JJ_z L L^\prime}
Y_{JJz;L1}(\hat k,\alpha\alpha^\prime) F_{JJ_z L1;JJ_z L^\prime 1}
\nn \\
&\times&
 Y^\dagger_{JJz;L^\prime 1}(\hat k,\beta\beta^\prime),
\eea
where we only keep the particle-hole triplet component.
The Landau matrix is diagonal with respect to the total angular
momentum $J$ and its $z$-component $J_z$, but may have
off-diagonal elements with $L\neq L^\prime$.
Constrained by the inversion symmetry, $L-L^\prime=0,2$.

Similarly to the non-relativistic case, when
an eigenvalue of the Landau interaction matrix is negatively large, i.e,
$\lambda<-1$, it triggers the Pomeranchuk instability in the corresponding
channel.
For example, the instability in the channel with
$J=1^-$, $L=S=1$, where $``-"$ means odd parity,
generates the 3D analogy of the Rashba spin-orbit coupling,
\bea
H_{so,1^-}=|n|\sum_k \psi^\dagger(\mathbf{k})
(\mathbf{k}\times \vec \sigma)\cdot \mathbf{\hat{l}}
\psi(\mathbf{k}),
\eea
where $\mathbf{\hat{l}}$ is a 3D unit direction,
$|n|$ is the magnitude of the spin-orbit order parameter.
The Pomeranchuk instability promotes it to the single particle
level by breaking the rotational symmetry and parity.

Let us still use the order parameter $n^{\mu b}$ defined in
Sect. \ref{sect:mulpolar} to represent the order parameters in
the sector $L=S=1$ for a 3D inversion invariant spin-orbit
coupled Fermi liquid theory.
The $3\times 3$ matrix of $n$ includes three sectors of
$J=0, 1, 2$, which corresponds to pseudo-scalar (gryotropic),
vector (Rashba), and tensor (Dresselhaus) type spin-orbit
coupling, respectively.

The Ginzburg-Landau free energy can be constructed as
$F=F_0+\Delta F$,
\bea
F_0&=&r_0 \tr (n^T n)+\beta_1 \left(\tr (n^T n)\right)^2 +\beta_2
\tr  \left (n^T n\right)^2,  \nn \\
\Delta F&=& \frac{r_1}{3} (\tr n)^2 +\frac{r_2}{4} \tr (n^T-n)^2.
\eea
Under $SO_L(3)$ and $SO_S(3)$ rotations, $n$ is transformed
as $n\to T_S n T_L^\dagger$, where $T_{L,S}$ is the rotation matrix
in the orbit and spin channels, respectively.
$F_0$ is invariant under independent $T_L$ and $T_S$, and $\Delta F$
is invariant under simultaneous spin-orbit rotations,
but not under a rotation in either channel.
The $r_{1,2}$ terms are an analogy to magnetic anisotropy for magnetic phase
transitions, which lead to different types of universal classes.
Then at the quadratic level, the eigenvalues of the pseudo-scalar,
vector, tensor channels are determined by $r_0+r_1$, $r_0+r_2$,
and $r_0$, respectively.
The actual ordering depends on which eigenvalue is negatively most
dominant.
If the pseudo-scalar channel instability dominates, the phase transition
only breaks parity, which is an Ising type transition without
the Goldstone mode.
If the vector channel instability dominates, there is also rotational
symmetry breaking with the Goldstone manifold $S^2$.
The symmetry breaking pattern for the tensor ordering channel is more
involved, the Goldstone manifold is formally denoted as $SO(3)/G$,
where $G$ is the residual symmetry group in the ordered state.
The nature of $G$ depends on in which tensor component the symmetry
breaking takes place.
Nevertheless, when $\beta_2$ is included, the situation is
complicated, and the analysis of the phase diagram is deferred to
another work.

\subsection{Discussions}
\label{sect:SUM}
We are not aware of conclusive evidence for the existence of
the unconventional magnets.
Taking into account the great discoveries of the unconventional
superconductivity and pairing superfluidity in  high T$_c$
cuprates and $^3$He, respectively, we are optimistic that
unconventional magnetic phases also exist in Nature.
We propose to systematically search for these new phases in
$^3$He, ultracold atomic systems, semiconductors, heavy fermion
materials and ruthenates, both in experiments and in
numerical simulations.

Unconventional magnetic orders are natural generalizations of
itinerant ferromagnetism, whose driving force is still the
exchange interaction.
But it needs to be in the non-local version, i.e., a
non-$s$-wave channel.
Nevertheless, interactions in the high angular momentum channels
are typically weak.
In Ref. \cite{leewu2009}, a heuristic argument is provided to employ the
orbital hybridized band structure to promote the Landau interaction
to high partial-wave channels.
Consider a $d_{xz}/d_{yz}$ hybridized orbital band.
Around the Fermi surface, the Bloch wavefunction takes the orbital
configuration as
\bea
|\Psi_\alpha(\mathbf{k})\rangle &=& e^{i\mathbf{k}\mathbf{r}}
\left( \cos\phi_k |d_{xz}\rangle +\sin\phi_k|d_{yz}\rangle
\right)\otimes \chi_\alpha,
\eea
where $\chi_\alpha$ is the spin eigenstate.
The Landau interaction at the Hartree-Fock level is
\bea
f_{\uparrow\uparrow}(\mathbf{k}_1\mathbf{k}_2)&=&
V(\mathbf{q}=0)-\frac{1}{2}(1+\cos 2\theta_{\mathbf{k}_1\mathbf{k}_2})
V(\mathbf{k}_1-\mathbf{k}_2), \nn \\
f_{\uparrow\downarrow}(\mathbf{k}_1\mathbf{k}_2)&=&V(\mathbf{q}=0).
\eea
The appearance of the $d$-wave form factor
$\cos 2\theta_{\mathbf{k}_1\mathbf{k}_2}$
is due to the orbital hybridization, i.e.,
even though two electrons possess the same spin,
they can still be distinguished by their orbital
components.
Hence, although $V(\mathbf{k}_1-\mathbf{k}_2)$ could be
dominated by the $s$-wave component, the angular form
factor shifts a significant part of the spectral
weight into the $d$-wave channel.
Based on this formalism, a possible explanation of the nematic
transition observed in Sr$_3$Ru$_2$O$_7$ was provided.

Below we summarize several possible directions for searching
unconventional magnetism.
Ferromagnetic fluctuations in the normal state of $^3$He are strong.
The values of $F_{1a}$ of $^3$He are measured as negative via
the normal-state spin diffusion constant, spin-wave spectrum,
and the temperature dependence of the specific heat
\cite{leggett1970,corruccini1971,osheroff1977,greywall1983}.
It varies from around $-0.5$ to $-1.2$ with increasing pressures
to the melting point, reasonably close to the instability point
$F^a_1=-3$.
We conjecture that $^3$He could support unconventional magnetism under
certain conditions or exhibit strong fluctuations of these orders.

An important direction to search for unconventional magnetism is the
so-called ``hidden-order" systems.
Hidden orders typically mean that thermodynamic quantity measurements
exhibit a transition to a low temperature ordered state.
However, the nature of the orders remains unknown since they do not
exhibit themselves in typical detections.
Unconventional magnetic orders neither break translation symmetry nor
exhibit magnetic orderings in real space.
They are multipolar orderings in momentum space, hence, they
are difficult to detect via typical experimental methods.
Hence, they are natural candidates for hidden orders.
In fact, multipolar orderings in real space are also popular
candidates for hidden orders in literature \cite{}.

For example, the well-known system of heavy fermion compound
URu$_2$Si$_2$ exhibits a mysterious phase transition at 17K
by showing a large anomaly in specific heat.
It also exhibits a jump in the non-linear magnetic susceptibility
at the transition.
However, even with efforts after a few decades, the nature of this
transition remains elusive \cite{mydosh2020,wolowiece2021}.
Varma proposed an order, which is essentially the $p$-wave
$\alpha$-phase in our language \cite{varma2005,varma2006}.
Calculations for thermodynamic quantities fit in experiment
measurements reasonably well.
Another hidden order compound Cd$_2$Re$_2$O$_7$ exhibits
heat capacity anomaly and a kink of DC resistivity around 200K.
Recently, it has been discovered that the hidden order phase
exhibits inversion symmetry breaking via the optical 2nd
harmonic generation measurements \cite{harter2017,harter2018}.
Since this is a heavy element compound, Pomeranchuk
instabilities of spin-orbit coupled Fermi liquid theory
may be a promising candidate \cite{fu2015,norman2020}.

An obstacle to identifying unconventional magnetism is the lack
of definitive experimental signatures and detection methods.
We know that antiferromagnetism is very common among transition
metal oxides, more common than itinerant ferromagnetism.
However, the experimental identification of the antiferromagnetic
ordering is only possible after the detection method
of neutron scattering spectroscopy became available.

Maybe unconventional magnetism already exists somewhere, but we need to
think about how to detect them.
In addition to the inelastic neutron scattering resonances
(Sect. \ref{sect:neutron}), we outline the following possible methods.

The $\beta$ phases exhibit effective spin-orbit coupling, hence,
standard methods to detect spin-orbit coupling still apply.
The distinctive feature is that the spin-orbit coupling effects
should turn on and off at a phase transition.

Transport measurements can be used to detect the dynamic
generation of spin-orbit coupling.
For example, the existence of the anomalous Hall effect (AHE) relies on
spin-orbit coupling.
Therefore, detecting the AHE signal turning on at a phase transition
would be an evidence of the onset of the entanglement of spin and
momentum.
As for the $d$-wave $\alpha$-phase, i.e., spin-$\uparrow$ and
spin-$\downarrow$ Fermi surfaces exhibit opposite quadrupolar
distortions.
Taking the principle axes of the quadrupolar distortion
as $x$ and $y$-axis, it is straightforward to show that
the spin and charge currents satisfy
\bea
\left( \begin{array}{c}
j_x^{sp}\\
j_y^{sp}
\end{array}
\right)\propto
\left( \begin{array}{cc}
1&0\\
0&-1
\end{array}
\right)
\left( \begin{array}{c}
j_x^{c}\\
j_y^{c}
\end{array}
\right).
\eea
A verification of this transport relation would be a
signature of the $d$-wave $\alpha$-phase \cite{wu2007}.

Methods that can detect Fermi surface splitting are useful.
The angular resolved photon emission spectroscopy (ARPES) can
be used to detect the band splitting.
In fact, such an experiment has been performed in the system with
relativistic spin-orbit coupling.
In the unconventional magnetic phases, ARPES in principle can
measure temperature-dependent Fermi surface splittings.
Fermi surface splitting also shows up in quantum oscillation experiments
(e.g. Shubunikov-de Haas (SdH) oscillations) as beat patterns.
Hence, a temperature-dependent beat pattern in this kind of experiments
would be a signature of the developing of unconventional magnetism.


\section{Conclusions}
\label{sect:conclude}

We have reviewed a few applications of the symmetry principle
in condensed matter and cold atom systems.

First, we reviewed the concept of ``space-time" group,
which provides a symmetry framework in studying transport and topological
properties in a variety of dynamic systems beyond the Floquet framework,
such as laser-driven solid state lattices, dynamic photonic crystals, and
optical lattices.
Various fundamental concepts are generalized, including space-time unit
cell, momentum-energy Brillouin zone, Bloch-Floquet theory.
Novel non-symmorphic space-time transformations are identified including
time-screw rotation, time-glide reflection, and time-shift
rotary reflection.
13 space-time groups are classified in 1+1D with 5 of them
non-symmorphic, and 275 space-time group are classified in 2+1D.
We expect that space-time group will play an important role for
studying dynamic systems, in a similar way to space group
for static crystals.

Second, we reviewed the progress of studying large-spin ultracold
fermions from the perspective of high symmetries.
Due to enhanced quantum spin fluctuations from
the large number of fermion components, such systems naturally
lie in the large-$N$ region instead of the large-$S$ region
which is typically studied in solids.
A generic Sp(4), or, isomorphically, SO(5) symmetry is proved
in spin-$\frac{3}{2}$ systems, which plays a similar role
of SU(2) in spin-$\frac{1}{2}$ systems.
This symmetry can be upgraded to SO(7) under certain conditions
which extends Yang's $\eta$-pairing to $\chi$-pairing as its
high rank Lie algebra counterpart.
The 7D vector and 21D adjoint representations of SO(7)
unify a variety of competing orders in both particle-particle
and particle-hole channels.
Large-spin systems can exhibit multi-particle clustering orderings
or correlations both in the superfluid state with attractions
and in the super-exchange physics with repulsions,
which is similar to 3-quark baryon (color singlet) formation
in high energy physics.
The competitions among quarteting superfluidity/density-wave
and pairing superfluidity/density-wave are investigated.
The SU(4) singlet plaquette states in a 3D cubic lattice can be
described by a quantum plaquette model, whose effective
description is mapped to a high order gauge theory.
We anticipate that research along this direction can bridge cold
atom physics, condensed matter matter, and high energy physics
together.
Along with the experimental progress, even
more exotic strong coupling physics that is not easily accessible
in usual solid state systems could be investigated.

At last, we reviewed the unconventional magnetism as a mechanism of
``spin from isospin" to generate spin-orbit coupling in
non-relativistic Fermi liquids.
They are also novel states of itinerant electrons generalzing
ferromagnetism to unconventional
symmetries based on the Fermi surface instabilities of the
Pomeranchuk type.
These states include the isotropic $\beta$-phase and the anisotropic
$\alpha$-phase, which are the particle-hole channel analogy to the
superfluid $^3$He-B and A phases, respectively.
Different from the relativistic spin-orbit coupling, these dynamically
generated spin-orbit couplings possess collective excitations
of Goldstone modes, whose dynamics couples to spin moment
and induces resonances in the inelastic neutron scattering
spectroscopy.
Possible realizations of ``unconventional magnetism" in hidden order
systems and experimental detections are discussed.

\section{Acknowledgments}
I am grateful to my Ph. D. advisor, S. C. Zhang (deceased),
who convinced me of the beauty and the power of symmetry.
The last two topics reviewed here were started in my Ph. D. period
and were continued with various new developments in my career.
I thank Jiangping Hu, E. H. Fradkin,
K. Sun, D. Arovas, W. C. Lee, H. H. Hung, Shu Chen,
Yupeng Wang,
C. K. Xu, Y. Li, D. Wang, S. L. Xu, Z. Q. Gao,
M. Pan, C. H. Ke,  Z. X. Lin
for collaborations on related topics, and
J. E. Hirsch, A. L. Fetter, T. L. Ho,
S. Das Sarma, S. Kivelson, L. J. Sham
for their warm encouragements and appreciations.
I also thank Ji Wang for  proofreading and polishing.
This work is supported by NSFC under the Grants
No. 12174317 and No. 11729402.
\bibliographystyle{apsrev4-2}
\bibliography{symmetry,spin32,DynSpOrbit}

\begin{thebibliography}{190}%
\makeatletter
\providecommand \@ifxundefined [1]{%
 \@ifx{#1\undefined}
}%
\providecommand \@ifnum [1]{%
 \ifnum #1\expandafter \@firstoftwo
 \else \expandafter \@secondoftwo
 \fi
}%
\providecommand \@ifx [1]{%
 \ifx #1\expandafter \@firstoftwo
 \else \expandafter \@secondoftwo
 \fi
}%
\providecommand \natexlab [1]{#1}%
\providecommand \enquote  [1]{``#1''}%
\providecommand \bibnamefont  [1]{#1}%
\providecommand \bibfnamefont [1]{#1}%
\providecommand \citenamefont [1]{#1}%
\providecommand \href@noop [0]{\@secondoftwo}%
\providecommand \href [0]{\begingroup \@sanitize@url \@href}%
\providecommand \@href[1]{\@@startlink{#1}\@@href}%
\providecommand \@@href[1]{\endgroup#1\@@endlink}%
\providecommand \@sanitize@url [0]{\catcode `\\12\catcode `\$12\catcode
  `\&12\catcode `\#12\catcode `\^12\catcode `\_12\catcode `\%12\relax}%
\providecommand \@@startlink[1]{}%
\providecommand \@@endlink[0]{}%
\providecommand \url  [0]{\begingroup\@sanitize@url \@url }%
\providecommand \@url [1]{\endgroup\@href {#1}{\urlprefix }}%
\providecommand \urlprefix  [0]{URL }%
\providecommand \Eprint [0]{\href }%
\providecommand \doibase [0]{https://doi.org/}%
\providecommand \selectlanguage [0]{\@gobble}%
\providecommand \bibinfo  [0]{\@secondoftwo}%
\providecommand \bibfield  [0]{\@secondoftwo}%
\providecommand \translation [1]{[#1]}%
\providecommand \BibitemOpen [0]{}%
\providecommand \bibitemStop [0]{}%
\providecommand \bibitemNoStop [0]{.\EOS\space}%
\providecommand \EOS [0]{\spacefactor3000\relax}%
\providecommand \BibitemShut  [1]{\csname bibitem#1\endcsname}%
\let\auto@bib@innerbib\@empty
\bibitem [{\citenamefont {Lee}\ and\ \citenamefont {Yang}(1956)}]{lee1956}%
  \BibitemOpen
  \bibfield  {author} {\bibinfo {author} {\bibfnamefont {T.~D.}\ \bibnamefont
  {Lee}}\ and\ \bibinfo {author} {\bibfnamefont {C.~N.}\ \bibnamefont {Yang}},\
  }\href {https://doi.org/10.1103/PhysRev.104.254} {\bibfield  {journal}
  {\bibinfo  {journal} {Phys. Rev.}\ }\textbf {\bibinfo {volume} {104}},\
  \bibinfo {pages} {254} (\bibinfo {year} {1956})}\BibitemShut {NoStop}%
\bibitem [{\citenamefont {Yang}\ and\ \citenamefont {Mills}(1954)}]{yang1954}%
  \BibitemOpen
  \bibfield  {author} {\bibinfo {author} {\bibfnamefont {C.~N.}\ \bibnamefont
  {Yang}}\ and\ \bibinfo {author} {\bibfnamefont {R.~L.}\ \bibnamefont
  {Mills}},\ }\href {https://doi.org/10.1103/PhysRev.96.191} {\bibfield
  {journal} {\bibinfo  {journal} {Phys. Rev.}\ }\textbf {\bibinfo {volume}
  {96}},\ \bibinfo {pages} {191} (\bibinfo {year} {1954})}\BibitemShut
  {NoStop}%
\bibitem [{\citenamefont {Yang}(1967)}]{yang1967}%
  \BibitemOpen
  \bibfield  {author} {\bibinfo {author} {\bibfnamefont {C.~N.}\ \bibnamefont
  {Yang}},\ }\href {https://doi.org/10.1103/PhysRevLett.19.1312} {\bibfield
  {journal} {\bibinfo  {journal} {Phys. Rev. Lett.}\ }\textbf {\bibinfo
  {volume} {19}},\ \bibinfo {pages} {1312} (\bibinfo {year}
  {1967})}\BibitemShut {NoStop}%
\bibitem [{\citenamefont {Wu}\ and\ \citenamefont {Yang}(1975)}]{wu1975}%
  \BibitemOpen
  \bibfield  {author} {\bibinfo {author} {\bibfnamefont {T.~T.}\ \bibnamefont
  {Wu}}\ and\ \bibinfo {author} {\bibfnamefont {C.~N.}\ \bibnamefont {Yang}},\
  }\href {https://doi.org/10.1103/PhysRevD.12.3845} {\bibfield  {journal}
  {\bibinfo  {journal} {Phys. Rev. D}\ }\textbf {\bibinfo {volume} {12}},\
  \bibinfo {pages} {3845} (\bibinfo {year} {1975})}\BibitemShut {NoStop}%
\bibitem [{\citenamefont {Wu}\ and\ \citenamefont {Yang}(1976)}]{wu1976}%
  \BibitemOpen
  \bibfield  {author} {\bibinfo {author} {\bibfnamefont {T.~T.}\ \bibnamefont
  {Wu}}\ and\ \bibinfo {author} {\bibfnamefont {C.~N.}\ \bibnamefont {Yang}},\
  }\href@noop {} {\bibfield  {journal} {\bibinfo  {journal} {Nucl. Phys. B}\
  }\textbf {\bibinfo {volume} {107}},\ \bibinfo {pages} {365} (\bibinfo {year}
  {1976})}\BibitemShut {NoStop}%
\bibitem [{\citenamefont {H.Weyl}(2016)}]{weyl}%
  \BibitemOpen
  \bibfield  {author} {\bibinfo {author} {\bibnamefont {H.Weyl}},\ }\href@noop
  {} {\emph {\bibinfo {title} {Symmetry}}}\ (\bibinfo  {publisher} {Princeton
  University Press; Reprint edition},\ \bibinfo {year} {2016})\BibitemShut
  {NoStop}%
\bibitem [{\citenamefont {Gross}(1996)}]{gross1996}%
  \BibitemOpen
  \bibfield  {author} {\bibinfo {author} {\bibfnamefont {D.~J.}\ \bibnamefont
  {Gross}},\ }\href {https://doi.org/10.1073/pnas.93.25.14256} {\bibfield
  {journal} {\bibinfo  {journal} {Proceedings of the National Academy of
  Sciences}\ }\textbf {\bibinfo {volume} {93}},\ \bibinfo {pages} {14256}
  (\bibinfo {year} {1996})},\ \Eprint
  {https://arxiv.org/abs/https://www.pnas.org/content/93/25/14256.full.pdf}
  {https://www.pnas.org/content/93/25/14256.full.pdf} \BibitemShut {NoStop}%
\bibitem [{\citenamefont {Yang}(1996)}]{yang1996}%
  \BibitemOpen
  \bibfield  {author} {\bibinfo {author} {\bibfnamefont {C.~N.}\ \bibnamefont
  {Yang}},\ }\href@noop {} {\bibfield  {journal} {\bibinfo  {journal}
  {Proceedings of the American Philosophical Society}\ }\textbf {\bibinfo
  {volume} {140}},\ \bibinfo {pages} {267} (\bibinfo {year}
  {1996})}\BibitemShut {NoStop}%
\bibitem [{\citenamefont {Lax}(2012)}]{lax2012}%
  \BibitemOpen
  \bibfield  {author} {\bibinfo {author} {\bibfnamefont {M.}~\bibnamefont
  {Lax}},\ }\href@noop {} {\emph {\bibinfo {title} {Symmetry Principles in
  Solid State and Molecular Physics}}}\ (\bibinfo  {publisher} {Dover
  Publications},\ \bibinfo {year} {2012})\BibitemShut {NoStop}%
\bibitem [{\citenamefont {Georgi}(1999)}]{georgi1999}%
  \BibitemOpen
  \bibfield  {author} {\bibinfo {author} {\bibfnamefont {H.}~\bibnamefont
  {Georgi}},\ }\href@noop {} {\emph {\bibinfo {title} {Lie Algebras In Particle
  Physics: from Isospin To Unified Theories}}}\ (\bibinfo  {publisher} {CRC
  Press; 1st edition},\ \bibinfo {year} {1999})\BibitemShut {NoStop}%
\bibitem [{\citenamefont {Noether}(1918)}]{noether1918}%
  \BibitemOpen
  \bibfield  {author} {\bibinfo {author} {\bibfnamefont {E.}~\bibnamefont
  {Noether}},\ }\href {http://eudml.org/doc/59024} {\bibfield  {journal}
  {\bibinfo  {journal} {Nachrichten von der Gesellschaft der Wissenschaften zu
  G\"ottingen, Mathematisch-Physikalische Klasse}\ }\textbf {\bibinfo {volume}
  {1918}},\ \bibinfo {pages} {235} (\bibinfo {year} {1918})}\BibitemShut
  {NoStop}%
\bibitem [{\citenamefont {Wigner}(1959)}]{wigner1959}%
  \BibitemOpen
  \bibfield  {author} {\bibinfo {author} {\bibfnamefont {E.~P.}\ \bibnamefont
  {Wigner}},\ }\href@noop {} {\emph {\bibinfo {title} {Group Theory and its
  Application to the Quantum Mechanics of Atomic Spectra}}}\ (\bibinfo
  {publisher} {Academic Press},\ \bibinfo {address} {New York},\ \bibinfo
  {year} {1959})\BibitemShut {NoStop}%
\bibitem [{\citenamefont {Weyl}(1950)}]{weyl1950}%
  \BibitemOpen
  \bibfield  {author} {\bibinfo {author} {\bibfnamefont {H.}~\bibnamefont
  {Weyl}},\ }\href@noop {} {\emph {\bibinfo {title} {The Theory of Groups and
  Quantum Mechanics}}}\ (\bibinfo  {publisher} {Dover},\ \bibinfo {address}
  {New York},\ \bibinfo {year} {1950})\BibitemShut {NoStop}%
\bibitem [{\citenamefont {Fock}(1935)}]{fock1935}%
  \BibitemOpen
  \bibfield  {author} {\bibinfo {author} {\bibfnamefont {V.}~\bibnamefont
  {Fock}},\ }\href@noop {} {\bibfield  {journal} {\bibinfo  {journal} {Z.
  Phys.}\ }\textbf {\bibinfo {volume} {98}},\ \bibinfo {pages} {145} (\bibinfo
  {year} {1935})}\BibitemShut {NoStop}%
\bibitem [{\citenamefont {Sakrai}\ and\ \citenamefont
  {Napolitano}(2010)}]{sakurai2010}%
  \BibitemOpen
  \bibfield  {author} {\bibinfo {author} {\bibfnamefont {J.~J.}\ \bibnamefont
  {Sakrai}}\ and\ \bibinfo {author} {\bibfnamefont {J.~J.}\ \bibnamefont
  {Napolitano}},\ }\href@noop {} {\emph {\bibinfo {title} {Modern Quantum
  Mechanics}}}\ (\bibinfo  {publisher} {Pearson},\ \bibinfo {year}
  {2010})\BibitemShut {NoStop}%
\bibitem [{\citenamefont {Glashow}(1959)}]{Glashow1959}%
  \BibitemOpen
  \bibfield  {author} {\bibinfo {author} {\bibfnamefont {S.}~\bibnamefont
  {Glashow}},\ }\href@noop {} {\bibfield  {journal} {\bibinfo  {journal} {Nucl.
  Phys. B}\ }\textbf {\bibinfo {volume} {10}},\ \bibinfo {pages} {107}
  (\bibinfo {year} {1959})}\BibitemShut {NoStop}%
\bibitem [{\citenamefont {Salam}\ and\ \citenamefont {Ward}(1959)}]{Salam1959}%
  \BibitemOpen
  \bibfield  {author} {\bibinfo {author} {\bibfnamefont {A.}~\bibnamefont
  {Salam}}\ and\ \bibinfo {author} {\bibfnamefont {J.~C.}\ \bibnamefont
  {Ward}},\ }\href@noop {} {\bibfield  {journal} {\bibinfo  {journal} {Nuovo
  Cimento.}\ }\textbf {\bibinfo {volume} {11}},\ \bibinfo {pages} {568}
  (\bibinfo {year} {1959})}\BibitemShut {NoStop}%
\bibitem [{\citenamefont {Weinberg}(1967)}]{Weinberg1967}%
  \BibitemOpen
  \bibfield  {author} {\bibinfo {author} {\bibfnamefont {S.}~\bibnamefont
  {Weinberg}},\ }\href {https://doi.org/10.1103/PhysRevLett.19.1264} {\bibfield
   {journal} {\bibinfo  {journal} {Phys. Rev. Lett.}\ }\textbf {\bibinfo
  {volume} {19}},\ \bibinfo {pages} {1264} (\bibinfo {year}
  {1967})}\BibitemShut {NoStop}%
\bibitem [{\citenamefont {Peskin}\ and\ \citenamefont
  {Schroeder}(1995)}]{peskin1995}%
  \BibitemOpen
  \bibfield  {author} {\bibinfo {author} {\bibfnamefont {M.}~\bibnamefont
  {Peskin}}\ and\ \bibinfo {author} {\bibfnamefont {D.}~\bibnamefont
  {Schroeder}},\ }\href {https://books.google.com.hk/books?id=EVeNNcslvX0C}
  {\emph {\bibinfo {title} {An Introduction To Quantum Field Theory}}},\
  Frontiers in Physics\ (\bibinfo  {publisher} {Avalon Publishing},\ \bibinfo
  {year} {1995})\BibitemShut {NoStop}%
\bibitem [{\citenamefont {Landau}(1937)}]{landau1937}%
  \BibitemOpen
  \bibfield  {author} {\bibinfo {author} {\bibfnamefont {L.~D.}\ \bibnamefont
  {Landau}},\ }\href@noop {} {\bibfield  {journal} {\bibinfo  {journal} {Zh.
  Eksp. Teor. Fiz.}\ }\textbf {\bibinfo {volume} {7}},\ \bibinfo {pages} {19}
  (\bibinfo {year} {1937})}\BibitemShut {NoStop}%
\bibitem [{\citenamefont {Ginzburg}\ and\ \citenamefont
  {Landau}(1950)}]{ginzburg1950}%
  \BibitemOpen
  \bibfield  {author} {\bibinfo {author} {\bibfnamefont {V.~L.}\ \bibnamefont
  {Ginzburg}}\ and\ \bibinfo {author} {\bibfnamefont {L.~D.}\ \bibnamefont
  {Landau}},\ }\href@noop {} {\bibfield  {journal} {\bibinfo  {journal} {Zh.
  Eksp. Teor. Fiz.}\ }\textbf {\bibinfo {volume} {20}},\ \bibinfo {pages}
  {1064} (\bibinfo {year} {1950})}\BibitemShut {NoStop}%
\bibitem [{\citenamefont {Landau}\ and\ \citenamefont
  {Lifshitz}(1980)}]{landau1980}%
  \BibitemOpen
  \bibfield  {author} {\bibinfo {author} {\bibfnamefont {L.~D.}\ \bibnamefont
  {Landau}}\ and\ \bibinfo {author} {\bibfnamefont {E.}~\bibnamefont
  {Lifshitz}},\ }\href@noop {} {\emph {\bibinfo {title} {Statistical
  Physics}}}\ (\bibinfo  {publisher} {Butterworth-Heinemann; 3rd edition},\
  \bibinfo {year} {1980})\BibitemShut {NoStop}%
\bibitem [{\citenamefont {Goldstone}\ \emph {et~al.}(1962)\citenamefont
  {Goldstone}, \citenamefont {Salam},\ and\ \citenamefont
  {Weinberg}}]{goldstone1962}%
  \BibitemOpen
  \bibfield  {author} {\bibinfo {author} {\bibfnamefont {J.}~\bibnamefont
  {Goldstone}}, \bibinfo {author} {\bibfnamefont {A.}~\bibnamefont {Salam}},\
  and\ \bibinfo {author} {\bibfnamefont {S.}~\bibnamefont {Weinberg}},\ }\href
  {https://doi.org/10.1103/PhysRev.127.965} {\bibfield  {journal} {\bibinfo
  {journal} {Phys. Rev.}\ }\textbf {\bibinfo {volume} {127}},\ \bibinfo {pages}
  {965} (\bibinfo {year} {1962})}\BibitemShut {NoStop}%
\bibitem [{\citenamefont {Anderson}(1963)}]{anderson1963}%
  \BibitemOpen
  \bibfield  {author} {\bibinfo {author} {\bibfnamefont {P.~W.}\ \bibnamefont
  {Anderson}},\ }\href {https://doi.org/10.1103/PhysRev.130.439} {\bibfield
  {journal} {\bibinfo  {journal} {Phys. Rev.}\ }\textbf {\bibinfo {volume}
  {130}},\ \bibinfo {pages} {439} (\bibinfo {year} {1963})}\BibitemShut
  {NoStop}%
\bibitem [{\citenamefont {Higgs}(1964)}]{higgs1964}%
  \BibitemOpen
  \bibfield  {author} {\bibinfo {author} {\bibfnamefont {P.~W.}\ \bibnamefont
  {Higgs}},\ }\href {https://doi.org/10.1103/PhysRevLett.13.508} {\bibfield
  {journal} {\bibinfo  {journal} {Phys. Rev. Lett.}\ }\textbf {\bibinfo
  {volume} {13}},\ \bibinfo {pages} {508} (\bibinfo {year} {1964})}\BibitemShut
  {NoStop}%
\bibitem [{\citenamefont {Yang}(1989)}]{yang1989}%
  \BibitemOpen
  \bibfield  {author} {\bibinfo {author} {\bibfnamefont {C.~N.}\ \bibnamefont
  {Yang}},\ }\href@noop {} {\bibfield  {journal} {\bibinfo  {journal} {Phys.
  Rev. Lett.}\ }\textbf {\bibinfo {volume} {63}},\ \bibinfo {pages} {2144}
  (\bibinfo {year} {1989})}\BibitemShut {NoStop}%
\bibitem [{\citenamefont {YANG}\ and\ \citenamefont {ZHANG}(1990)}]{yang1990}%
  \BibitemOpen
  \bibfield  {author} {\bibinfo {author} {\bibfnamefont {C.~N.}\ \bibnamefont
  {YANG}}\ and\ \bibinfo {author} {\bibfnamefont {S.}~\bibnamefont {ZHANG}},\
  }\href {https://doi.org/10.1142/S0217984990000933} {\bibfield  {journal}
  {\bibinfo  {journal} {Modern Physics Letters B}\ }\textbf {\bibinfo {volume}
  {04}},\ \bibinfo {pages} {759} (\bibinfo {year} {1990})},\ \Eprint
  {https://arxiv.org/abs/http://www.worldscientific.com/doi/pdf/10.1142/S0217984990000933}
  {http://www.worldscientific.com/doi/pdf/10.1142/S0217984990000933}
  \BibitemShut {NoStop}%
\bibitem [{\citenamefont {Zhang}(1991)}]{zhang1991}%
  \BibitemOpen
  \bibfield  {author} {\bibinfo {author} {\bibfnamefont {S.~C.}\ \bibnamefont
  {Zhang}},\ }\href@noop {} {\bibfield  {journal} {\bibinfo  {journal}
  {International Journal of Modern Physics B}\ }\textbf {\bibinfo {volume}
  {5}},\ \bibinfo {pages} {153} (\bibinfo {year} {1991})}\BibitemShut {NoStop}%
\bibitem [{\citenamefont {Zhang}(1997)}]{zhang1997}%
  \BibitemOpen
  \bibfield  {author} {\bibinfo {author} {\bibfnamefont {S.~C.}\ \bibnamefont
  {Zhang}},\ }\href@noop {} {\bibfield  {journal} {\bibinfo  {journal}
  {Science}\ }\textbf {\bibinfo {volume} {275}},\ \bibinfo {pages} {1089}
  (\bibinfo {year} {1997})}\BibitemShut {NoStop}%
\bibitem [{\citenamefont {Demler}\ \emph {et~al.}(2004)\citenamefont {Demler},
  \citenamefont {Hanke},\ and\ \citenamefont {Zhang}}]{demler2004}%
  \BibitemOpen
  \bibfield  {author} {\bibinfo {author} {\bibfnamefont {E.}~\bibnamefont
  {Demler}}, \bibinfo {author} {\bibfnamefont {W.}~\bibnamefont {Hanke}},\ and\
  \bibinfo {author} {\bibfnamefont {S.-C.}\ \bibnamefont {Zhang}},\ }\href
  {https://doi.org/10.1103/RevModPhys.76.909} {\bibfield  {journal} {\bibinfo
  {journal} {Rev. Mod. Phys.}\ }\textbf {\bibinfo {volume} {76}},\ \bibinfo
  {pages} {909} (\bibinfo {year} {2004})}\BibitemShut {NoStop}%
\bibitem [{\citenamefont {Demler}\ and\ \citenamefont
  {Zhang}(1995)}]{demler1995}%
  \BibitemOpen
  \bibfield  {author} {\bibinfo {author} {\bibfnamefont {E.}~\bibnamefont
  {Demler}}\ and\ \bibinfo {author} {\bibfnamefont {S.~C.}\ \bibnamefont
  {Zhang}},\ }\href@noop {} {\bibfield  {journal} {\bibinfo  {journal} {Phys.
  Rev. Lett.}\ }\textbf {\bibinfo {volume} {75}},\ \bibinfo {pages} {4126}
  (\bibinfo {year} {1995})}\BibitemShut {NoStop}%
\bibitem [{\citenamefont {Kittel}(1987)}]{kittel1987}%
  \BibitemOpen
  \bibfield  {author} {\bibinfo {author} {\bibfnamefont {C.}~\bibnamefont
  {Kittel}},\ }\href@noop {} {\emph {\bibinfo {title} {Quantum theory of
  Solids}}}\ (\bibinfo  {publisher} {Wiley},\ \bibinfo {address} {New York},\
  \bibinfo {year} {1987})\BibitemShut {NoStop}%
\bibitem [{\citenamefont {Wang}\ \emph {et~al.}(2013)\citenamefont {Wang},
  \citenamefont {Steinberg}, \citenamefont {Jarillo-Herrero},\ and\
  \citenamefont {Gedik}}]{gedik2013}%
  \BibitemOpen
  \bibfield  {author} {\bibinfo {author} {\bibfnamefont {Y.~H.}\ \bibnamefont
  {Wang}}, \bibinfo {author} {\bibfnamefont {H.}~\bibnamefont {Steinberg}},
  \bibinfo {author} {\bibfnamefont {P.}~\bibnamefont {Jarillo-Herrero}},\ and\
  \bibinfo {author} {\bibfnamefont {N.}~\bibnamefont {Gedik}},\ }\href
  {https://doi.org/10.1126/science.1239834} {\bibfield  {journal} {\bibinfo
  {journal} {Science}\ }\textbf {\bibinfo {volume} {342}},\ \bibinfo {pages}
  {453} (\bibinfo {year} {2013})}\BibitemShut {NoStop}%
\bibitem [{\citenamefont {Zhu}\ \emph {et~al.}(2018)\citenamefont {Zhu},
  \citenamefont {Yi}, \citenamefont {Li}, \citenamefont {Xiao}, \citenamefont
  {Zhang}, \citenamefont {Yang}, \citenamefont {Kaindl}, \citenamefont {Li},
  \citenamefont {Wang},\ and\ \citenamefont {Zhang}}]{zhangxiang2018}%
  \BibitemOpen
  \bibfield  {author} {\bibinfo {author} {\bibfnamefont {H.}~\bibnamefont
  {Zhu}}, \bibinfo {author} {\bibfnamefont {J.}~\bibnamefont {Yi}}, \bibinfo
  {author} {\bibfnamefont {M.-Y.}\ \bibnamefont {Li}}, \bibinfo {author}
  {\bibfnamefont {J.}~\bibnamefont {Xiao}}, \bibinfo {author} {\bibfnamefont
  {L.}~\bibnamefont {Zhang}}, \bibinfo {author} {\bibfnamefont {C.-W.}\
  \bibnamefont {Yang}}, \bibinfo {author} {\bibfnamefont {R.~A.}\ \bibnamefont
  {Kaindl}}, \bibinfo {author} {\bibfnamefont {L.-J.}\ \bibnamefont {Li}},
  \bibinfo {author} {\bibfnamefont {Y.}~\bibnamefont {Wang}},\ and\ \bibinfo
  {author} {\bibfnamefont {X.}~\bibnamefont {Zhang}},\ }\href@noop {}
  {\bibfield  {journal} {\bibinfo  {journal} {Science}\ }\textbf {\bibinfo
  {volume} {359}},\ \bibinfo {pages} {579} (\bibinfo {year}
  {2018})}\BibitemShut {NoStop}%
\bibitem [{\citenamefont {Parker}\ \emph {et~al.}(2013)\citenamefont {Parker},
  \citenamefont {Ha},\ and\ \citenamefont {Chin}}]{parker2013}%
  \BibitemOpen
  \bibfield  {author} {\bibinfo {author} {\bibfnamefont {C.~V.}\ \bibnamefont
  {Parker}}, \bibinfo {author} {\bibfnamefont {L.-C.}\ \bibnamefont {Ha}},\
  and\ \bibinfo {author} {\bibfnamefont {C.}~\bibnamefont {Chin}},\ }\href
  {https://doi.org/10.1038/nphys2789} {\bibfield  {journal} {\bibinfo
  {journal} {Nature Physics}\ }\textbf {\bibinfo {volume} {9}},\ \bibinfo
  {pages} {769每774} (\bibinfo {year} {2013})}\BibitemShut {NoStop}%
\bibitem [{\citenamefont {Anderson}\ \emph {et~al.}(2017)\citenamefont
  {Anderson}, \citenamefont {Clark}, \citenamefont {Crawford}, \citenamefont
  {Glatz}, \citenamefont {Aranson}, \citenamefont {Scherpelz}, \citenamefont
  {Feng}, \citenamefont {Chin},\ and\ \citenamefont {Levin}}]{anderson2017}%
  \BibitemOpen
  \bibfield  {author} {\bibinfo {author} {\bibfnamefont {B.~M.}\ \bibnamefont
  {Anderson}}, \bibinfo {author} {\bibfnamefont {L.~W.}\ \bibnamefont {Clark}},
  \bibinfo {author} {\bibfnamefont {J.}~\bibnamefont {Crawford}}, \bibinfo
  {author} {\bibfnamefont {A.}~\bibnamefont {Glatz}}, \bibinfo {author}
  {\bibfnamefont {I.~S.}\ \bibnamefont {Aranson}}, \bibinfo {author}
  {\bibfnamefont {P.}~\bibnamefont {Scherpelz}}, \bibinfo {author}
  {\bibfnamefont {L.}~\bibnamefont {Feng}}, \bibinfo {author} {\bibfnamefont
  {C.}~\bibnamefont {Chin}},\ and\ \bibinfo {author} {\bibfnamefont
  {K.}~\bibnamefont {Levin}},\ }\bibfield  {journal} {\bibinfo  {journal}
  {Physical Review Letters}\ }\textbf {\bibinfo {volume} {118}},\ \href
  {https://doi.org/10.1103/physrevlett.118.220401}
  {10.1103/physrevlett.118.220401} (\bibinfo {year} {2017})\BibitemShut
  {NoStop}%
\bibitem [{\citenamefont {Rechtsman}\ \emph {et~al.}(2013)\citenamefont
  {Rechtsman}, \citenamefont {Zeuner}, \citenamefont {Plotnik}, \citenamefont
  {Lumer}, \citenamefont {Podolsky}, \citenamefont {Dreisow}, \citenamefont
  {Nolte}, \citenamefont {Segev},\ and\ \citenamefont
  {Szameit}}]{rechtsman2013a}%
  \BibitemOpen
  \bibfield  {author} {\bibinfo {author} {\bibfnamefont {M.~C.}\ \bibnamefont
  {Rechtsman}}, \bibinfo {author} {\bibfnamefont {J.~M.}\ \bibnamefont
  {Zeuner}}, \bibinfo {author} {\bibfnamefont {Y.}~\bibnamefont {Plotnik}},
  \bibinfo {author} {\bibfnamefont {Y.}~\bibnamefont {Lumer}}, \bibinfo
  {author} {\bibfnamefont {D.}~\bibnamefont {Podolsky}}, \bibinfo {author}
  {\bibfnamefont {F.}~\bibnamefont {Dreisow}}, \bibinfo {author} {\bibfnamefont
  {S.}~\bibnamefont {Nolte}}, \bibinfo {author} {\bibfnamefont
  {M.}~\bibnamefont {Segev}},\ and\ \bibinfo {author} {\bibfnamefont
  {A.}~\bibnamefont {Szameit}},\ }\href {https://doi.org/10.1038/nature12066}
  {\bibfield  {journal} {\bibinfo  {journal} {Nature}\ }\textbf {\bibinfo
  {volume} {496}},\ \bibinfo {pages} {196} (\bibinfo {year}
  {2013})}\BibitemShut {NoStop}%
\bibitem [{\citenamefont {Rudner}\ \emph {et~al.}(2013)\citenamefont {Rudner},
  \citenamefont {Lindner}, \citenamefont {Berg},\ and\ \citenamefont
  {Levin}}]{rudner2013}%
  \BibitemOpen
  \bibfield  {author} {\bibinfo {author} {\bibfnamefont {M.~S.}\ \bibnamefont
  {Rudner}}, \bibinfo {author} {\bibfnamefont {N.~H.}\ \bibnamefont {Lindner}},
  \bibinfo {author} {\bibfnamefont {E.}~\bibnamefont {Berg}},\ and\ \bibinfo
  {author} {\bibfnamefont {M.}~\bibnamefont {Levin}},\ }\href
  {https://doi.org/10.1103/PhysRevX.3.031005} {\bibfield  {journal} {\bibinfo
  {journal} {Phys. Rev. X}\ }\textbf {\bibinfo {volume} {3}},\ \bibinfo {pages}
  {031005} (\bibinfo {year} {2013})},\ \Eprint
  {https://arxiv.org/abs/arXiv:1212.3324v1} {arXiv:arXiv:1212.3324v1}
  \BibitemShut {NoStop}%
\bibitem [{\citenamefont {Thakurathi}\ \emph {et~al.}(2013)\citenamefont
  {Thakurathi}, \citenamefont {Patel}, \citenamefont {Sen},\ and\ \citenamefont
  {Dutta}}]{thakurathi2013a}%
  \BibitemOpen
  \bibfield  {author} {\bibinfo {author} {\bibfnamefont {M.}~\bibnamefont
  {Thakurathi}}, \bibinfo {author} {\bibfnamefont {A.~A.}\ \bibnamefont
  {Patel}}, \bibinfo {author} {\bibfnamefont {D.}~\bibnamefont {Sen}},\ and\
  \bibinfo {author} {\bibfnamefont {A.}~\bibnamefont {Dutta}},\ }\href
  {https://doi.org/10.1103/PhysRevB.88.155133} {\bibfield  {journal} {\bibinfo
  {journal} {Phys. Rev. B - Condens. Matter Mater. Phys.}\ }\textbf {\bibinfo
  {volume} {88}},\ \bibinfo {pages} {155133} (\bibinfo {year} {2013})},\
  \Eprint {https://arxiv.org/abs/arXiv:1303.2300v1} {arXiv:arXiv:1303.2300v1}
  \BibitemShut {NoStop}%
\bibitem [{\citenamefont {von Keyserlingk}\ and\ \citenamefont
  {Sondhi}(2016)}]{vonKeyserlingk2016b}%
  \BibitemOpen
  \bibfield  {author} {\bibinfo {author} {\bibfnamefont {C.~W.}\ \bibnamefont
  {von Keyserlingk}}\ and\ \bibinfo {author} {\bibfnamefont {S.~L.}\
  \bibnamefont {Sondhi}},\ }\href {https://doi.org/10.1103/PhysRevB.93.245146}
  {\bibfield  {journal} {\bibinfo  {journal} {Phys. Rev. B}\ }\textbf {\bibinfo
  {volume} {93}},\ \bibinfo {pages} {245146} (\bibinfo {year} {2016})},\
  \Eprint {https://arxiv.org/abs/1602.06949} {arXiv:1602.06949} \BibitemShut
  {NoStop}%
\bibitem [{\citenamefont {Gu}\ \emph {et~al.}(2011)\citenamefont {Gu},
  \citenamefont {Fertig}, \citenamefont {Arovas},\ and\ \citenamefont
  {Auerbach}}]{gu2011}%
  \BibitemOpen
  \bibfield  {author} {\bibinfo {author} {\bibfnamefont {Z.}~\bibnamefont
  {Gu}}, \bibinfo {author} {\bibfnamefont {H.~A.}\ \bibnamefont {Fertig}},
  \bibinfo {author} {\bibfnamefont {D.~P.}\ \bibnamefont {Arovas}},\ and\
  \bibinfo {author} {\bibfnamefont {A.}~\bibnamefont {Auerbach}},\ }\href
  {https://doi.org/10.1103/PhysRevLett.107.216601} {\bibfield  {journal}
  {\bibinfo  {journal} {Phys. Rev. Lett.}\ }\textbf {\bibinfo {volume} {107}},\
  \bibinfo {pages} {216601} (\bibinfo {year} {2011})},\ \Eprint
  {https://arxiv.org/abs/1106.0302} {arXiv:1106.0302} \BibitemShut {NoStop}%
\bibitem [{\citenamefont {Lindner}\ \emph {et~al.}(2011)\citenamefont
  {Lindner}, \citenamefont {Refael},\ and\ \citenamefont
  {Galitski}}]{lindner2011}%
  \BibitemOpen
  \bibfield  {author} {\bibinfo {author} {\bibfnamefont {N.~H.}\ \bibnamefont
  {Lindner}}, \bibinfo {author} {\bibfnamefont {G.}~\bibnamefont {Refael}},\
  and\ \bibinfo {author} {\bibfnamefont {V.}~\bibnamefont {Galitski}},\ }\href
  {https://doi.org/10.1038/nphys1926} {\bibfield  {journal} {\bibinfo
  {journal} {Nat. Phys.}\ }\textbf {\bibinfo {volume} {7}},\ \bibinfo {pages}
  {490} (\bibinfo {year} {2011})},\ \Eprint {https://arxiv.org/abs/1008.1792}
  {arXiv:1008.1792} \BibitemShut {NoStop}%
\bibitem [{\citenamefont {Else}\ \emph {et~al.}(2016)\citenamefont {Else},
  \citenamefont {Bauer},\ and\ \citenamefont {Nayak}}]{else2016}%
  \BibitemOpen
  \bibfield  {author} {\bibinfo {author} {\bibfnamefont {D.~V.}\ \bibnamefont
  {Else}}, \bibinfo {author} {\bibfnamefont {B.}~\bibnamefont {Bauer}},\ and\
  \bibinfo {author} {\bibfnamefont {C.}~\bibnamefont {Nayak}},\ }\href
  {https://doi.org/10.1103/PhysRevLett.117.090402} {\bibfield  {journal}
  {\bibinfo  {journal} {Phys. Rev. Lett.}\ }\textbf {\bibinfo {volume} {117}},\
  \bibinfo {pages} {090402} (\bibinfo {year} {2016})}\BibitemShut {NoStop}%
\bibitem [{\citenamefont {{Von Keyserlingk}}\ and\ \citenamefont
  {Sondhi}(2016)}]{vonKeyserlingk2016a}%
  \BibitemOpen
  \bibfield  {author} {\bibinfo {author} {\bibfnamefont {C.~W.}\ \bibnamefont
  {{Von Keyserlingk}}}\ and\ \bibinfo {author} {\bibfnamefont {S.~L.}\
  \bibnamefont {Sondhi}},\ }\href {https://doi.org/10.1103/PhysRevB.93.245145}
  {\bibfield  {journal} {\bibinfo  {journal} {Phys. Rev. B - Condens. Matter
  Mater. Phys.}\ }\textbf {\bibinfo {volume} {93}},\ \bibinfo {pages} {245145}
  (\bibinfo {year} {2016})},\ \Eprint {https://arxiv.org/abs/1602.02157}
  {arXiv:1602.02157} \BibitemShut {NoStop}%
\bibitem [{\citenamefont {Potter}\ \emph {et~al.}(2016)\citenamefont {Potter},
  \citenamefont {Morimoto},\ and\ \citenamefont {Vishwanath}}]{potter2016a}%
  \BibitemOpen
  \bibfield  {author} {\bibinfo {author} {\bibfnamefont {A.~C.}\ \bibnamefont
  {Potter}}, \bibinfo {author} {\bibfnamefont {T.}~\bibnamefont {Morimoto}},\
  and\ \bibinfo {author} {\bibfnamefont {A.}~\bibnamefont {Vishwanath}},\
  }\href {https://doi.org/10.1103/PhysRevX.6.041001} {\bibfield  {journal}
  {\bibinfo  {journal} {Phys. Rev. X}\ }\textbf {\bibinfo {volume} {6}},\
  \bibinfo {pages} {041001} (\bibinfo {year} {2016})},\ \Eprint
  {https://arxiv.org/abs/1602.05194} {arXiv:1602.05194} \BibitemShut {NoStop}%
\bibitem [{\citenamefont {Roy}\ and\ \citenamefont {Harper}(2016)}]{roy2016}%
  \BibitemOpen
  \bibfield  {author} {\bibinfo {author} {\bibfnamefont {R.}~\bibnamefont
  {Roy}}\ and\ \bibinfo {author} {\bibfnamefont {F.}~\bibnamefont {Harper}},\
  }\href {http://arxiv.org/abs/1603.06944} {\bibfield  {journal} {\bibinfo
  {journal} {arXiv:1603.06944}\ } (\bibinfo {year} {2016})},\ \Eprint
  {https://arxiv.org/abs/1603.06944} {arXiv:1603.06944} \BibitemShut {NoStop}%
\bibitem [{\citenamefont {Nathan}\ and\ \citenamefont
  {Rudner}(2015)}]{nathan2015}%
  \BibitemOpen
  \bibfield  {author} {\bibinfo {author} {\bibfnamefont {F.}~\bibnamefont
  {Nathan}}\ and\ \bibinfo {author} {\bibfnamefont {M.~S.}\ \bibnamefont
  {Rudner}},\ }\href {https://doi.org/10.1088/1367-2630/17/12/125014}
  {\bibfield  {journal} {\bibinfo  {journal} {New J. Phys.}\ }\textbf {\bibinfo
  {volume} {17}},\ \bibinfo {pages} {125014} (\bibinfo {year} {2015})},\
  \Eprint {https://arxiv.org/abs/1506.07647} {arXiv:1506.07647} \BibitemShut
  {NoStop}%
\bibitem [{\citenamefont {Fl{\"{a}}schner}\ \emph {et~al.}(2015)\citenamefont
  {Fl{\"{a}}schner}, \citenamefont {Rem}, \citenamefont {Tarnowski},
  \citenamefont {Vogel}, \citenamefont {L{\"{u}}hmann}, \citenamefont
  {Sengstock},\ and\ \citenamefont {Weitenberg}}]{flaschner2016}%
  \BibitemOpen
  \bibfield  {author} {\bibinfo {author} {\bibfnamefont {N.}~\bibnamefont
  {Fl{\"{a}}schner}}, \bibinfo {author} {\bibfnamefont {B.~S.}\ \bibnamefont
  {Rem}}, \bibinfo {author} {\bibfnamefont {M.}~\bibnamefont {Tarnowski}},
  \bibinfo {author} {\bibfnamefont {D.}~\bibnamefont {Vogel}}, \bibinfo
  {author} {\bibfnamefont {D.-S.}\ \bibnamefont {L{\"{u}}hmann}}, \bibinfo
  {author} {\bibfnamefont {K.}~\bibnamefont {Sengstock}},\ and\ \bibinfo
  {author} {\bibfnamefont {C.}~\bibnamefont {Weitenberg}},\ }\href
  {https://doi.org/10.1126/science.aad4568} {\bibfield  {journal} {\bibinfo
  {journal} {1509.05763}\ }\textbf {\bibinfo {volume} {05882}},\ \bibinfo
  {pages} {1} (\bibinfo {year} {2015})},\ \Eprint
  {https://arxiv.org/abs/1509.05763} {arXiv:1509.05763} \BibitemShut {NoStop}%
\bibitem [{\citenamefont {Xu}\ and\ \citenamefont {Wu}(2018)}]{xu2018}%
  \BibitemOpen
  \bibfield  {author} {\bibinfo {author} {\bibfnamefont {S.}~\bibnamefont
  {Xu}}\ and\ \bibinfo {author} {\bibfnamefont {C.}~\bibnamefont {Wu}},\ }\href
  {https://doi.org/10.1103/PhysRevLett.120.096401} {\bibfield  {journal}
  {\bibinfo  {journal} {Phys. Rev. Lett.}\ }\textbf {\bibinfo {volume} {120}},\
  \bibinfo {pages} {096401} (\bibinfo {year} {2018})}\BibitemShut {NoStop}%
\bibitem [{\citenamefont {Morimoto}\ \emph {et~al.}(2017)\citenamefont
  {Morimoto}, \citenamefont {Po},\ and\ \citenamefont
  {Vishwanath}}]{morimoto2017}%
  \BibitemOpen
  \bibfield  {author} {\bibinfo {author} {\bibfnamefont {T.}~\bibnamefont
  {Morimoto}}, \bibinfo {author} {\bibfnamefont {H.~C.}\ \bibnamefont {Po}},\
  and\ \bibinfo {author} {\bibfnamefont {A.}~\bibnamefont {Vishwanath}},\
  }\href {https://doi.org/10.1103/PhysRevB.95.195155} {\bibfield  {journal}
  {\bibinfo  {journal} {Phys. Rev. B}\ }\textbf {\bibinfo {volume} {95}},\
  \bibinfo {pages} {195155} (\bibinfo {year} {2017})}\BibitemShut {NoStop}%
\bibitem [{\citenamefont {Wu}(2010)}]{wu2010}%
  \BibitemOpen
  \bibfield  {author} {\bibinfo {author} {\bibfnamefont {C.}~\bibnamefont
  {Wu}},\ }\href {https://doi.org/10.1103/Physics.3.92} {\bibfield  {journal}
  {\bibinfo  {journal} {Physics}\ }\textbf {\bibinfo {volume} {3}},\ \bibinfo
  {pages} {92} (\bibinfo {year} {2010})}\BibitemShut {NoStop}%
\bibitem [{\citenamefont {Affleck}(1985)}]{affleck1985}%
  \BibitemOpen
  \bibfield  {author} {\bibinfo {author} {\bibfnamefont {I.}~\bibnamefont
  {Affleck}},\ }\href {https://doi.org/10.1103/PhysRevLett.54.966} {\bibfield
  {journal} {\bibinfo  {journal} {Phys. Rev. Lett.}\ }\textbf {\bibinfo
  {volume} {54}},\ \bibinfo {pages} {966} (\bibinfo {year} {1985})}\BibitemShut
  {NoStop}%
\bibitem [{\citenamefont {Arovas}\ and\ \citenamefont
  {Auerbach}(1988)}]{arovas1988}%
  \BibitemOpen
  \bibfield  {author} {\bibinfo {author} {\bibfnamefont {D.~P.}\ \bibnamefont
  {Arovas}}\ and\ \bibinfo {author} {\bibfnamefont {A.}~\bibnamefont
  {Auerbach}},\ }\href {https://doi.org/10.1103/PhysRevB.38.316} {\bibfield
  {journal} {\bibinfo  {journal} {Phys. Rev. B}\ }\textbf {\bibinfo {volume}
  {38}},\ \bibinfo {pages} {316} (\bibinfo {year} {1988})}\BibitemShut
  {NoStop}%
\bibitem [{\citenamefont {Affleck}\ and\ \citenamefont
  {Marston}(1988)}]{affleck1988}%
  \BibitemOpen
  \bibfield  {author} {\bibinfo {author} {\bibfnamefont {I.}~\bibnamefont
  {Affleck}}\ and\ \bibinfo {author} {\bibfnamefont {J.~B.}\ \bibnamefont
  {Marston}},\ }\href {https://doi.org/10.1103/PhysRevB.37.3774} {\bibfield
  {journal} {\bibinfo  {journal} {Phys. Rev. B}\ }\textbf {\bibinfo {volume}
  {37}},\ \bibinfo {pages} {3774} (\bibinfo {year} {1988})}\BibitemShut
  {NoStop}%
\bibitem [{\citenamefont {Read}\ and\ \citenamefont
  {Sachdev}(1990)}]{read1990}%
  \BibitemOpen
  \bibfield  {author} {\bibinfo {author} {\bibfnamefont {N.}~\bibnamefont
  {Read}}\ and\ \bibinfo {author} {\bibfnamefont {S.}~\bibnamefont {Sachdev}},\
  }\href {https://doi.org/10.1103/PhysRevB.42.4568} {\bibfield  {journal}
  {\bibinfo  {journal} {Phys. Rev. B}\ }\textbf {\bibinfo {volume} {42}},\
  \bibinfo {pages} {4568} (\bibinfo {year} {1990})}\BibitemShut {NoStop}%
\bibitem [{\citenamefont {Wu}\ \emph {et~al.}(2003)\citenamefont {Wu},
  \citenamefont {Hu},\ and\ \citenamefont {Zhang}}]{wu2003}%
  \BibitemOpen
  \bibfield  {author} {\bibinfo {author} {\bibfnamefont {C.}~\bibnamefont
  {Wu}}, \bibinfo {author} {\bibfnamefont {J.-p.}\ \bibnamefont {Hu}},\ and\
  \bibinfo {author} {\bibfnamefont {S.-c.}\ \bibnamefont {Zhang}},\ }\href
  {https://doi.org/10.1103/PhysRevLett.91.186402} {\bibfield  {journal}
  {\bibinfo  {journal} {Phys. Rev. Lett.}\ }\textbf {\bibinfo {volume} {91}},\
  \bibinfo {pages} {186402} (\bibinfo {year} {2003})}\BibitemShut {NoStop}%
\bibitem [{\citenamefont {Chen}\ \emph {et~al.}(2005)\citenamefont {Chen},
  \citenamefont {Wu}, \citenamefont {Zhang},\ and\ \citenamefont
  {Wang}}]{chen2005}%
  \BibitemOpen
  \bibfield  {author} {\bibinfo {author} {\bibfnamefont {S.}~\bibnamefont
  {Chen}}, \bibinfo {author} {\bibfnamefont {C.}~\bibnamefont {Wu}}, \bibinfo
  {author} {\bibfnamefont {S.-C.}\ \bibnamefont {Zhang}},\ and\ \bibinfo
  {author} {\bibfnamefont {Y.}~\bibnamefont {Wang}},\ }\href
  {https://doi.org/10.1103/PhysRevB.72.214428} {\bibfield  {journal} {\bibinfo
  {journal} {Phys. Rev. B}\ }\textbf {\bibinfo {volume} {72}},\ \bibinfo
  {pages} {214428} (\bibinfo {year} {2005})}\BibitemShut {NoStop}%
\bibitem [{\citenamefont {Wu}(2005)}]{wu2005}%
  \BibitemOpen
  \bibfield  {author} {\bibinfo {author} {\bibfnamefont {C.}~\bibnamefont
  {Wu}},\ }\href {https://doi.org/10.1103/PhysRevLett.95.266404} {\bibfield
  {journal} {\bibinfo  {journal} {Phys. Rev. Lett.}\ }\textbf {\bibinfo
  {volume} {95}},\ \bibinfo {pages} {266404} (\bibinfo {year}
  {2005})}\BibitemShut {NoStop}%
\bibitem [{\citenamefont {{Wu}}(2006)}]{wu2006}%
  \BibitemOpen
  \bibfield  {author} {\bibinfo {author} {\bibfnamefont {C.}~\bibnamefont
  {{Wu}}},\ }\href {https://doi.org/10.1142/S0217984906012213} {\bibfield
  {journal} {\bibinfo  {journal} {Modern Physics Letters B}\ }\textbf {\bibinfo
  {volume} {20}},\ \bibinfo {pages} {1707} (\bibinfo {year} {2006})},\ \Eprint
  {https://arxiv.org/abs/arXiv:cond-mat/0608690} {arXiv:cond-mat/0608690}
  \BibitemShut {NoStop}%
\bibitem [{\citenamefont {Xu}\ and\ \citenamefont {Wu}(2008)}]{xu2008}%
  \BibitemOpen
  \bibfield  {author} {\bibinfo {author} {\bibfnamefont {C.}~\bibnamefont
  {Xu}}\ and\ \bibinfo {author} {\bibfnamefont {C.}~\bibnamefont {Wu}},\ }\href
  {https://doi.org/10.1103/PhysRevB.77.134449} {\bibfield  {journal} {\bibinfo
  {journal} {Phys. Rev. B}\ }\textbf {\bibinfo {volume} {77}},\ \bibinfo
  {pages} {134449} (\bibinfo {year} {2008})}\BibitemShut {NoStop}%
\bibitem [{\citenamefont {WU}\ \emph {et~al.}(2010)\citenamefont {WU},
  \citenamefont {HU},\ and\ \citenamefont {ZHANG}}]{wu2010a}%
  \BibitemOpen
  \bibfield  {author} {\bibinfo {author} {\bibfnamefont {C.}~\bibnamefont
  {WU}}, \bibinfo {author} {\bibfnamefont {J.}~\bibnamefont {HU}},\ and\
  \bibinfo {author} {\bibfnamefont {S.-C.}\ \bibnamefont {ZHANG}},\ }\href
  {https://doi.org/10.1142/S0217979210054968} {\bibfield  {journal} {\bibinfo
  {journal} {International Journal of Modern Physics B}\ }\textbf {\bibinfo
  {volume} {24}},\ \bibinfo {pages} {311} (\bibinfo {year} {2010})},\ \Eprint
  {https://arxiv.org/abs/http://www.worldscientific.com/doi/pdf/10.1142/S0217979210054968}
  {http://www.worldscientific.com/doi/pdf/10.1142/S0217979210054968}
  \BibitemShut {NoStop}%
\bibitem [{\citenamefont {Hung}\ \emph {et~al.}(2011)\citenamefont {Hung},
  \citenamefont {Wang},\ and\ \citenamefont {Wu}}]{hung2011}%
  \BibitemOpen
  \bibfield  {author} {\bibinfo {author} {\bibfnamefont {H.-H.}\ \bibnamefont
  {Hung}}, \bibinfo {author} {\bibfnamefont {Y.}~\bibnamefont {Wang}},\ and\
  \bibinfo {author} {\bibfnamefont {C.}~\bibnamefont {Wu}},\ }\href
  {https://doi.org/10.1103/PhysRevB.84.054406} {\bibfield  {journal} {\bibinfo
  {journal} {Phys. Rev. B}\ }\textbf {\bibinfo {volume} {84}},\ \bibinfo
  {pages} {054406} (\bibinfo {year} {2011})}\BibitemShut {NoStop}%
\bibitem [{\citenamefont {Gao}\ and\ \citenamefont {Wu}(2020)}]{gao2020}%
  \BibitemOpen
  \bibfield  {author} {\bibinfo {author} {\bibfnamefont {Z.-Q.}\ \bibnamefont
  {Gao}}\ and\ \bibinfo {author} {\bibfnamefont {C.}~\bibnamefont {Wu}},\
  }\href@noop {} {\bibinfo {title} {Exceptional symmetry of g$_2$ in
  spin-$\frac{3}{2}$ fermion systems}} (\bibinfo {year} {2020}),\ \Eprint
  {https://arxiv.org/abs/2010.14126} {arXiv:2010.14126 [cond-mat.str-el]}
  \BibitemShut {NoStop}%
\bibitem [{\citenamefont {Gorshkov}\ \emph {et~al.}(2010)\citenamefont
  {Gorshkov}, \citenamefont {Hermele}, \citenamefont {Gurarie}, \citenamefont
  {Xu}, \citenamefont {Julienne}, \citenamefont {Ye}, \citenamefont {Zoller},
  \citenamefont {Demler}, \citenamefont {Lukin},\ and\ \citenamefont
  {Rey}}]{gorshkov2010}%
  \BibitemOpen
  \bibfield  {author} {\bibinfo {author} {\bibfnamefont {A.~V.}\ \bibnamefont
  {Gorshkov}}, \bibinfo {author} {\bibfnamefont {M.}~\bibnamefont {Hermele}},
  \bibinfo {author} {\bibfnamefont {V.}~\bibnamefont {Gurarie}}, \bibinfo
  {author} {\bibfnamefont {C.}~\bibnamefont {Xu}}, \bibinfo {author}
  {\bibfnamefont {P.~S.}\ \bibnamefont {Julienne}}, \bibinfo {author}
  {\bibfnamefont {J.}~\bibnamefont {Ye}}, \bibinfo {author} {\bibfnamefont
  {P.}~\bibnamefont {Zoller}}, \bibinfo {author} {\bibfnamefont
  {E.}~\bibnamefont {Demler}}, \bibinfo {author} {\bibfnamefont {M.~D.}\
  \bibnamefont {Lukin}},\ and\ \bibinfo {author} {\bibfnamefont {A.~M.}\
  \bibnamefont {Rey}},\ }\href {http://dx.doi.org/10.1038/nphys1535} {\bibfield
   {journal} {\bibinfo  {journal} {Nat Phys}\ }\textbf {\bibinfo {volume}
  {6}},\ \bibinfo {pages} {289} (\bibinfo {year} {2010})}\BibitemShut {NoStop}%
\bibitem [{\citenamefont {Hermele}\ \emph {et~al.}(2009)\citenamefont
  {Hermele}, \citenamefont {Gurarie},\ and\ \citenamefont {Rey}}]{hermele2009}%
  \BibitemOpen
  \bibfield  {author} {\bibinfo {author} {\bibfnamefont {M.}~\bibnamefont
  {Hermele}}, \bibinfo {author} {\bibfnamefont {V.}~\bibnamefont {Gurarie}},\
  and\ \bibinfo {author} {\bibfnamefont {A.~M.}\ \bibnamefont {Rey}},\ }\href
  {https://doi.org/10.1103/PhysRevLett.103.135301} {\bibfield  {journal}
  {\bibinfo  {journal} {Phys. Rev. Lett.}\ }\textbf {\bibinfo {volume} {103}},\
  \bibinfo {pages} {135301} (\bibinfo {year} {2009})}\BibitemShut {NoStop}%
\bibitem [{\citenamefont {Xu}(2010)}]{xu2010}%
  \BibitemOpen
  \bibfield  {author} {\bibinfo {author} {\bibfnamefont {C.}~\bibnamefont
  {Xu}},\ }\href {https://doi.org/10.1103/PhysRevB.81.144431} {\bibfield
  {journal} {\bibinfo  {journal} {Phys. Rev. B}\ }\textbf {\bibinfo {volume}
  {81}},\ \bibinfo {pages} {144431} (\bibinfo {year} {2010})}\BibitemShut
  {NoStop}%
\bibitem [{\citenamefont {Cazalilla}\ \emph {et~al.}(2009)\citenamefont
  {Cazalilla}, \citenamefont {Ho},\ and\ \citenamefont {Ueda}}]{cazalilla2009}%
  \BibitemOpen
  \bibfield  {author} {\bibinfo {author} {\bibfnamefont {M.~A.}\ \bibnamefont
  {Cazalilla}}, \bibinfo {author} {\bibfnamefont {A.~F.}\ \bibnamefont {Ho}},\
  and\ \bibinfo {author} {\bibfnamefont {M.}~\bibnamefont {Ueda}},\ }\href
  {http://stacks.iop.org/1367-2630/11/i=10/a=103033} {\bibfield  {journal}
  {\bibinfo  {journal} {New Journal of Physics}\ }\textbf {\bibinfo {volume}
  {11}},\ \bibinfo {pages} {103033} (\bibinfo {year} {2009})}\BibitemShut
  {NoStop}%
\bibitem [{\citenamefont {Controzzi}\ and\ \citenamefont
  {Tsvelik}(2006)}]{controzzi2006}%
  \BibitemOpen
  \bibfield  {author} {\bibinfo {author} {\bibfnamefont {D.}~\bibnamefont
  {Controzzi}}\ and\ \bibinfo {author} {\bibfnamefont {A.~M.}\ \bibnamefont
  {Tsvelik}},\ }\href {https://doi.org/10.1103/PhysRevLett.96.097205}
  {\bibfield  {journal} {\bibinfo  {journal} {Phys. Rev. Lett.}\ }\textbf
  {\bibinfo {volume} {96}},\ \bibinfo {pages} {097205} (\bibinfo {year}
  {2006})}\BibitemShut {NoStop}%
\bibitem [{\citenamefont {Lecheminant}\ \emph {et~al.}(2005)\citenamefont
  {Lecheminant}, \citenamefont {Boulat},\ and\ \citenamefont
  {Azaria}}]{lecheminant2005}%
  \BibitemOpen
  \bibfield  {author} {\bibinfo {author} {\bibfnamefont {P.}~\bibnamefont
  {Lecheminant}}, \bibinfo {author} {\bibfnamefont {E.}~\bibnamefont
  {Boulat}},\ and\ \bibinfo {author} {\bibfnamefont {P.}~\bibnamefont
  {Azaria}},\ }\href {https://doi.org/10.1103/PhysRevLett.95.240402} {\bibfield
   {journal} {\bibinfo  {journal} {Phys. Rev. Lett.}\ }\textbf {\bibinfo
  {volume} {95}},\ \bibinfo {pages} {240402} (\bibinfo {year}
  {2005})}\BibitemShut {NoStop}%
\bibitem [{\citenamefont {Hattori}(2005)}]{hattori2005}%
  \BibitemOpen
  \bibfield  {author} {\bibinfo {author} {\bibfnamefont {K.}~\bibnamefont
  {Hattori}},\ }\href {https://doi.org/10.1143/JPSJ.74.3135} {\bibfield
  {journal} {\bibinfo  {journal} {Journal of the Physical Society of Japan}\
  }\textbf {\bibinfo {volume} {74}},\ \bibinfo {pages} {3135} (\bibinfo {year}
  {2005})}\BibitemShut {NoStop}%
\bibitem [{\citenamefont {Tu}\ \emph {et~al.}(2006)\citenamefont {Tu},
  \citenamefont {Zhang},\ and\ \citenamefont {Yu}}]{tu2006}%
  \BibitemOpen
  \bibfield  {author} {\bibinfo {author} {\bibfnamefont {H.-H.}\ \bibnamefont
  {Tu}}, \bibinfo {author} {\bibfnamefont {G.-M.}\ \bibnamefont {Zhang}},\ and\
  \bibinfo {author} {\bibfnamefont {L.}~\bibnamefont {Yu}},\ }\href
  {https://doi.org/10.1103/PhysRevB.74.174404} {\bibfield  {journal} {\bibinfo
  {journal} {Phys. Rev. B}\ }\textbf {\bibinfo {volume} {74}},\ \bibinfo
  {pages} {174404} (\bibinfo {year} {2006})}\BibitemShut {NoStop}%
\bibitem [{\citenamefont {Tu}\ \emph {et~al.}(2007)\citenamefont {Tu},
  \citenamefont {Zhang},\ and\ \citenamefont {Yu}}]{tu2007}%
  \BibitemOpen
  \bibfield  {author} {\bibinfo {author} {\bibfnamefont {H.-H.}\ \bibnamefont
  {Tu}}, \bibinfo {author} {\bibfnamefont {G.-M.}\ \bibnamefont {Zhang}},\ and\
  \bibinfo {author} {\bibfnamefont {L.}~\bibnamefont {Yu}},\ }\href
  {https://doi.org/10.1103/PhysRevB.76.014438} {\bibfield  {journal} {\bibinfo
  {journal} {Phys. Rev. B}\ }\textbf {\bibinfo {volume} {76}},\ \bibinfo
  {pages} {014438} (\bibinfo {year} {2007})}\BibitemShut {NoStop}%
\bibitem [{\citenamefont {\"Ostlund}\ \emph {et~al.}(2005)\citenamefont
  {\"Ostlund}, \citenamefont {Hansson},\ and\ \citenamefont
  {Karlhede}}]{ostlund2005}%
  \BibitemOpen
  \bibfield  {author} {\bibinfo {author} {\bibfnamefont {S.}~\bibnamefont
  {\"Ostlund}}, \bibinfo {author} {\bibfnamefont {T.~H.}\ \bibnamefont
  {Hansson}},\ and\ \bibinfo {author} {\bibfnamefont {A.}~\bibnamefont
  {Karlhede}},\ }\href {https://doi.org/10.1103/PhysRevB.71.165121} {\bibfield
  {journal} {\bibinfo  {journal} {Phys. Rev. B}\ }\textbf {\bibinfo {volume}
  {71}},\ \bibinfo {pages} {165121} (\bibinfo {year} {2005})}\BibitemShut
  {NoStop}%
\bibitem [{\citenamefont {Bartenstein}\ \emph {et~al.}(2005)\citenamefont
  {Bartenstein}, \citenamefont {Altmeyer}, \citenamefont {Riedl}, \citenamefont
  {Geursen}, \citenamefont {Jochim}, \citenamefont {Chin}, \citenamefont
  {Denschlag}, \citenamefont {Grimm}, \citenamefont {Simoni}, \citenamefont
  {Tiesinga}, \citenamefont {Williams},\ and\ \citenamefont
  {Julienne}}]{bartenstein2005}%
  \BibitemOpen
  \bibfield  {author} {\bibinfo {author} {\bibfnamefont {M.}~\bibnamefont
  {Bartenstein}}, \bibinfo {author} {\bibfnamefont {A.}~\bibnamefont
  {Altmeyer}}, \bibinfo {author} {\bibfnamefont {S.}~\bibnamefont {Riedl}},
  \bibinfo {author} {\bibfnamefont {R.}~\bibnamefont {Geursen}}, \bibinfo
  {author} {\bibfnamefont {S.}~\bibnamefont {Jochim}}, \bibinfo {author}
  {\bibfnamefont {C.}~\bibnamefont {Chin}}, \bibinfo {author} {\bibfnamefont
  {J.~H.}\ \bibnamefont {Denschlag}}, \bibinfo {author} {\bibfnamefont
  {R.}~\bibnamefont {Grimm}}, \bibinfo {author} {\bibfnamefont
  {A.}~\bibnamefont {Simoni}}, \bibinfo {author} {\bibfnamefont
  {E.}~\bibnamefont {Tiesinga}}, \bibinfo {author} {\bibfnamefont {C.~J.}\
  \bibnamefont {Williams}},\ and\ \bibinfo {author} {\bibfnamefont {P.~S.}\
  \bibnamefont {Julienne}},\ }\href
  {https://doi.org/10.1103/PhysRevLett.94.103201} {\bibfield  {journal}
  {\bibinfo  {journal} {Phys. Rev. Lett.}\ }\textbf {\bibinfo {volume} {94}},\
  \bibinfo {pages} {103201} (\bibinfo {year} {2005})}\BibitemShut {NoStop}%
\bibitem [{\citenamefont {Rapp}\ \emph {et~al.}(2007)\citenamefont {Rapp},
  \citenamefont {Zar\'and}, \citenamefont {Honerkamp},\ and\ \citenamefont
  {Hofstetter}}]{rapp2007}%
  \BibitemOpen
  \bibfield  {author} {\bibinfo {author} {\bibfnamefont {A.}~\bibnamefont
  {Rapp}}, \bibinfo {author} {\bibfnamefont {G.}~\bibnamefont {Zar\'and}},
  \bibinfo {author} {\bibfnamefont {C.}~\bibnamefont {Honerkamp}},\ and\
  \bibinfo {author} {\bibfnamefont {W.}~\bibnamefont {Hofstetter}},\ }\href
  {https://doi.org/10.1103/PhysRevLett.98.160405} {\bibfield  {journal}
  {\bibinfo  {journal} {Phys. Rev. Lett.}\ }\textbf {\bibinfo {volume} {98}},\
  \bibinfo {pages} {160405} (\bibinfo {year} {2007})}\BibitemShut {NoStop}%
\bibitem [{\citenamefont {{Rapp}}\ \emph {et~al.}(2008)\citenamefont {{Rapp}},
  \citenamefont {{Hofstetter}},\ and\ \citenamefont {{Zar{\'a}nd}}}]{rapp2008}%
  \BibitemOpen
  \bibfield  {author} {\bibinfo {author} {\bibfnamefont {{\'A}.}~\bibnamefont
  {{Rapp}}}, \bibinfo {author} {\bibfnamefont {W.}~\bibnamefont
  {{Hofstetter}}},\ and\ \bibinfo {author} {\bibfnamefont {G.}~\bibnamefont
  {{Zar{\'a}nd}}},\ }\href {https://doi.org/10.1103/PhysRevB.77.144520}
  {\bibfield  {journal} {\bibinfo  {journal} {Phys. Rev. B}\ }\textbf {\bibinfo
  {volume} {77}},\ \bibinfo {eid} {144520} (\bibinfo {year} {2008})},\ \Eprint
  {https://arxiv.org/abs/0707.2378} {arXiv:0707.2378 [cond-mat.supr-con]}
  \BibitemShut {NoStop}%
\bibitem [{\citenamefont {Taie}\ \emph {et~al.}(2010)\citenamefont {Taie},
  \citenamefont {Takasu}, \citenamefont {Sugawa}, \citenamefont {Yamazaki},
  \citenamefont {Tsujimoto}, \citenamefont {Murakami},\ and\ \citenamefont
  {Takahashi}}]{taie2010}%
  \BibitemOpen
  \bibfield  {author} {\bibinfo {author} {\bibfnamefont {S.}~\bibnamefont
  {Taie}}, \bibinfo {author} {\bibfnamefont {Y.}~\bibnamefont {Takasu}},
  \bibinfo {author} {\bibfnamefont {S.}~\bibnamefont {Sugawa}}, \bibinfo
  {author} {\bibfnamefont {R.}~\bibnamefont {Yamazaki}}, \bibinfo {author}
  {\bibfnamefont {T.}~\bibnamefont {Tsujimoto}}, \bibinfo {author}
  {\bibfnamefont {R.}~\bibnamefont {Murakami}},\ and\ \bibinfo {author}
  {\bibfnamefont {Y.}~\bibnamefont {Takahashi}},\ }\href
  {https://doi.org/10.1103/PhysRevLett.105.190401} {\bibfield  {journal}
  {\bibinfo  {journal} {Phys. Rev. Lett.}\ }\textbf {\bibinfo {volume} {105}},\
  \bibinfo {pages} {190401} (\bibinfo {year} {2010})}\BibitemShut {NoStop}%
\bibitem [{\citenamefont {Taie}\ \emph {et~al.}(2012)\citenamefont {Taie},
  \citenamefont {Yamazaki}, \citenamefont {Sugawa},\ and\ \citenamefont
  {Takahashi}}]{taie2012}%
  \BibitemOpen
  \bibfield  {author} {\bibinfo {author} {\bibfnamefont {S.}~\bibnamefont
  {Taie}}, \bibinfo {author} {\bibfnamefont {R.}~\bibnamefont {Yamazaki}},
  \bibinfo {author} {\bibfnamefont {S.}~\bibnamefont {Sugawa}},\ and\ \bibinfo
  {author} {\bibfnamefont {Y.}~\bibnamefont {Takahashi}},\ }\href
  {http://dx.doi.org/10.1038/nphys2430} {\bibfield  {journal} {\bibinfo
  {journal} {Nat Phys}\ }\textbf {\bibinfo {volume} {8}},\ \bibinfo {pages}
  {825} (\bibinfo {year} {2012})}\BibitemShut {NoStop}%
\bibitem [{\citenamefont {{Sugawa}}\ \emph {et~al.}(2011)\citenamefont
  {{Sugawa}}, \citenamefont {{Inaba}}, \citenamefont {{Taie}}, \citenamefont
  {{Yamazaki}}, \citenamefont {{Yamashita}},\ and\ \citenamefont
  {{Takahashi}}}]{sugawa2011}%
  \BibitemOpen
  \bibfield  {author} {\bibinfo {author} {\bibfnamefont {S.}~\bibnamefont
  {{Sugawa}}}, \bibinfo {author} {\bibfnamefont {K.}~\bibnamefont {{Inaba}}},
  \bibinfo {author} {\bibfnamefont {S.}~\bibnamefont {{Taie}}}, \bibinfo
  {author} {\bibfnamefont {R.}~\bibnamefont {{Yamazaki}}}, \bibinfo {author}
  {\bibfnamefont {M.}~\bibnamefont {{Yamashita}}},\ and\ \bibinfo {author}
  {\bibfnamefont {Y.}~\bibnamefont {{Takahashi}}},\ }\href
  {https://doi.org/10.1038/nphys2028} {\bibfield  {journal} {\bibinfo
  {journal} {Nature Physics}\ }\textbf {\bibinfo {volume} {7}},\ \bibinfo
  {pages} {642} (\bibinfo {year} {2011})},\ \Eprint
  {https://arxiv.org/abs/1011.4503} {arXiv:1011.4503 [cond-mat.quant-gas]}
  \BibitemShut {NoStop}%
\bibitem [{\citenamefont {Hara}\ \emph {et~al.}(2011)\citenamefont {Hara},
  \citenamefont {Takasu}, \citenamefont {Yamaoka}, \citenamefont {Doyle},\ and\
  \citenamefont {Takahashi}}]{hara2011}%
  \BibitemOpen
  \bibfield  {author} {\bibinfo {author} {\bibfnamefont {H.}~\bibnamefont
  {Hara}}, \bibinfo {author} {\bibfnamefont {Y.}~\bibnamefont {Takasu}},
  \bibinfo {author} {\bibfnamefont {Y.}~\bibnamefont {Yamaoka}}, \bibinfo
  {author} {\bibfnamefont {J.~M.}\ \bibnamefont {Doyle}},\ and\ \bibinfo
  {author} {\bibfnamefont {Y.}~\bibnamefont {Takahashi}},\ }\href
  {https://doi.org/10.1103/PhysRevLett.106.205304} {\bibfield  {journal}
  {\bibinfo  {journal} {Phys. Rev. Lett.}\ }\textbf {\bibinfo {volume} {106}},\
  \bibinfo {pages} {205304} (\bibinfo {year} {2011})}\BibitemShut {NoStop}%
\bibitem [{\citenamefont {Pagano}\ \emph {et~al.}(2014)\citenamefont {Pagano},
  \citenamefont {Mancini}, \citenamefont {Cappellini}, \citenamefont
  {Lombardi}, \citenamefont {Sch?fer}, \citenamefont {Hu}, \citenamefont {Liu},
  \citenamefont {Catani}, \citenamefont {Sias}, \citenamefont {Inguscio},\ and\
  \citenamefont {Fallani}}]{fallani2014}%
  \BibitemOpen
  \bibfield  {author} {\bibinfo {author} {\bibfnamefont {G.}~\bibnamefont
  {Pagano}}, \bibinfo {author} {\bibfnamefont {M.}~\bibnamefont {Mancini}},
  \bibinfo {author} {\bibfnamefont {G.}~\bibnamefont {Cappellini}}, \bibinfo
  {author} {\bibfnamefont {P.}~\bibnamefont {Lombardi}}, \bibinfo {author}
  {\bibfnamefont {F.}~\bibnamefont {Sch?fer}}, \bibinfo {author} {\bibfnamefont
  {H.}~\bibnamefont {Hu}}, \bibinfo {author} {\bibfnamefont {X.-J.}\
  \bibnamefont {Liu}}, \bibinfo {author} {\bibfnamefont {J.}~\bibnamefont
  {Catani}}, \bibinfo {author} {\bibfnamefont {C.}~\bibnamefont {Sias}},
  \bibinfo {author} {\bibfnamefont {M.}~\bibnamefont {Inguscio}},\ and\
  \bibinfo {author} {\bibfnamefont {L.}~\bibnamefont {Fallani}},\ }\href
  {https://doi.org/10.1038/nphys2878} {\bibfield  {journal} {\bibinfo
  {journal} {Nature Physics}\ }\textbf {\bibinfo {volume} {10}},\ \bibinfo
  {pages} {198每201} (\bibinfo {year} {2014})}\BibitemShut {NoStop}%
\bibitem [{\citenamefont {DeSalvo}\ \emph {et~al.}(2010)\citenamefont
  {DeSalvo}, \citenamefont {Yan}, \citenamefont {Mickelson}, \citenamefont
  {Martinez~de Escobar},\ and\ \citenamefont {Killian}}]{desalvo2010}%
  \BibitemOpen
  \bibfield  {author} {\bibinfo {author} {\bibfnamefont {B.~J.}\ \bibnamefont
  {DeSalvo}}, \bibinfo {author} {\bibfnamefont {M.}~\bibnamefont {Yan}},
  \bibinfo {author} {\bibfnamefont {P.~G.}\ \bibnamefont {Mickelson}}, \bibinfo
  {author} {\bibfnamefont {Y.~N.}\ \bibnamefont {Martinez~de Escobar}},\ and\
  \bibinfo {author} {\bibfnamefont {T.~C.}\ \bibnamefont {Killian}},\ }\href
  {https://doi.org/10.1103/PhysRevLett.105.030402} {\bibfield  {journal}
  {\bibinfo  {journal} {Phys. Rev. Lett.}\ }\textbf {\bibinfo {volume} {105}},\
  \bibinfo {pages} {030402} (\bibinfo {year} {2010})}\BibitemShut {NoStop}%
\bibitem [{\citenamefont {{Mickelson}}\ \emph {et~al.}(2010)\citenamefont
  {{Mickelson}}, \citenamefont {{Martinez de Escobar}}, \citenamefont {{Yan}},
  \citenamefont {{Desalvo}},\ and\ \citenamefont {{Killian}}}]{mickelson2010}%
  \BibitemOpen
  \bibfield  {author} {\bibinfo {author} {\bibfnamefont {P.~G.}\ \bibnamefont
  {{Mickelson}}}, \bibinfo {author} {\bibfnamefont {Y.~N.}\ \bibnamefont
  {{Martinez de Escobar}}}, \bibinfo {author} {\bibfnamefont {M.}~\bibnamefont
  {{Yan}}}, \bibinfo {author} {\bibfnamefont {B.~J.}\ \bibnamefont
  {{Desalvo}}},\ and\ \bibinfo {author} {\bibfnamefont {T.~C.}\ \bibnamefont
  {{Killian}}},\ }\href {https://doi.org/10.1103/PhysRevA.81.051601} {\bibfield
   {journal} {\bibinfo  {journal} {Phys. Rev. A}\ }\textbf {\bibinfo {volume}
  {81}},\ \bibinfo {eid} {051601} (\bibinfo {year} {2010})},\ \Eprint
  {https://arxiv.org/abs/1003.3867} {arXiv:1003.3867 [cond-mat.quant-gas]}
  \BibitemShut {NoStop}%
\bibitem [{\citenamefont {{Heinze}}\ \emph {et~al.}(2013)\citenamefont
  {{Heinze}}, \citenamefont {{Krauser}}, \citenamefont {{Fl{\"a}schner}},
  \citenamefont {{Sengstock}}, \citenamefont {{Becker}}, \citenamefont
  {{Ebling}}, \citenamefont {{Eckardt}},\ and\ \citenamefont
  {{Lewenstein}}}]{heinze2013}%
  \BibitemOpen
  \bibfield  {author} {\bibinfo {author} {\bibfnamefont {J.}~\bibnamefont
  {{Heinze}}}, \bibinfo {author} {\bibfnamefont {J.~S.}\ \bibnamefont
  {{Krauser}}}, \bibinfo {author} {\bibfnamefont {N.}~\bibnamefont
  {{Fl{\"a}schner}}}, \bibinfo {author} {\bibfnamefont {K.}~\bibnamefont
  {{Sengstock}}}, \bibinfo {author} {\bibfnamefont {C.}~\bibnamefont
  {{Becker}}}, \bibinfo {author} {\bibfnamefont {U.}~\bibnamefont {{Ebling}}},
  \bibinfo {author} {\bibfnamefont {A.}~\bibnamefont {{Eckardt}}},\ and\
  \bibinfo {author} {\bibfnamefont {M.}~\bibnamefont {{Lewenstein}}},\
  }\href@noop {} {\bibfield  {journal} {\bibinfo  {journal} {ArXiv e-prints}\ }
  (\bibinfo {year} {2013})},\ \Eprint {https://arxiv.org/abs/1302.4323}
  {arXiv:1302.4323 [cond-mat.quant-gas]} \BibitemShut {NoStop}%
\bibitem [{\citenamefont {Krauser}\ \emph {et~al.}(2012)\citenamefont
  {Krauser}, \citenamefont {Heinze}, \citenamefont {Flaschner}, \citenamefont
  {Gotze}, \citenamefont {Jurgensen}, \citenamefont {Luhmann}, \citenamefont
  {Becker},\ and\ \citenamefont {Sengstock}}]{krauser2012}%
  \BibitemOpen
  \bibfield  {author} {\bibinfo {author} {\bibfnamefont {J.~S.}\ \bibnamefont
  {Krauser}}, \bibinfo {author} {\bibfnamefont {J.}~\bibnamefont {Heinze}},
  \bibinfo {author} {\bibfnamefont {N.}~\bibnamefont {Flaschner}}, \bibinfo
  {author} {\bibfnamefont {S.}~\bibnamefont {Gotze}}, \bibinfo {author}
  {\bibfnamefont {O.}~\bibnamefont {Jurgensen}}, \bibinfo {author}
  {\bibfnamefont {D.-S.}\ \bibnamefont {Luhmann}}, \bibinfo {author}
  {\bibfnamefont {C.}~\bibnamefont {Becker}},\ and\ \bibinfo {author}
  {\bibfnamefont {K.}~\bibnamefont {Sengstock}},\ }\href
  {http://dx.doi.org/10.1038/nphys2409} {\bibfield  {journal} {\bibinfo
  {journal} {Nat Phys}\ }\textbf {\bibinfo {volume} {8}},\ \bibinfo {pages}
  {813} (\bibinfo {year} {2012})}\BibitemShut {NoStop}%
\bibitem [{\citenamefont {{Bishof}}\ \emph
  {et~al.}(2011{\natexlab{a}})\citenamefont {{Bishof}}, \citenamefont
  {{Martin}}, \citenamefont {{Swallows}}, \citenamefont {{Benko}},
  \citenamefont {{Lin}}, \citenamefont {{Qu{\'e}m{\'e}ner}}, \citenamefont
  {{Rey}},\ and\ \citenamefont {{Ye}}}]{bishof2011}%
  \BibitemOpen
  \bibfield  {author} {\bibinfo {author} {\bibfnamefont {M.}~\bibnamefont
  {{Bishof}}}, \bibinfo {author} {\bibfnamefont {M.~J.}\ \bibnamefont
  {{Martin}}}, \bibinfo {author} {\bibfnamefont {M.~D.}\ \bibnamefont
  {{Swallows}}}, \bibinfo {author} {\bibfnamefont {C.}~\bibnamefont {{Benko}}},
  \bibinfo {author} {\bibfnamefont {Y.}~\bibnamefont {{Lin}}}, \bibinfo
  {author} {\bibfnamefont {G.}~\bibnamefont {{Qu{\'e}m{\'e}ner}}}, \bibinfo
  {author} {\bibfnamefont {A.~M.}\ \bibnamefont {{Rey}}},\ and\ \bibinfo
  {author} {\bibfnamefont {J.}~\bibnamefont {{Ye}}},\ }\href
  {https://doi.org/10.1103/PhysRevA.84.052716} {\bibfield  {journal} {\bibinfo
  {journal} {Phys. Rev. A}\ }\textbf {\bibinfo {volume} {84}},\ \bibinfo {eid}
  {052716} (\bibinfo {year} {2011}{\natexlab{a}})},\ \Eprint
  {https://arxiv.org/abs/1108.1431} {arXiv:1108.1431 [physics.atom-ph]}
  \BibitemShut {NoStop}%
\bibitem [{\citenamefont {{Bishof}}\ \emph
  {et~al.}(2011{\natexlab{b}})\citenamefont {{Bishof}}, \citenamefont {{Lin}},
  \citenamefont {{Swallows}}, \citenamefont {{Gorshkov}}, \citenamefont
  {{Ye}},\ and\ \citenamefont {{Rey}}}]{bishof2011a}%
  \BibitemOpen
  \bibfield  {author} {\bibinfo {author} {\bibfnamefont {M.}~\bibnamefont
  {{Bishof}}}, \bibinfo {author} {\bibfnamefont {Y.}~\bibnamefont {{Lin}}},
  \bibinfo {author} {\bibfnamefont {M.~D.}\ \bibnamefont {{Swallows}}},
  \bibinfo {author} {\bibfnamefont {A.~V.}\ \bibnamefont {{Gorshkov}}},
  \bibinfo {author} {\bibfnamefont {J.}~\bibnamefont {{Ye}}},\ and\ \bibinfo
  {author} {\bibfnamefont {A.~M.}\ \bibnamefont {{Rey}}},\ }\href
  {https://doi.org/10.1103/PhysRevLett.106.250801} {\bibfield  {journal}
  {\bibinfo  {journal} {Physical Review Letters}\ }\textbf {\bibinfo {volume}
  {106}},\ \bibinfo {eid} {250801} (\bibinfo {year} {2011}{\natexlab{b}})},\
  \Eprint {https://arxiv.org/abs/1102.1016} {arXiv:1102.1016 [quant-ph]}
  \BibitemShut {NoStop}%
\bibitem [{\citenamefont {{Martin}}\ \emph {et~al.}(2012)\citenamefont
  {{Martin}}, \citenamefont {{Bishof}}, \citenamefont {{Swallows}},
  \citenamefont {{Zhang}}, \citenamefont {{Benko}}, \citenamefont
  {{von-Stecher}}, \citenamefont {{Gorshkov}}, \citenamefont {{Rey}},\ and\
  \citenamefont {{Ye}}}]{martin2012}%
  \BibitemOpen
  \bibfield  {author} {\bibinfo {author} {\bibfnamefont {M.~J.}\ \bibnamefont
  {{Martin}}}, \bibinfo {author} {\bibfnamefont {M.}~\bibnamefont {{Bishof}}},
  \bibinfo {author} {\bibfnamefont {M.~D.}\ \bibnamefont {{Swallows}}},
  \bibinfo {author} {\bibfnamefont {X.}~\bibnamefont {{Zhang}}}, \bibinfo
  {author} {\bibfnamefont {C.}~\bibnamefont {{Benko}}}, \bibinfo {author}
  {\bibfnamefont {J.}~\bibnamefont {{von-Stecher}}}, \bibinfo {author}
  {\bibfnamefont {A.~V.}\ \bibnamefont {{Gorshkov}}}, \bibinfo {author}
  {\bibfnamefont {A.~M.}\ \bibnamefont {{Rey}}},\ and\ \bibinfo {author}
  {\bibfnamefont {J.}~\bibnamefont {{Ye}}},\ }\href@noop {} {\bibfield
  {journal} {\bibinfo  {journal} {ArXiv e-prints}\ } (\bibinfo {year}
  {2012})},\ \Eprint {https://arxiv.org/abs/1212.6291} {arXiv:1212.6291
  [physics.atom-ph]} \BibitemShut {NoStop}%
\bibitem [{\citenamefont {Wu}(2012)}]{wu2012}%
  \BibitemOpen
  \bibfield  {author} {\bibinfo {author} {\bibfnamefont {C.}~\bibnamefont
  {Wu}},\ }\href {http://dx.doi.org/10.1038/nphys2432} {\bibfield  {journal}
  {\bibinfo  {journal} {Nat Phys}\ }\textbf {\bibinfo {volume} {8}},\ \bibinfo
  {pages} {784} (\bibinfo {year} {2012})}\BibitemShut {NoStop}%
\bibitem [{\citenamefont {Schlottmann}(1994)}]{schlottmann1994}%
  \BibitemOpen
  \bibfield  {author} {\bibinfo {author} {\bibfnamefont {P.}~\bibnamefont
  {Schlottmann}},\ }\href {http://stacks.iop.org/0953-8984/6/i=7/a=008}
  {\bibfield  {journal} {\bibinfo  {journal} {Journal of Physics: Condensed
  Matter}\ }\textbf {\bibinfo {volume} {6}},\ \bibinfo {pages} {1359} (\bibinfo
  {year} {1994})}\BibitemShut {NoStop}%
\bibitem [{\citenamefont {{Stepanenko}}\ and\ \citenamefont
  {{Gunn}}(1999)}]{stepanenko1999}%
  \BibitemOpen
  \bibfield  {author} {\bibinfo {author} {\bibfnamefont {A.~S.}\ \bibnamefont
  {{Stepanenko}}}\ and\ \bibinfo {author} {\bibfnamefont {J.~M.~F.}\
  \bibnamefont {{Gunn}}},\ }\href@noop {} {\bibfield  {journal} {\bibinfo
  {journal} {eprint arXiv:cond-mat/9901317}\ } (\bibinfo {year} {1999})},\
  \Eprint {https://arxiv.org/abs/arXiv:cond-mat/9901317}
  {arXiv:cond-mat/9901317} \BibitemShut {NoStop}%
\bibitem [{\citenamefont {Wang}\ \emph {et~al.}(2014)\citenamefont {Wang},
  \citenamefont {Li}, \citenamefont {Cai}, \citenamefont {Zhou}, \citenamefont
  {Wang},\ and\ \citenamefont {Wu}}]{wang2014}%
  \BibitemOpen
  \bibfield  {author} {\bibinfo {author} {\bibfnamefont {D.}~\bibnamefont
  {Wang}}, \bibinfo {author} {\bibfnamefont {Y.}~\bibnamefont {Li}}, \bibinfo
  {author} {\bibfnamefont {Z.}~\bibnamefont {Cai}}, \bibinfo {author}
  {\bibfnamefont {Z.}~\bibnamefont {Zhou}}, \bibinfo {author} {\bibfnamefont
  {Y.}~\bibnamefont {Wang}},\ and\ \bibinfo {author} {\bibfnamefont
  {C.}~\bibnamefont {Wu}},\ }\href
  {https://doi.org/10.1103/PhysRevLett.112.156403} {\bibfield  {journal}
  {\bibinfo  {journal} {Phys. Rev. Lett.}\ }\textbf {\bibinfo {volume} {112}},\
  \bibinfo {pages} {156403} (\bibinfo {year} {2014})}\BibitemShut {NoStop}%
\bibitem [{\citenamefont {Wang}\ \emph {et~al.}(2019)\citenamefont {Wang},
  \citenamefont {Wang},\ and\ \citenamefont {Wu}}]{wang2019}%
  \BibitemOpen
  \bibfield  {author} {\bibinfo {author} {\bibfnamefont {D.}~\bibnamefont
  {Wang}}, \bibinfo {author} {\bibfnamefont {L.}~\bibnamefont {Wang}},\ and\
  \bibinfo {author} {\bibfnamefont {C.}~\bibnamefont {Wu}},\ }\href
  {https://doi.org/10.1103/PhysRevB.100.115155} {\bibfield  {journal} {\bibinfo
   {journal} {Phys. Rev. B}\ }\textbf {\bibinfo {volume} {100}},\ \bibinfo
  {pages} {115155} (\bibinfo {year} {2019})}\BibitemShut {NoStop}%
\bibitem [{\citenamefont {Zhou}\ \emph {et~al.}(2016)\citenamefont {Zhou},
  \citenamefont {Wang}, \citenamefont {Meng}, \citenamefont {Wang},\ and\
  \citenamefont {Wu}}]{zhou2016}%
  \BibitemOpen
  \bibfield  {author} {\bibinfo {author} {\bibfnamefont {Z.}~\bibnamefont
  {Zhou}}, \bibinfo {author} {\bibfnamefont {D.}~\bibnamefont {Wang}}, \bibinfo
  {author} {\bibfnamefont {Z.~Y.}\ \bibnamefont {Meng}}, \bibinfo {author}
  {\bibfnamefont {Y.}~\bibnamefont {Wang}},\ and\ \bibinfo {author}
  {\bibfnamefont {C.}~\bibnamefont {Wu}},\ }\href
  {https://doi.org/10.1103/PhysRevB.93.245157} {\bibfield  {journal} {\bibinfo
  {journal} {Phys. Rev. B}\ }\textbf {\bibinfo {volume} {93}},\ \bibinfo
  {pages} {245157} (\bibinfo {year} {2016})}\BibitemShut {NoStop}%
\bibitem [{\citenamefont {Xu}\ \emph {et~al.}(2019)\citenamefont {Xu},
  \citenamefont {Wang}, \citenamefont {Zhou},\ and\ \citenamefont
  {Wu}}]{xu2019}%
  \BibitemOpen
  \bibfield  {author} {\bibinfo {author} {\bibfnamefont {H.}~\bibnamefont
  {Xu}}, \bibinfo {author} {\bibfnamefont {Y.}~\bibnamefont {Wang}}, \bibinfo
  {author} {\bibfnamefont {Z.}~\bibnamefont {Zhou}},\ and\ \bibinfo {author}
  {\bibfnamefont {C.}~\bibnamefont {Wu}},\ }\href@noop {} {\bibinfo {title}
  {Mott insulating states of the anisotropic su(4) dirac fermions}} (\bibinfo
  {year} {2019}),\ \Eprint {https://arxiv.org/abs/1912.11791} {arXiv:1912.11791
  [cond-mat.quant-gas]} \BibitemShut {NoStop}%
\bibitem [{\citenamefont {Xu}\ \emph {et~al.}(2018)\citenamefont {Xu},
  \citenamefont {Barreiro}, \citenamefont {Wang},\ and\ \citenamefont
  {Wu}}]{xu2018a}%
  \BibitemOpen
  \bibfield  {author} {\bibinfo {author} {\bibfnamefont {S.}~\bibnamefont
  {Xu}}, \bibinfo {author} {\bibfnamefont {J.~T.}\ \bibnamefont {Barreiro}},
  \bibinfo {author} {\bibfnamefont {Y.}~\bibnamefont {Wang}},\ and\ \bibinfo
  {author} {\bibfnamefont {C.}~\bibnamefont {Wu}},\ }\href
  {https://doi.org/10.1103/PhysRevLett.121.167205} {\bibfield  {journal}
  {\bibinfo  {journal} {Phys. Rev. Lett.}\ }\textbf {\bibinfo {volume} {121}},\
  \bibinfo {pages} {167205} (\bibinfo {year} {2018})}\BibitemShut {NoStop}%
\bibitem [{\citenamefont {Pomeranchuk}(1959)}]{pomeranchuk1959}%
  \BibitemOpen
  \bibfield  {author} {\bibinfo {author} {\bibfnamefont {I.~I.}\ \bibnamefont
  {Pomeranchuk}},\ }\href@noop {} {\bibfield  {journal} {\bibinfo  {journal}
  {Soviet Physics Jetp-Ussr}\ }\textbf {\bibinfo {volume} {8}},\ \bibinfo
  {pages} {361} (\bibinfo {year} {1959})}\BibitemShut {NoStop}%
\bibitem [{\citenamefont {Wu}\ and\ \citenamefont {Zhang}(2004)}]{wu2004a}%
  \BibitemOpen
  \bibfield  {author} {\bibinfo {author} {\bibfnamefont {C.}~\bibnamefont
  {Wu}}\ and\ \bibinfo {author} {\bibfnamefont {S.~C.}\ \bibnamefont {Zhang}},\
  }\href {doi:10.1103/PhysRevLett.93.036403} {\bibfield  {journal} {\bibinfo
  {journal} {Physical Review Letters}\ }\textbf {\bibinfo {volume} {93}},\
  \bibinfo {pages} {36403} (\bibinfo {year} {2004})}\BibitemShut {NoStop}%
\bibitem [{\citenamefont {Wu}\ \emph {et~al.}(2007)\citenamefont {Wu},
  \citenamefont {Sun}, \citenamefont {Fradkin},\ and\ \citenamefont
  {Zhang}}]{wu2007}%
  \BibitemOpen
  \bibfield  {author} {\bibinfo {author} {\bibfnamefont {C.}~\bibnamefont
  {Wu}}, \bibinfo {author} {\bibfnamefont {K.}~\bibnamefont {Sun}}, \bibinfo
  {author} {\bibfnamefont {E.}~\bibnamefont {Fradkin}},\ and\ \bibinfo {author}
  {\bibfnamefont {S.-C.}\ \bibnamefont {Zhang}},\ }\href
  {https://doi.org/10.1103/PhysRevB.75.115103} {\bibfield  {journal} {\bibinfo
  {journal} {Phys. Rev. B}\ }\textbf {\bibinfo {volume} {75}},\ \bibinfo
  {pages} {115103} (\bibinfo {year} {2007})}\BibitemShut {NoStop}%
\bibitem [{\citenamefont {Xiang}\ and\ \citenamefont {Wu}(2022)}]{xiang2022}%
  \BibitemOpen
  \bibfield  {author} {\bibinfo {author} {\bibfnamefont {T.}~\bibnamefont
  {Xiang}}\ and\ \bibinfo {author} {\bibfnamefont {C.}~\bibnamefont {Wu}},\
  }\href@noop {} {\emph {\bibinfo {title} {D-Wave Superconductivity}}}\
  (\bibinfo  {publisher} {Peking University Press and Cambrige University
  Press},\ \bibinfo {year} {2022})\BibitemShut {NoStop}%
\bibitem [{\citenamefont {Leggett}(1975)}]{leggett1975}%
  \BibitemOpen
  \bibfield  {author} {\bibinfo {author} {\bibfnamefont {A.~J.}\ \bibnamefont
  {Leggett}},\ }\href@noop {} {\bibfield  {journal} {\bibinfo  {journal} {Rev.
  Mod. Phys.}\ }\textbf {\bibinfo {volume} {47}},\ \bibinfo {pages} {331}
  (\bibinfo {year} {1975})}\BibitemShut {NoStop}%
\bibitem [{\citenamefont {Sigrist}\ and\ \citenamefont
  {Ueda}(1991)}]{sigrist1991}%
  \BibitemOpen
  \bibfield  {author} {\bibinfo {author} {\bibfnamefont {M.}~\bibnamefont
  {Sigrist}}\ and\ \bibinfo {author} {\bibfnamefont {K.}~\bibnamefont {Ueda}},\
  }\href {https://doi.org/10.1103/RevModPhys.63.239} {\bibfield  {journal}
  {\bibinfo  {journal} {Rev. Mod. Phys.}\ }\textbf {\bibinfo {volume} {63}},\
  \bibinfo {pages} {239} (\bibinfo {year} {1991})}\BibitemShut {NoStop}%
\bibitem [{\citenamefont {Fradkin}\ \emph {et~al.}(2010)\citenamefont
  {Fradkin}, \citenamefont {Kivelson}, \citenamefont {Lawler}, \citenamefont
  {Eisenstein},\ and\ \citenamefont {Mackenzie}}]{fradkin2010}%
  \BibitemOpen
  \bibfield  {author} {\bibinfo {author} {\bibfnamefont {E.}~\bibnamefont
  {Fradkin}}, \bibinfo {author} {\bibfnamefont {S.~A.}\ \bibnamefont
  {Kivelson}}, \bibinfo {author} {\bibfnamefont {M.~J.}\ \bibnamefont
  {Lawler}}, \bibinfo {author} {\bibfnamefont {J.~P.}\ \bibnamefont
  {Eisenstein}},\ and\ \bibinfo {author} {\bibfnamefont {A.~P.}\ \bibnamefont
  {Mackenzie}},\ }\href
  {https://doi.org/10.1146/annurev-conmatphys-070909-103925} {\bibfield
  {journal} {\bibinfo  {journal} {Annual Review of Condensed Matter Physics}\
  }\textbf {\bibinfo {volume} {1}},\ \bibinfo {pages} {153每178} (\bibinfo
  {year} {2010})}\BibitemShut {NoStop}%
\bibitem [{\citenamefont {Fu}(2011)}]{fu2011}%
  \BibitemOpen
  \bibfield  {author} {\bibinfo {author} {\bibfnamefont {L.}~\bibnamefont
  {Fu}},\ }\href {https://doi.org/10.1103/PhysRevLett.106.106802} {\bibfield
  {journal} {\bibinfo  {journal} {Phys. Rev. Lett.}\ }\textbf {\bibinfo
  {volume} {106}},\ \bibinfo {pages} {106802} (\bibinfo {year}
  {2011})}\BibitemShut {NoStop}%
\bibitem [{\citenamefont {Parameswaran}\ \emph {et~al.}(2013)\citenamefont
  {Parameswaran}, \citenamefont {Turner}, \citenamefont {Arovas},\ and\
  \citenamefont {Vishwanath}}]{parameswaran2013}%
  \BibitemOpen
  \bibfield  {author} {\bibinfo {author} {\bibfnamefont {S.~A.}\ \bibnamefont
  {Parameswaran}}, \bibinfo {author} {\bibfnamefont {A.~M.}\ \bibnamefont
  {Turner}}, \bibinfo {author} {\bibfnamefont {D.~P.}\ \bibnamefont {Arovas}},\
  and\ \bibinfo {author} {\bibfnamefont {A.}~\bibnamefont {Vishwanath}},\
  }\href {https://doi.org/10.1038/nphys2600} {\bibfield  {journal} {\bibinfo
  {journal} {Nat. Phys.}\ }\textbf {\bibinfo {volume} {9}},\ \bibinfo {pages}
  {299} (\bibinfo {year} {2013})}\BibitemShut {NoStop}%
\bibitem [{\citenamefont {Young}\ and\ \citenamefont {Kane}(2015)}]{young2015}%
  \BibitemOpen
  \bibfield  {author} {\bibinfo {author} {\bibfnamefont {S.~M.}\ \bibnamefont
  {Young}}\ and\ \bibinfo {author} {\bibfnamefont {C.~L.}\ \bibnamefont
  {Kane}},\ }\href {https://doi.org/10.1103/PhysRevLett.115.126803} {\bibfield
  {journal} {\bibinfo  {journal} {Phys. Rev. Lett.}\ }\textbf {\bibinfo
  {volume} {115}},\ \bibinfo {pages} {126803} (\bibinfo {year}
  {2015})}\BibitemShut {NoStop}%
\bibitem [{\citenamefont {Wang}\ \emph {et~al.}(2016)\citenamefont {Wang},
  \citenamefont {Alexandradinata}, \citenamefont {Cava},\ and\ \citenamefont
  {Bernevig}}]{wang2016a}%
  \BibitemOpen
  \bibfield  {author} {\bibinfo {author} {\bibfnamefont {Z.}~\bibnamefont
  {Wang}}, \bibinfo {author} {\bibfnamefont {A.}~\bibnamefont
  {Alexandradinata}}, \bibinfo {author} {\bibfnamefont {R.~J.}\ \bibnamefont
  {Cava}},\ and\ \bibinfo {author} {\bibfnamefont {B.~A.}\ \bibnamefont
  {Bernevig}},\ }\href {https://doi.org/10.1038/nature17410} {\bibfield
  {journal} {\bibinfo  {journal} {Nature}\ }\textbf {\bibinfo {volume} {532}},\
  \bibinfo {pages} {189} (\bibinfo {year} {2016})}\BibitemShut {NoStop}%
\bibitem [{\citenamefont {Watanabe}\ \emph {et~al.}(2016)\citenamefont
  {Watanabe}, \citenamefont {Po}, \citenamefont {Zaletel},\ and\ \citenamefont
  {Vishwanath}}]{watanabe2016}%
  \BibitemOpen
  \bibfield  {author} {\bibinfo {author} {\bibfnamefont {H.}~\bibnamefont
  {Watanabe}}, \bibinfo {author} {\bibfnamefont {H.~C.}\ \bibnamefont {Po}},
  \bibinfo {author} {\bibfnamefont {M.~P.}\ \bibnamefont {Zaletel}},\ and\
  \bibinfo {author} {\bibfnamefont {A.}~\bibnamefont {Vishwanath}},\ }\href
  {https://doi.org/10.1103/PhysRevLett.117.096404} {\bibfield  {journal}
  {\bibinfo  {journal} {Phys. Rev. Lett.}\ }\textbf {\bibinfo {volume} {117}},\
  \bibinfo {pages} {096404} (\bibinfo {year} {2016})}\BibitemShut {NoStop}%
\bibitem [{\citenamefont {Kadic}\ \emph {et~al.}(2019)\citenamefont {Kadic},
  \citenamefont {van Hecke},\ and\ \citenamefont {Wegener}}]{milton2019}%
  \BibitemOpen
  \bibfield  {author} {\bibinfo {author} {\bibfnamefont {G.}~\bibnamefont
  {Kadic}, \bibfnamefont {M.and~Milton}}, \bibinfo {author} {\bibfnamefont
  {M.}~\bibnamefont {van Hecke}},\ and\ \bibinfo {author} {\bibfnamefont
  {M.}~\bibnamefont {Wegener}},\ }\href
  {https://doi.org/doi.org/10.1038/s42254-018-0018-y} {\bibfield  {journal}
  {\bibinfo  {journal} {Nature Reviews Physics}\ }\textbf {\bibinfo {volume}
  {1}},\ \bibinfo {pages} {198} (\bibinfo {year} {2019})}\BibitemShut {NoStop}%
\bibitem [{\citenamefont {Rudner}\ and\ \citenamefont
  {Linder}(2020)}]{rudner2020}%
  \BibitemOpen
  \bibfield  {author} {\bibinfo {author} {\bibfnamefont {M.~S.}\ \bibnamefont
  {Rudner}}\ and\ \bibinfo {author} {\bibfnamefont {N.~H.}\ \bibnamefont
  {Linder}},\ }\href {https://doi.org/doi.org/10.1038/s42254-020-0170-z}
  {\bibfield  {journal} {\bibinfo  {journal} {Nature Reviews Physics}\ }\textbf
  {\bibinfo {volume} {2}},\ \bibinfo {pages} {229} (\bibinfo {year}
  {2020})}\BibitemShut {NoStop}%
\bibitem [{\citenamefont {Harper}\ \emph {et~al.}(2020)\citenamefont {Harper},
  \citenamefont {Roy}, \citenamefont {Rudner},\ and\ \citenamefont
  {Sondhi}}]{harper2020}%
  \BibitemOpen
  \bibfield  {author} {\bibinfo {author} {\bibfnamefont {F.}~\bibnamefont
  {Harper}}, \bibinfo {author} {\bibfnamefont {R.}~\bibnamefont {Roy}},
  \bibinfo {author} {\bibfnamefont {M.~S.}\ \bibnamefont {Rudner}},\ and\
  \bibinfo {author} {\bibfnamefont {S.}~\bibnamefont {Sondhi}},\ }\href
  {https://doi.org/10.1146/annurev-conmatphys-031218-013721} {\bibfield
  {journal} {\bibinfo  {journal} {Annual Review of Condensed Matter Physics}\
  }\textbf {\bibinfo {volume} {11}},\ \bibinfo {pages} {345} (\bibinfo {year}
  {2020})}\BibitemShut {NoStop}%
\bibitem [{\citenamefont {Guo}\ and\ \citenamefont {Liang}(2021)}]{guo2021}%
  \BibitemOpen
  \bibfield  {author} {\bibinfo {author} {\bibfnamefont {L.}~\bibnamefont
  {Guo}}\ and\ \bibinfo {author} {\bibfnamefont {P.}~\bibnamefont {Liang}},\
  }\href@noop {} {\bibfield  {journal} {\bibinfo  {journal} {New. J. Phys.}\
  }\textbf {\bibinfo {volume} {22}},\ \bibinfo {pages} {075003} (\bibinfo
  {year} {2021})}\BibitemShut {NoStop}%
\bibitem [{\citenamefont {Yu}\ \emph {et~al.}(2021)\citenamefont {Yu},
  \citenamefont {Zhang},\ and\ \citenamefont {Song}}]{yu2021}%
  \BibitemOpen
  \bibfield  {author} {\bibinfo {author} {\bibfnamefont {J.}~\bibnamefont
  {Yu}}, \bibinfo {author} {\bibfnamefont {R.~X.}\ \bibnamefont {Zhang}},\ and\
  \bibinfo {author} {\bibfnamefont {Z.~D.}\ \bibnamefont {Song}},\ }\href@noop
  {} {\bibfield  {journal} {\bibinfo  {journal} {Nat. Comm.}\ }\textbf
  {\bibinfo {volume} {12}},\ \bibinfo {pages} {5985} (\bibinfo {year}
  {2021})}\BibitemShut {NoStop}%
\bibitem [{\citenamefont {Giergiel1}\ \emph {et~al.}(2019)\citenamefont
  {Giergiel1}, \citenamefont {Dauphin}, \citenamefont {Zakrzewski},\ and\
  \citenamefont {Sacha}}]{giergiel2019}%
  \BibitemOpen
  \bibfield  {author} {\bibinfo {author} {\bibfnamefont {K.}~\bibnamefont
  {Giergiel1}}, \bibinfo {author} {\bibfnamefont {M.}~\bibnamefont {Dauphin},
  \bibfnamefont {A.and~Lewenstein}}, \bibinfo {author} {\bibfnamefont
  {J.}~\bibnamefont {Zakrzewski}},\ and\ \bibinfo {author} {\bibfnamefont
  {K.}~\bibnamefont {Sacha}},\ }\href@noop {} {\bibfield  {journal} {\bibinfo
  {journal} {New. J. Phys.}\ }\textbf {\bibinfo {volume} {21}},\ \bibinfo
  {pages} {052003} (\bibinfo {year} {2019})}\BibitemShut {NoStop}%
\bibitem [{\citenamefont {Peng}\ and\ \citenamefont {Refael}(2019)}]{peng2019}%
  \BibitemOpen
  \bibfield  {author} {\bibinfo {author} {\bibfnamefont {Y.}~\bibnamefont
  {Peng}}\ and\ \bibinfo {author} {\bibfnamefont {G.}~\bibnamefont {Refael}},\
  }\href {https://doi.org/10.1103/PhysRevLett.123.016806} {\bibfield  {journal}
  {\bibinfo  {journal} {Phys. Rev. Lett.}\ }\textbf {\bibinfo {volume} {123}},\
  \bibinfo {pages} {016806} (\bibinfo {year} {2019})}\BibitemShut {NoStop}%
\bibitem [{\citenamefont {Kleiner}\ \emph {et~al.}(2021)\citenamefont
  {Kleiner}, \citenamefont {Zhou}, \citenamefont {Dorsch}, \citenamefont
  {Zhang}, \citenamefont {Koelle},\ and\ \citenamefont {Jin}}]{kleiner2021}%
  \BibitemOpen
  \bibfield  {author} {\bibinfo {author} {\bibfnamefont {R.}~\bibnamefont
  {Kleiner}}, \bibinfo {author} {\bibfnamefont {X.}~\bibnamefont {Zhou}},
  \bibinfo {author} {\bibfnamefont {E.}~\bibnamefont {Dorsch}}, \bibinfo
  {author} {\bibfnamefont {X.}~\bibnamefont {Zhang}}, \bibinfo {author}
  {\bibfnamefont {D.}~\bibnamefont {Koelle}},\ and\ \bibinfo {author}
  {\bibfnamefont {D.}~\bibnamefont {Jin}},\ }\bibfield  {journal} {\bibinfo
  {journal} {{NATURE COMMUNICATIONS}}\ }\textbf {\bibinfo {volume} {{12}}},\
  \href {https://doi.org/{10.1038/s41467-021-26132-y}}
  {{10.1038/s41467-021-26132-y}} (\bibinfo {year} {{2021}})\BibitemShut
  {NoStop}%
\bibitem [{\citenamefont {Gao}\ and\ \citenamefont {Niu}(2021)}]{gao2021}%
  \BibitemOpen
  \bibfield  {author} {\bibinfo {author} {\bibfnamefont {Q.}~\bibnamefont
  {Gao}}\ and\ \bibinfo {author} {\bibfnamefont {Q.}~\bibnamefont {Niu}},\
  }\href {https://doi.org/10.1103/PhysRevLett.127.036401} {\bibfield  {journal}
  {\bibinfo  {journal} {Phys. Rev. Lett.}\ }\textbf {\bibinfo {volume} {127}},\
  \bibinfo {pages} {036401} (\bibinfo {year} {2021})}\BibitemShut {NoStop}%
\bibitem [{\citenamefont {Muenchinger}\ \emph {et~al.}(2022)\citenamefont
  {Muenchinger}, \citenamefont {Hahn}, \citenamefont {Beutel}, \citenamefont
  {Woska}, \citenamefont {Monti}, \citenamefont {Rockstuhl}, \citenamefont
  {Blasco},\ and\ \citenamefont {Wegener}}]{muenchinger2022}%
  \BibitemOpen
  \bibfield  {author} {\bibinfo {author} {\bibfnamefont {A.}~\bibnamefont
  {Muenchinger}}, \bibinfo {author} {\bibfnamefont {V.}~\bibnamefont {Hahn}},
  \bibinfo {author} {\bibfnamefont {D.}~\bibnamefont {Beutel}}, \bibinfo
  {author} {\bibfnamefont {S.}~\bibnamefont {Woska}}, \bibinfo {author}
  {\bibfnamefont {J.}~\bibnamefont {Monti}}, \bibinfo {author} {\bibfnamefont
  {C.}~\bibnamefont {Rockstuhl}}, \bibinfo {author} {\bibfnamefont
  {E.}~\bibnamefont {Blasco}},\ and\ \bibinfo {author} {\bibfnamefont
  {M.}~\bibnamefont {Wegener}},\ }\bibfield  {journal} {\bibinfo  {journal}
  {{ADVANCED MATERIALS TECHNOLOGIES}}\ }\textbf {\bibinfo {volume} {{7}}},\
  \href {https://doi.org/{10.1002/admt.202100944}} {{10.1002/admt.202100944}}
  (\bibinfo {year} {{2022}})\BibitemShut {NoStop}%
\bibitem [{\citenamefont {Zhang}\ and\ \citenamefont
  {Niu}(2015)}]{zhangniu2015}%
  \BibitemOpen
  \bibfield  {author} {\bibinfo {author} {\bibfnamefont {L.}~\bibnamefont
  {Zhang}}\ and\ \bibinfo {author} {\bibfnamefont {Q.}~\bibnamefont {Niu}},\
  }\href {https://doi.org/10.1103/PhysRevLett.115.115502} {\bibfield  {journal}
  {\bibinfo  {journal} {Phys. Rev. Lett.}\ }\textbf {\bibinfo {volume} {115}},\
  \bibinfo {pages} {115502} (\bibinfo {year} {2015})}\BibitemShut {NoStop}%
\bibitem [{\citenamefont {Xiao}\ \emph {et~al.}(2010)\citenamefont {Xiao},
  \citenamefont {Chang},\ and\ \citenamefont {Niu}}]{xiao2010}%
  \BibitemOpen
  \bibfield  {author} {\bibinfo {author} {\bibfnamefont {D.}~\bibnamefont
  {Xiao}}, \bibinfo {author} {\bibfnamefont {M.-C.}\ \bibnamefont {Chang}},\
  and\ \bibinfo {author} {\bibfnamefont {Q.}~\bibnamefont {Niu}},\ }\href
  {https://doi.org/10.1103/RevModPhys.82.1959} {\bibfield  {journal} {\bibinfo
  {journal} {Rev. Mod. Phys.}\ }\textbf {\bibinfo {volume} {82}},\ \bibinfo
  {pages} {1959} (\bibinfo {year} {2010})}\BibitemShut {NoStop}%
\bibitem [{\citenamefont {Chiu}\ \emph {et~al.}(2016)\citenamefont {Chiu},
  \citenamefont {Teo}, \citenamefont {Schnyder},\ and\ \citenamefont
  {Ryu}}]{chiu2016}%
  \BibitemOpen
  \bibfield  {author} {\bibinfo {author} {\bibfnamefont {C.~K.}\ \bibnamefont
  {Chiu}}, \bibinfo {author} {\bibfnamefont {J.~C.~Y.}\ \bibnamefont {Teo}},
  \bibinfo {author} {\bibfnamefont {A.~P.}\ \bibnamefont {Schnyder}},\ and\
  \bibinfo {author} {\bibfnamefont {S.}~\bibnamefont {Ryu}},\ }\href
  {https://doi.org/10.1103/RevModPhys.88.035005} {\bibfield  {journal}
  {\bibinfo  {journal} {Rev. Mod. Phys.}\ }\textbf {\bibinfo {volume} {88}},\
  \bibinfo {pages} {1} (\bibinfo {year} {2016})},\ \Eprint
  {https://arxiv.org/abs/1505.03535} {arXiv:1505.03535} \BibitemShut {NoStop}%
\bibitem [{\citenamefont {Kyprianidis}\ \emph {et~al.}(2021)\citenamefont
  {Kyprianidis}, \citenamefont {Machado}, \citenamefont {Morong}, \citenamefont
  {Becker}, \citenamefont {Collins}, \citenamefont {Else}, \citenamefont
  {Feng}, \citenamefont {Hess}, \citenamefont {Nayak}, \citenamefont {Pagano},
  \citenamefont {Yao},\ and\ \citenamefont {Monroe}}]{Kyp2021}%
  \BibitemOpen
  \bibfield  {author} {\bibinfo {author} {\bibfnamefont {A.}~\bibnamefont
  {Kyprianidis}}, \bibinfo {author} {\bibfnamefont {F.}~\bibnamefont
  {Machado}}, \bibinfo {author} {\bibfnamefont {W.}~\bibnamefont {Morong}},
  \bibinfo {author} {\bibfnamefont {P.}~\bibnamefont {Becker}}, \bibinfo
  {author} {\bibfnamefont {K.~S.}\ \bibnamefont {Collins}}, \bibinfo {author}
  {\bibfnamefont {D.~V.}\ \bibnamefont {Else}}, \bibinfo {author}
  {\bibfnamefont {L.}~\bibnamefont {Feng}}, \bibinfo {author} {\bibfnamefont
  {P.~W.}\ \bibnamefont {Hess}}, \bibinfo {author} {\bibfnamefont
  {C.}~\bibnamefont {Nayak}}, \bibinfo {author} {\bibfnamefont
  {G.}~\bibnamefont {Pagano}}, \bibinfo {author} {\bibfnamefont {N.~Y.}\
  \bibnamefont {Yao}},\ and\ \bibinfo {author} {\bibfnamefont {C.}~\bibnamefont
  {Monroe}},\ }\href@noop {} {\bibfield  {journal} {\bibinfo  {journal}
  {Science}\ }\textbf {\bibinfo {volume} {372}},\ \bibinfo {pages} {1192每1196}
  (\bibinfo {year} {2021})}\BibitemShut {NoStop}%
\bibitem [{\citenamefont {Shapere}\ and\ \citenamefont
  {Wilczek}(2012)}]{shapere2012}%
  \BibitemOpen
  \bibfield  {author} {\bibinfo {author} {\bibfnamefont {A.}~\bibnamefont
  {Shapere}}\ and\ \bibinfo {author} {\bibfnamefont {F.}~\bibnamefont
  {Wilczek}},\ }\href {https://doi.org/10.1103/PhysRevLett.109.160402}
  {\bibfield  {journal} {\bibinfo  {journal} {Phys. Rev. Lett.}\ }\textbf
  {\bibinfo {volume} {109}},\ \bibinfo {pages} {160402} (\bibinfo {year}
  {2012})}\BibitemShut {NoStop}%
\bibitem [{\citenamefont {Wilczek}(2012)}]{wilczek2012}%
  \BibitemOpen
  \bibfield  {author} {\bibinfo {author} {\bibfnamefont {F.}~\bibnamefont
  {Wilczek}},\ }\href {https://doi.org/10.1103/PhysRevLett.109.160401}
  {\bibfield  {journal} {\bibinfo  {journal} {Phys. Rev. Lett.}\ }\textbf
  {\bibinfo {volume} {109}},\ \bibinfo {pages} {160401} (\bibinfo {year}
  {2012})}\BibitemShut {NoStop}%
\bibitem [{\citenamefont {Bruno}(2013)}]{bruno2013}%
  \BibitemOpen
  \bibfield  {author} {\bibinfo {author} {\bibfnamefont {P.}~\bibnamefont
  {Bruno}},\ }\href {https://doi.org/10.1103/PhysRevLett.111.070402} {\bibfield
   {journal} {\bibinfo  {journal} {Phys. Rev. Lett.}\ }\textbf {\bibinfo
  {volume} {111}},\ \bibinfo {pages} {070402} (\bibinfo {year}
  {2013})}\BibitemShut {NoStop}%
\bibitem [{\citenamefont {Watanabe}\ and\ \citenamefont
  {Oshikawa}(2015)}]{watanabe2015}%
  \BibitemOpen
  \bibfield  {author} {\bibinfo {author} {\bibfnamefont {H.}~\bibnamefont
  {Watanabe}}\ and\ \bibinfo {author} {\bibfnamefont {M.}~\bibnamefont
  {Oshikawa}},\ }\href {https://doi.org/10.1103/PhysRevLett.114.251603}
  {\bibfield  {journal} {\bibinfo  {journal} {Phys. Rev. Lett.}\ }\textbf
  {\bibinfo {volume} {114}},\ \bibinfo {pages} {251603} (\bibinfo {year}
  {2015})}\BibitemShut {NoStop}%
\bibitem [{\citenamefont {Sacha}(2015)}]{sacha2015}%
  \BibitemOpen
  \bibfield  {author} {\bibinfo {author} {\bibfnamefont {K.}~\bibnamefont
  {Sacha}},\ }\href {https://doi.org/10.1103/PhysRevA.91.033617} {\bibfield
  {journal} {\bibinfo  {journal} {Phys. Rev. A}\ }\textbf {\bibinfo {volume}
  {91}},\ \bibinfo {pages} {033617} (\bibinfo {year} {2015})}\BibitemShut
  {NoStop}%
\bibitem [{\citenamefont {Yao}\ \emph {et~al.}(2017)\citenamefont {Yao},
  \citenamefont {Potter}, \citenamefont {Potirniche},\ and\ \citenamefont
  {Vishwanath}}]{yao2017}%
  \BibitemOpen
  \bibfield  {author} {\bibinfo {author} {\bibfnamefont {N.~Y.}\ \bibnamefont
  {Yao}}, \bibinfo {author} {\bibfnamefont {A.~C.}\ \bibnamefont {Potter}},
  \bibinfo {author} {\bibfnamefont {I.-D.}\ \bibnamefont {Potirniche}},\ and\
  \bibinfo {author} {\bibfnamefont {A.}~\bibnamefont {Vishwanath}},\ }\href
  {https://doi.org/10.1103/PhysRevLett.118.030401} {\bibfield  {journal}
  {\bibinfo  {journal} {Phys. Rev. Lett.}\ }\textbf {\bibinfo {volume} {118}},\
  \bibinfo {pages} {030401} (\bibinfo {year} {2017})}\BibitemShut {NoStop}%
\bibitem [{\citenamefont {Khemani}\ \emph {et~al.}(2016)\citenamefont
  {Khemani}, \citenamefont {Lazarides}, \citenamefont {Moessner},\ and\
  \citenamefont {Sondhi}}]{khemani2016}%
  \BibitemOpen
  \bibfield  {author} {\bibinfo {author} {\bibfnamefont {V.}~\bibnamefont
  {Khemani}}, \bibinfo {author} {\bibfnamefont {A.}~\bibnamefont {Lazarides}},
  \bibinfo {author} {\bibfnamefont {R.}~\bibnamefont {Moessner}},\ and\
  \bibinfo {author} {\bibfnamefont {S.~L.}\ \bibnamefont {Sondhi}},\ }\href
  {https://doi.org/10.1103/PhysRevLett.116.250401} {\bibfield  {journal}
  {\bibinfo  {journal} {Phys. Rev. Lett.}\ }\textbf {\bibinfo {volume} {116}},\
  \bibinfo {pages} {250401} (\bibinfo {year} {2016})}\BibitemShut {NoStop}%
\bibitem [{\citenamefont {AI}\ and\ \citenamefont
  {collaborations}(2021)}]{google2021}%
  \BibitemOpen
  \bibfield  {author} {\bibinfo {author} {\bibfnamefont {G.~Q.}\ \bibnamefont
  {AI}}\ and\ \bibinfo {author} {\bibnamefont {collaborations}},\ }\href@noop
  {} {\bibinfo {title} {Observation of time-crystalline eigenstate order on a
  quantum processor}} (\bibinfo {year} {2021}),\ \Eprint
  {https://arxiv.org/abs/2107.13571} {arXiv:2107.13571 [quant-ph]} \BibitemShut
  {NoStop}%
\bibitem [{\citenamefont {Randall}\ \emph {et~al.}(2021)\citenamefont
  {Randall}, \citenamefont {Bradley}, \citenamefont {van~der Gronden},
  \citenamefont {Galicia}, \citenamefont {Abobeih}, \citenamefont {Markham},
  \citenamefont {Twitchen}, \citenamefont {Machado}, \citenamefont {Yao},\ and\
  \citenamefont {Taminiau}}]{randall2021}%
  \BibitemOpen
  \bibfield  {author} {\bibinfo {author} {\bibfnamefont {J.}~\bibnamefont
  {Randall}}, \bibinfo {author} {\bibfnamefont {C.~E.}\ \bibnamefont
  {Bradley}}, \bibinfo {author} {\bibfnamefont {F.~V.}\ \bibnamefont {van~der
  Gronden}}, \bibinfo {author} {\bibfnamefont {A.}~\bibnamefont {Galicia}},
  \bibinfo {author} {\bibfnamefont {M.~H.}\ \bibnamefont {Abobeih}}, \bibinfo
  {author} {\bibfnamefont {M.}~\bibnamefont {Markham}}, \bibinfo {author}
  {\bibfnamefont {D.~J.}\ \bibnamefont {Twitchen}}, \bibinfo {author}
  {\bibfnamefont {F.}~\bibnamefont {Machado}}, \bibinfo {author} {\bibfnamefont
  {N.~Y.}\ \bibnamefont {Yao}},\ and\ \bibinfo {author} {\bibfnamefont {T.~H.}\
  \bibnamefont {Taminiau}},\ }\href {https://doi.org/10.1126/science.abk0603}
  {\bibfield  {journal} {\bibinfo  {journal} {Science}\ }\textbf {\bibinfo
  {volume} {374}},\ \bibinfo {pages} {1474每1478} (\bibinfo {year}
  {2021})}\BibitemShut {NoStop}%
\bibitem [{\citenamefont {Ho}\ and\ \citenamefont {Yip}(1999)}]{ho1999}%
  \BibitemOpen
  \bibfield  {author} {\bibinfo {author} {\bibfnamefont {T.-L.}\ \bibnamefont
  {Ho}}\ and\ \bibinfo {author} {\bibfnamefont {S.}~\bibnamefont {Yip}},\
  }\href {https://doi.org/10.1103/PhysRevLett.82.247} {\bibfield  {journal}
  {\bibinfo  {journal} {Phys. Rev. Lett.}\ }\textbf {\bibinfo {volume} {82}},\
  \bibinfo {pages} {247} (\bibinfo {year} {1999})}\BibitemShut {NoStop}%
\bibitem [{\citenamefont {Yip}\ and\ \citenamefont {Ho}(1999)}]{yip1999}%
  \BibitemOpen
  \bibfield  {author} {\bibinfo {author} {\bibfnamefont {S.-K.}\ \bibnamefont
  {Yip}}\ and\ \bibinfo {author} {\bibfnamefont {T.-L.}\ \bibnamefont {Ho}},\
  }\href {https://doi.org/10.1103/PhysRevA.59.4653} {\bibfield  {journal}
  {\bibinfo  {journal} {Phys. Rev. A}\ }\textbf {\bibinfo {volume} {59}},\
  \bibinfo {pages} {4653} (\bibinfo {year} {1999})}\BibitemShut {NoStop}%
\bibitem [{\citenamefont {Wu}\ \emph {et~al.}(2006)\citenamefont {Wu},
  \citenamefont {Bernevig},\ and\ \citenamefont {Zhang}}]{wu2006a}%
  \BibitemOpen
  \bibfield  {author} {\bibinfo {author} {\bibfnamefont {C.}~\bibnamefont
  {Wu}}, \bibinfo {author} {\bibfnamefont {B.~A.}\ \bibnamefont {Bernevig}},\
  and\ \bibinfo {author} {\bibfnamefont {S.-C.}\ \bibnamefont {Zhang}},\ }\href
  {https://doi.org/10.1103/PhysRevLett.96.106401} {\bibfield  {journal}
  {\bibinfo  {journal} {Phys. Rev. Lett.}\ }\textbf {\bibinfo {volume} {96}},\
  \bibinfo {pages} {106401} (\bibinfo {year} {2006})}\BibitemShut {NoStop}%
\bibitem [{\citenamefont {Lin}\ \emph {et~al.}(1998)\citenamefont {Lin},
  \citenamefont {Balents},\ and\ \citenamefont {Fisher}}]{lin1998}%
  \BibitemOpen
  \bibfield  {author} {\bibinfo {author} {\bibfnamefont {H.-H.}\ \bibnamefont
  {Lin}}, \bibinfo {author} {\bibfnamefont {L.}~\bibnamefont {Balents}},\ and\
  \bibinfo {author} {\bibfnamefont {M.~P.~A.}\ \bibnamefont {Fisher}},\ }\href
  {https://doi.org/10.1103/PhysRevB.58.1794} {\bibfield  {journal} {\bibinfo
  {journal} {Phys. Rev. B}\ }\textbf {\bibinfo {volume} {58}},\ \bibinfo
  {pages} {1794} (\bibinfo {year} {1998})}\BibitemShut {NoStop}%
\bibitem [{\citenamefont {Berg}\ \emph {et~al.}(2009)\citenamefont {Berg},
  \citenamefont {Fradkin},\ and\ \citenamefont {Kivelson}}]{berg2009}%
  \BibitemOpen
  \bibfield  {author} {\bibinfo {author} {\bibfnamefont {E.}~\bibnamefont
  {Berg}}, \bibinfo {author} {\bibfnamefont {E.}~\bibnamefont {Fradkin}},\ and\
  \bibinfo {author} {\bibfnamefont {S.~A.}\ \bibnamefont {Kivelson}},\ }\href
  {https://doi.org/10.1038/nphys1389} {\bibfield  {journal} {\bibinfo
  {journal} {Nature Physics}\ }\textbf {\bibinfo {volume} {5}},\ \bibinfo
  {pages} {830每833} (\bibinfo {year} {2009})}\BibitemShut {NoStop}%
\bibitem [{\citenamefont {Agterberg}\ \emph {et~al.}(2020)\citenamefont
  {Agterberg}, \citenamefont {Davis}, \citenamefont {Edkins}, \citenamefont
  {Fradkin}, \citenamefont {Van~Harlingen}, \citenamefont {Kivelson},
  \citenamefont {Lee}, \citenamefont {Radzihovsky}, \citenamefont {Tranquada},\
  and\ \citenamefont {Wang}}]{agterberg2020}%
  \BibitemOpen
  \bibfield  {author} {\bibinfo {author} {\bibfnamefont {D.~F.}\ \bibnamefont
  {Agterberg}}, \bibinfo {author} {\bibfnamefont {J.~S.}\ \bibnamefont
  {Davis}}, \bibinfo {author} {\bibfnamefont {S.~D.}\ \bibnamefont {Edkins}},
  \bibinfo {author} {\bibfnamefont {E.}~\bibnamefont {Fradkin}}, \bibinfo
  {author} {\bibfnamefont {D.~J.}\ \bibnamefont {Van~Harlingen}}, \bibinfo
  {author} {\bibfnamefont {S.~A.}\ \bibnamefont {Kivelson}}, \bibinfo {author}
  {\bibfnamefont {P.~A.}\ \bibnamefont {Lee}}, \bibinfo {author} {\bibfnamefont
  {L.}~\bibnamefont {Radzihovsky}}, \bibinfo {author} {\bibfnamefont {J.~M.}\
  \bibnamefont {Tranquada}},\ and\ \bibinfo {author} {\bibfnamefont
  {Y.}~\bibnamefont {Wang}},\ }\href
  {https://doi.org/10.1146/annurev-conmatphys-031119-050711} {\bibfield
  {journal} {\bibinfo  {journal} {Annual Review of Condensed Matter Physics}\
  }\textbf {\bibinfo {volume} {11}},\ \bibinfo {pages} {231每270} (\bibinfo
  {year} {2020})}\BibitemShut {NoStop}%
\bibitem [{\citenamefont {Jian}\ \emph {et~al.}(2021)\citenamefont {Jian},
  \citenamefont {Huang},\ and\ \citenamefont {Yao}}]{jian2021}%
  \BibitemOpen
  \bibfield  {author} {\bibinfo {author} {\bibfnamefont {S.-K.}\ \bibnamefont
  {Jian}}, \bibinfo {author} {\bibfnamefont {Y.}~\bibnamefont {Huang}},\ and\
  \bibinfo {author} {\bibfnamefont {H.}~\bibnamefont {Yao}},\ }\href
  {https://doi.org/10.1103/PhysRevLett.127.227001} {\bibfield  {journal}
  {\bibinfo  {journal} {Phys. Rev. Lett.}\ }\textbf {\bibinfo {volume} {127}},\
  \bibinfo {pages} {227001} (\bibinfo {year} {2021})}\BibitemShut {NoStop}%
\bibitem [{\citenamefont {Fernandes}\ and\ \citenamefont
  {Fu}(2021)}]{fernandes2021}%
  \BibitemOpen
  \bibfield  {author} {\bibinfo {author} {\bibfnamefont {R.~M.}\ \bibnamefont
  {Fernandes}}\ and\ \bibinfo {author} {\bibfnamefont {L.}~\bibnamefont {Fu}},\
  }\href {https://doi.org/10.1103/PhysRevLett.127.047001} {\bibfield  {journal}
  {\bibinfo  {journal} {Phys. Rev. Lett.}\ }\textbf {\bibinfo {volume} {127}},\
  \bibinfo {pages} {047001} (\bibinfo {year} {2021})}\BibitemShut {NoStop}%
\bibitem [{\citenamefont {Zeng}\ \emph {et~al.}(6158)\citenamefont {Zeng},
  \citenamefont {Hu}, \citenamefont {Hu}, \citenamefont {You},\ and\
  \citenamefont {Wu}}]{zeng2021}%
  \BibitemOpen
  \bibfield  {author} {\bibinfo {author} {\bibfnamefont {M.}~\bibnamefont
  {Zeng}}, \bibinfo {author} {\bibfnamefont {L.-H.}\ \bibnamefont {Hu}},
  \bibinfo {author} {\bibfnamefont {H.-Y.}\ \bibnamefont {Hu}}, \bibinfo
  {author} {\bibfnamefont {Y.-Z.}\ \bibnamefont {You}},\ and\ \bibinfo {author}
  {\bibfnamefont {C.}~\bibnamefont {Wu}},\ }\href@noop {} {\bibinfo {title}
  {Phase-fluctuation induced time-reversal symmetry breaking normal state}}
  (\bibinfo {year} {arXiv:2102.06158}),\ \Eprint
  {https://arxiv.org/abs/2102.06158} {arXiv:2102.06158 [cond-mat.supr-con]}
  \BibitemShut {NoStop}%
\bibitem [{\citenamefont {Ge}\ \emph {et~al.}(0352)\citenamefont {Ge},
  \citenamefont {Wang}, \citenamefont {Xing}, \citenamefont {Yin},
  \citenamefont {Lei}, \citenamefont {Wang},\ and\ \citenamefont
  {Wang}}]{gewang2022}%
  \BibitemOpen
  \bibfield  {author} {\bibinfo {author} {\bibfnamefont {J.}~\bibnamefont
  {Ge}}, \bibinfo {author} {\bibfnamefont {P.}~\bibnamefont {Wang}}, \bibinfo
  {author} {\bibfnamefont {Y.}~\bibnamefont {Xing}}, \bibinfo {author}
  {\bibfnamefont {Q.}~\bibnamefont {Yin}}, \bibinfo {author} {\bibfnamefont
  {H.}~\bibnamefont {Lei}}, \bibinfo {author} {\bibfnamefont {Z.}~\bibnamefont
  {Wang}},\ and\ \bibinfo {author} {\bibfnamefont {J.}~\bibnamefont {Wang}},\
  }\href@noop {} {\bibinfo {title} {Discovery of charge-4e and charge-6e
  superconductivity in kagome superconductor csv3sb5}} (\bibinfo {year}
  {arXiv:2201.10352}),\ \Eprint {https://arxiv.org/abs/2201.10352}
  {arXiv:2201.10352 [cond-mat.supr-con]} \BibitemShut {NoStop}%
\bibitem [{Note1()}]{Note1}%
  \BibitemOpen
  \bibinfo {note} {In terms of SO(6), which equals SU(4)$/$Z$_2$, they form the
  6-vector representation.}\BibitemShut {Stop}%
\bibitem [{\citenamefont {van~den Bossche}\ \emph {et~al.}(2000)\citenamefont
  {van~den Bossche}, \citenamefont {Zhang},\ and\ \citenamefont
  {Mila}}]{bossche2000}%
  \BibitemOpen
  \bibfield  {author} {\bibinfo {author} {\bibfnamefont {M.}~\bibnamefont
  {van~den Bossche}}, \bibinfo {author} {\bibfnamefont {F.-C.}\ \bibnamefont
  {Zhang}},\ and\ \bibinfo {author} {\bibfnamefont {F.}~\bibnamefont {Mila}},\
  }\href {https://doi.org/10.1007/PL00011085} {\bibfield  {journal} {\bibinfo
  {journal} {The European Physical Journal B - Condensed Matter and Complex
  Systems}\ }\textbf {\bibinfo {volume} {17}},\ \bibinfo {pages} {367}
  (\bibinfo {year} {2000})}\BibitemShut {NoStop}%
\bibitem [{\citenamefont {Pankov}\ \emph {et~al.}(2007)\citenamefont {Pankov},
  \citenamefont {Moessner},\ and\ \citenamefont {Sondhi}}]{pankov2007}%
  \BibitemOpen
  \bibfield  {author} {\bibinfo {author} {\bibfnamefont {S.}~\bibnamefont
  {Pankov}}, \bibinfo {author} {\bibfnamefont {R.}~\bibnamefont {Moessner}},\
  and\ \bibinfo {author} {\bibfnamefont {S.~L.}\ \bibnamefont {Sondhi}},\
  }\href {https://doi.org/10.1103/PhysRevB.76.104436} {\bibfield  {journal}
  {\bibinfo  {journal} {Phys. Rev. B}\ }\textbf {\bibinfo {volume} {76}},\
  \bibinfo {pages} {104436} (\bibinfo {year} {2007})}\BibitemShut {NoStop}%
\bibitem [{\citenamefont {Rokhsar}\ and\ \citenamefont
  {Kivelson}(1988)}]{rokhsar1988}%
  \BibitemOpen
  \bibfield  {author} {\bibinfo {author} {\bibfnamefont {D.~S.}\ \bibnamefont
  {Rokhsar}}\ and\ \bibinfo {author} {\bibfnamefont {S.~A.}\ \bibnamefont
  {Kivelson}},\ }\href {https://doi.org/10.1103/PhysRevLett.61.2376} {\bibfield
   {journal} {\bibinfo  {journal} {Phys. Rev. Lett.}\ }\textbf {\bibinfo
  {volume} {61}},\ \bibinfo {pages} {2376} (\bibinfo {year}
  {1988})}\BibitemShut {NoStop}%
\bibitem [{\citenamefont {Nandkishore}\ and\ \citenamefont
  {Hermele}(2019)}]{nandkishore2019}%
  \BibitemOpen
  \bibfield  {author} {\bibinfo {author} {\bibfnamefont {R.~M.}\ \bibnamefont
  {Nandkishore}}\ and\ \bibinfo {author} {\bibfnamefont {M.}~\bibnamefont
  {Hermele}},\ }\href
  {https://doi.org/10.1146/annurev-conmatphys-031218-013604} {\bibfield
  {journal} {\bibinfo  {journal} {Annual Review of Condensed Matter Physics}\
  }\textbf {\bibinfo {volume} {10}},\ \bibinfo {pages} {295} (\bibinfo {year}
  {2019})},\ \Eprint
  {https://arxiv.org/abs/https://doi.org/10.1146/annurev-conmatphys-031218-013604}
  {https://doi.org/10.1146/annurev-conmatphys-031218-013604} \BibitemShut
  {NoStop}%
\bibitem [{\citenamefont {Hirsch}(1983)}]{hirsch1983}%
  \BibitemOpen
  \bibfield  {author} {\bibinfo {author} {\bibfnamefont {J.~E.}\ \bibnamefont
  {Hirsch}},\ }\href {https://doi.org/10.1103/PhysRevB.28.4059} {\bibfield
  {journal} {\bibinfo  {journal} {Phys. Rev. B}\ }\textbf {\bibinfo {volume}
  {28}},\ \bibinfo {pages} {4059} (\bibinfo {year} {1983})}\BibitemShut
  {NoStop}%
\bibitem [{\citenamefont {Hirsch}(1985)}]{hirsch1985}%
  \BibitemOpen
  \bibfield  {author} {\bibinfo {author} {\bibfnamefont {J.~E.}\ \bibnamefont
  {Hirsch}},\ }\href {https://doi.org/10.1103/PhysRevB.31.4403} {\bibfield
  {journal} {\bibinfo  {journal} {Phys. Rev. B}\ }\textbf {\bibinfo {volume}
  {31}},\ \bibinfo {pages} {4403} (\bibinfo {year} {1985})}\BibitemShut
  {NoStop}%
\bibitem [{\citenamefont {Hirsch}\ and\ \citenamefont
  {Tang}(1989)}]{hirsch1989}%
  \BibitemOpen
  \bibfield  {author} {\bibinfo {author} {\bibfnamefont {J.~E.}\ \bibnamefont
  {Hirsch}}\ and\ \bibinfo {author} {\bibfnamefont {S.}~\bibnamefont {Tang}},\
  }\href {https://doi.org/10.1103/PhysRevLett.62.591} {\bibfield  {journal}
  {\bibinfo  {journal} {Phys. Rev. Lett.}\ }\textbf {\bibinfo {volume} {62}},\
  \bibinfo {pages} {591} (\bibinfo {year} {1989})}\BibitemShut {NoStop}%
\bibitem [{\citenamefont {Read}\ and\ \citenamefont
  {Sachdev}(1991)}]{read1991}%
  \BibitemOpen
  \bibfield  {author} {\bibinfo {author} {\bibfnamefont {N.}~\bibnamefont
  {Read}}\ and\ \bibinfo {author} {\bibfnamefont {S.}~\bibnamefont {Sachdev}},\
  }\href {https://doi.org/10.1103/PhysRevLett.66.1773} {\bibfield  {journal}
  {\bibinfo  {journal} {Phys. Rev. Lett.}\ }\textbf {\bibinfo {volume} {66}},\
  \bibinfo {pages} {1773} (\bibinfo {year} {1991})}\BibitemShut {NoStop}%
\bibitem [{\citenamefont {Jackiw}\ and\ \citenamefont
  {Rebbi}(1976)}]{jackiw1976}%
  \BibitemOpen
  \bibfield  {author} {\bibinfo {author} {\bibfnamefont {R.}~\bibnamefont
  {Jackiw}}\ and\ \bibinfo {author} {\bibfnamefont {C.}~\bibnamefont {Rebbi}},\
  }\href {https://doi.org/10.1103/PhysRevLett.36.1116} {\bibfield  {journal}
  {\bibinfo  {journal} {Phys. Rev. Lett.}\ }\textbf {\bibinfo {volume} {36}},\
  \bibinfo {pages} {1116} (\bibinfo {year} {1976})}\BibitemShut {NoStop}%
\bibitem [{\citenamefont {Wu}\ \emph {et~al.}(2004)\citenamefont {Wu},
  \citenamefont {Chen}, \citenamefont {Hu},\ and\ \citenamefont
  {Zhang}}]{wu2004}%
  \BibitemOpen
  \bibfield  {author} {\bibinfo {author} {\bibfnamefont {C.}~\bibnamefont
  {Wu}}, \bibinfo {author} {\bibfnamefont {H.-d.}\ \bibnamefont {Chen}},
  \bibinfo {author} {\bibfnamefont {J.-p.}\ \bibnamefont {Hu}},\ and\ \bibinfo
  {author} {\bibfnamefont {S.-C.}\ \bibnamefont {Zhang}},\ }\href
  {https://doi.org/10.1103/PhysRevA.69.043609} {\bibfield  {journal} {\bibinfo
  {journal} {Phys. Rev. A}\ }\textbf {\bibinfo {volume} {69}},\ \bibinfo
  {pages} {043609} (\bibinfo {year} {2004})}\BibitemShut {NoStop}%
\bibitem [{\citenamefont {Baym}\ and\ \citenamefont
  {Pethick}(1984)}]{baym1991}%
  \BibitemOpen
  \bibfield  {author} {\bibinfo {author} {\bibfnamefont {G.}~\bibnamefont
  {Baym}}\ and\ \bibinfo {author} {\bibfnamefont {C.}~\bibnamefont {Pethick}},\
  }\href@noop {} {\emph {\bibinfo {title} {Landau Fermi-liquid theory}}}\
  (\bibinfo  {publisher} {John Wiley \& Sons, Inc.},\ \bibinfo {year}
  {1984})\BibitemShut {NoStop}%
\bibitem [{\citenamefont {Hirsch}(1990{\natexlab{a}})}]{hirsch1990}%
  \BibitemOpen
  \bibfield  {author} {\bibinfo {author} {\bibfnamefont {J.~E.}\ \bibnamefont
  {Hirsch}},\ }\href {https://doi.org/10.1103/PhysRevB.41.6820} {\bibfield
  {journal} {\bibinfo  {journal} {Phys. Rev. B}\ }\textbf {\bibinfo {volume}
  {41}},\ \bibinfo {pages} {6820} (\bibinfo {year}
  {1990}{\natexlab{a}})}\BibitemShut {NoStop}%
\bibitem [{\citenamefont {Hirsch}(1990{\natexlab{b}})}]{hirsch1990a}%
  \BibitemOpen
  \bibfield  {author} {\bibinfo {author} {\bibfnamefont {J.~E.}\ \bibnamefont
  {Hirsch}},\ }\href {https://doi.org/10.1103/PhysRevB.41.6828} {\bibfield
  {journal} {\bibinfo  {journal} {Phys. Rev. B}\ }\textbf {\bibinfo {volume}
  {41}},\ \bibinfo {pages} {6828} (\bibinfo {year}
  {1990}{\natexlab{b}})}\BibitemShut {NoStop}%
\bibitem [{\citenamefont {Barci}\ and\ \citenamefont
  {Oxman}(2003)}]{barci2003}%
  \BibitemOpen
  \bibfield  {author} {\bibinfo {author} {\bibfnamefont {D.~G.}\ \bibnamefont
  {Barci}}\ and\ \bibinfo {author} {\bibfnamefont {L.~E.}\ \bibnamefont
  {Oxman}},\ }\href {https://doi.org/10.1103/PhysRevB.67.205108} {\bibfield
  {journal} {\bibinfo  {journal} {Phys. Rev. B}\ }\textbf {\bibinfo {volume}
  {67}},\ \bibinfo {pages} {205108} (\bibinfo {year} {2003})}\BibitemShut
  {NoStop}%
\bibitem [{\citenamefont {Lawler}\ \emph {et~al.}(2006)\citenamefont {Lawler},
  \citenamefont {Barci}, \citenamefont {Fernandez}, \citenamefont {Fradkin},\
  and\ \citenamefont {Oxman}}]{lawler2006}%
  \BibitemOpen
  \bibfield  {author} {\bibinfo {author} {\bibfnamefont {M.~J.}\ \bibnamefont
  {Lawler}}, \bibinfo {author} {\bibfnamefont {D.~G.}\ \bibnamefont {Barci}},
  \bibinfo {author} {\bibfnamefont {V.}~\bibnamefont {Fernandez}}, \bibinfo
  {author} {\bibfnamefont {E.}~\bibnamefont {Fradkin}},\ and\ \bibinfo {author}
  {\bibfnamefont {L.}~\bibnamefont {Oxman}},\ }\href@noop {} {\bibfield
  {journal} {\bibinfo  {journal} {Phys. Rev. B}\ }\textbf {\bibinfo {volume}
  {73}},\ \bibinfo {pages} {085101} (\bibinfo {year} {2006})}\BibitemShut
  {NoStop}%
\bibitem [{\citenamefont {Halboth}\ and\ \citenamefont
  {Metzner}(2000)}]{halboth2000}%
  \BibitemOpen
  \bibfield  {author} {\bibinfo {author} {\bibfnamefont {C.~J.}\ \bibnamefont
  {Halboth}}\ and\ \bibinfo {author} {\bibfnamefont {W.}~\bibnamefont
  {Metzner}},\ }\href {https://doi.org/10.1103/PhysRevLett.85.5162} {\bibfield
  {journal} {\bibinfo  {journal} {Phys. Rev. Lett.}\ }\textbf {\bibinfo
  {volume} {85}},\ \bibinfo {pages} {5162} (\bibinfo {year}
  {2000})}\BibitemShut {NoStop}%
\bibitem [{\citenamefont {Dell'Anna}\ and\ \citenamefont
  {Metzner}(2006)}]{dellanna2006}%
  \BibitemOpen
  \bibfield  {author} {\bibinfo {author} {\bibfnamefont {L.}~\bibnamefont
  {Dell'Anna}}\ and\ \bibinfo {author} {\bibfnamefont {W.}~\bibnamefont
  {Metzner}},\ }\href@noop {} {\bibfield  {journal} {\bibinfo  {journal} {Phys.
  Rev. B}\ }\textbf {\bibinfo {volume} {73}},\ \bibinfo {pages} {045127}
  (\bibinfo {year} {2006})}\BibitemShut {NoStop}%
\bibitem [{\citenamefont {Kee}(2003)}]{kee2003}%
  \BibitemOpen
  \bibfield  {author} {\bibinfo {author} {\bibfnamefont {H.-Y.}\ \bibnamefont
  {Kee}},\ }\href {https://doi.org/10.1103/PhysRevB.67.073105} {\bibfield
  {journal} {\bibinfo  {journal} {Phys. Rev. B}\ }\textbf {\bibinfo {volume}
  {67}},\ \bibinfo {pages} {073105} (\bibinfo {year} {2003})}\BibitemShut
  {NoStop}%
\bibitem [{\citenamefont {Kivelson}\ \emph {et~al.}(2003)\citenamefont
  {Kivelson}, \citenamefont {Bindloss}, \citenamefont {Fradkin}, \citenamefont
  {Oganesyan}, \citenamefont {Tranquada}, \citenamefont {Kapitulnik},\ and\
  \citenamefont {Howald}}]{kivelson2003}%
  \BibitemOpen
  \bibfield  {author} {\bibinfo {author} {\bibfnamefont {S.~A.}\ \bibnamefont
  {Kivelson}}, \bibinfo {author} {\bibfnamefont {I.~P.}\ \bibnamefont
  {Bindloss}}, \bibinfo {author} {\bibfnamefont {E.}~\bibnamefont {Fradkin}},
  \bibinfo {author} {\bibfnamefont {V.}~\bibnamefont {Oganesyan}}, \bibinfo
  {author} {\bibfnamefont {J.~M.}\ \bibnamefont {Tranquada}}, \bibinfo {author}
  {\bibfnamefont {A.}~\bibnamefont {Kapitulnik}},\ and\ \bibinfo {author}
  {\bibfnamefont {C.}~\bibnamefont {Howald}},\ }\href
  {https://doi.org/10.1103/RevModPhys.75.1201} {\bibfield  {journal} {\bibinfo
  {journal} {Rev. Mod. Phys.}\ }\textbf {\bibinfo {volume} {75}},\ \bibinfo
  {pages} {1201} (\bibinfo {year} {2003})}\BibitemShut {NoStop}%
\bibitem [{\citenamefont {Gorkov}\ and\ \citenamefont
  {Sokol}(1999)}]{gorkov1992}%
  \BibitemOpen
  \bibfield  {author} {\bibinfo {author} {\bibfnamefont {L.~P.}\ \bibnamefont
  {Gorkov}}\ and\ \bibinfo {author} {\bibfnamefont {A.}~\bibnamefont {Sokol}},\
  }\href@noop {} {\bibfield  {journal} {\bibinfo  {journal} {Phys. Rev. Lett.}\
  }\textbf {\bibinfo {volume} {69}},\ \bibinfo {pages} {2586} (\bibinfo {year}
  {1999})}\BibitemShut {NoStop}%
\bibitem [{\citenamefont {Varma}(2005)}]{varma2005}%
  \BibitemOpen
  \bibfield  {author} {\bibinfo {author} {\bibfnamefont {C.}~\bibnamefont
  {Varma}},\ }\href@noop {} {\bibfield  {journal} {\bibinfo  {journal}
  {Philosophical Magazine}\ }\textbf {\bibinfo {volume} {85}},\ \bibinfo
  {pages} {1657} (\bibinfo {year} {2005})}\BibitemShut {NoStop}%
\bibitem [{\citenamefont {Varma}\ and\ \citenamefont {Zhu}(2006)}]{varma2006}%
  \BibitemOpen
  \bibfield  {author} {\bibinfo {author} {\bibfnamefont {C.~M.}\ \bibnamefont
  {Varma}}\ and\ \bibinfo {author} {\bibfnamefont {L.~J.}\ \bibnamefont
  {Zhu}},\ }\href@noop {} {\bibfield  {journal} {\bibinfo  {journal} {Phys.
  Rev. Lett.}\ }\textbf {\bibinfo {volume} {96}},\ \bibinfo {pages} {036405}
  (\bibinfo {year} {2006})}\BibitemShut {NoStop}%
\bibitem [{\citenamefont {Kee}\ and\ \citenamefont {Kim}(2005)}]{kee2005}%
  \BibitemOpen
  \bibfield  {author} {\bibinfo {author} {\bibfnamefont {H.~Y.}\ \bibnamefont
  {Kee}}\ and\ \bibinfo {author} {\bibfnamefont {E.~H.}\ \bibnamefont {Kim}},\
  }\href@noop {} {\bibfield  {journal} {\bibinfo  {journal} {Phys. Rev. B}\
  }\textbf {\bibinfo {volume} {71}},\ \bibinfo {pages} {184402} (\bibinfo
  {year} {2005})}\BibitemShut {NoStop}%
\bibitem [{\citenamefont {Honerkamp}(2005)}]{honerkamp2005}%
  \BibitemOpen
  \bibfield  {author} {\bibinfo {author} {\bibfnamefont {C.}~\bibnamefont
  {Honerkamp}},\ }\href {doi:10.1103/PhysRevB.72.115103} {\bibfield  {journal}
  {\bibinfo  {journal} {Physical Review B}\ }\textbf {\bibinfo {volume} {72}},\
  \bibinfo {pages} {115103} (\bibinfo {year} {2005})}\BibitemShut {NoStop}%
\bibitem [{\citenamefont {Oganesyan}\ \emph {et~al.}(2001)\citenamefont
  {Oganesyan}, \citenamefont {Kivelson},\ and\ \citenamefont
  {Fradkin}}]{oganesyan2001}%
  \BibitemOpen
  \bibfield  {author} {\bibinfo {author} {\bibfnamefont {V.}~\bibnamefont
  {Oganesyan}}, \bibinfo {author} {\bibfnamefont {S.~A.}\ \bibnamefont
  {Kivelson}},\ and\ \bibinfo {author} {\bibfnamefont {E.}~\bibnamefont
  {Fradkin}},\ }\href {https://doi.org/10.1103/PhysRevB.64.195109} {\bibfield
  {journal} {\bibinfo  {journal} {Phys. Rev. B}\ }\textbf {\bibinfo {volume}
  {64}},\ \bibinfo {pages} {195109} (\bibinfo {year} {2001})}\BibitemShut
  {NoStop}%
\bibitem [{\citenamefont {Kivelson}\ \emph {et~al.}(1998)\citenamefont
  {Kivelson}, \citenamefont {Fradkin},\ and\ \citenamefont
  {Emery}}]{kivelson1998}%
  \BibitemOpen
  \bibfield  {author} {\bibinfo {author} {\bibfnamefont {S.~A.}\ \bibnamefont
  {Kivelson}}, \bibinfo {author} {\bibfnamefont {E.}~\bibnamefont {Fradkin}},\
  and\ \bibinfo {author} {\bibfnamefont {V.~J.}\ \bibnamefont {Emery}},\
  }\href@noop {} {\bibfield  {journal} {\bibinfo  {journal} {Nature}\ }\textbf
  {\bibinfo {volume} {393}},\ \bibinfo {pages} {550} (\bibinfo {year}
  {1998})}\BibitemShut {NoStop}%
\bibitem [{\citenamefont {Lilly}\ \emph {et~al.}(1999)\citenamefont {Lilly},
  \citenamefont {Cooper}, \citenamefont {Eisenstein}, \citenamefont
  {Pfeiffer},\ and\ \citenamefont {West}}]{lilly1999}%
  \BibitemOpen
  \bibfield  {author} {\bibinfo {author} {\bibfnamefont {M.~P.}\ \bibnamefont
  {Lilly}}, \bibinfo {author} {\bibfnamefont {K.~B.}\ \bibnamefont {Cooper}},
  \bibinfo {author} {\bibfnamefont {J.~P.}\ \bibnamefont {Eisenstein}},
  \bibinfo {author} {\bibfnamefont {L.~N.}\ \bibnamefont {Pfeiffer}},\ and\
  \bibinfo {author} {\bibfnamefont {K.~W.}\ \bibnamefont {West}},\ }\href@noop
  {} {\bibfield  {journal} {\bibinfo  {journal} {Phy. Rev. Lett.}\ }\textbf
  {\bibinfo {volume} {82}},\ \bibinfo {pages} {394} (\bibinfo {year}
  {1999})}\BibitemShut {NoStop}%
\bibitem [{\citenamefont {Du}\ \emph {et~al.}(1999)\citenamefont {Du},
  \citenamefont {Tsui}, \citenamefont {Stormer}, \citenamefont {Pfeiffer},
  \citenamefont {Baldwin},\ and\ \citenamefont {West}}]{du1999}%
  \BibitemOpen
  \bibfield  {author} {\bibinfo {author} {\bibfnamefont {R.~R.}\ \bibnamefont
  {Du}}, \bibinfo {author} {\bibfnamefont {D.~C.}\ \bibnamefont {Tsui}},
  \bibinfo {author} {\bibfnamefont {H.~L.}\ \bibnamefont {Stormer}}, \bibinfo
  {author} {\bibfnamefont {L.~N.}\ \bibnamefont {Pfeiffer}}, \bibinfo {author}
  {\bibfnamefont {K.~W.}\ \bibnamefont {Baldwin}},\ and\ \bibinfo {author}
  {\bibfnamefont {K.~W.}\ \bibnamefont {West}},\ }\href@noop {} {\bibfield
  {journal} {\bibinfo  {journal} {Sol. Stat. Comm.}\ }\textbf {\bibinfo
  {volume} {109}},\ \bibinfo {pages} {389} (\bibinfo {year}
  {1999})}\BibitemShut {NoStop}%
\bibitem [{\citenamefont {Grigera}\ \emph {et~al.}(2001)\citenamefont
  {Grigera}, \citenamefont {Perry}, \citenamefont {Schofield}, \citenamefont
  {Chiao}, \citenamefont {Julian}, \citenamefont {Lonzarich}, \citenamefont
  {Ikeda}, \citenamefont {Maeno}, \citenamefont {Millis},\ and\ \citenamefont
  {Mackenzie}}]{grigera2001}%
  \BibitemOpen
  \bibfield  {author} {\bibinfo {author} {\bibfnamefont {S.~A.}\ \bibnamefont
  {Grigera}}, \bibinfo {author} {\bibfnamefont {R.~S.}\ \bibnamefont {Perry}},
  \bibinfo {author} {\bibfnamefont {A.~J.}\ \bibnamefont {Schofield}}, \bibinfo
  {author} {\bibfnamefont {M.}~\bibnamefont {Chiao}}, \bibinfo {author}
  {\bibfnamefont {S.~R.}\ \bibnamefont {Julian}}, \bibinfo {author}
  {\bibfnamefont {G.~G.}\ \bibnamefont {Lonzarich}}, \bibinfo {author}
  {\bibfnamefont {S.~I.}\ \bibnamefont {Ikeda}}, \bibinfo {author}
  {\bibfnamefont {Y.}~\bibnamefont {Maeno}}, \bibinfo {author} {\bibfnamefont
  {A.~J.}\ \bibnamefont {Millis}},\ and\ \bibinfo {author} {\bibfnamefont
  {A.~P.}\ \bibnamefont {Mackenzie}},\ }\href@noop {} {\bibfield  {journal}
  {\bibinfo  {journal} {Science}\ }\textbf {\bibinfo {volume} {294}},\ \bibinfo
  {pages} {329} (\bibinfo {year} {2001})}\BibitemShut {NoStop}%
\bibitem [{\citenamefont {Grigera}\ \emph {et~al.}(2004)\citenamefont
  {Grigera}, \citenamefont {Gegenwart}, \citenamefont {Borzi}, \citenamefont
  {Weickert}, \citenamefont {Schofield}, \citenamefont {Perry}, \citenamefont
  {Tayama}, \citenamefont {Sakakibara}, \citenamefont {Maeno}, \citenamefont
  {Green},\ and\ \citenamefont {Mackenzie}}]{grigera2004}%
  \BibitemOpen
  \bibfield  {author} {\bibinfo {author} {\bibfnamefont {S.~A.}\ \bibnamefont
  {Grigera}}, \bibinfo {author} {\bibfnamefont {P.}~\bibnamefont {Gegenwart}},
  \bibinfo {author} {\bibfnamefont {R.~A.}\ \bibnamefont {Borzi}}, \bibinfo
  {author} {\bibfnamefont {F.}~\bibnamefont {Weickert}}, \bibinfo {author}
  {\bibfnamefont {A.~J.}\ \bibnamefont {Schofield}}, \bibinfo {author}
  {\bibfnamefont {R.~S.}\ \bibnamefont {Perry}}, \bibinfo {author}
  {\bibfnamefont {T.}~\bibnamefont {Tayama}}, \bibinfo {author} {\bibfnamefont
  {T.}~\bibnamefont {Sakakibara}}, \bibinfo {author} {\bibfnamefont
  {Y.}~\bibnamefont {Maeno}}, \bibinfo {author} {\bibfnamefont {A.~G.}\
  \bibnamefont {Green}},\ and\ \bibinfo {author} {\bibfnamefont {A.~P.}\
  \bibnamefont {Mackenzie}},\ }\href@noop {} {\bibfield  {journal} {\bibinfo
  {journal} {Science}\ }\textbf {\bibinfo {volume} {306}},\ \bibinfo {pages}
  {1154} (\bibinfo {year} {2004})}\BibitemShut {NoStop}%
\bibitem [{\citenamefont {Borzi}\ \emph {et~al.}(2007)\citenamefont {Borzi},
  \citenamefont {Grigera}, \citenamefont {Farrell}, \citenamefont {Perry},
  \citenamefont {Lister}, \citenamefont {Lee}, \citenamefont {Tennant},
  \citenamefont {Maeno},\ and\ \citenamefont {Mackenzie}}]{borzi2007}%
  \BibitemOpen
  \bibfield  {author} {\bibinfo {author} {\bibfnamefont {R.~A.}\ \bibnamefont
  {Borzi}}, \bibinfo {author} {\bibfnamefont {S.~A.}\ \bibnamefont {Grigera}},
  \bibinfo {author} {\bibfnamefont {J.}~\bibnamefont {Farrell}}, \bibinfo
  {author} {\bibfnamefont {R.~S.}\ \bibnamefont {Perry}}, \bibinfo {author}
  {\bibfnamefont {S.~J.~S.}\ \bibnamefont {Lister}}, \bibinfo {author}
  {\bibfnamefont {S.~L.}\ \bibnamefont {Lee}}, \bibinfo {author} {\bibfnamefont
  {D.~A.}\ \bibnamefont {Tennant}}, \bibinfo {author} {\bibfnamefont
  {Y.}~\bibnamefont {Maeno}},\ and\ \bibinfo {author} {\bibfnamefont {A.~P.}\
  \bibnamefont {Mackenzie}},\ }\href@noop {} {\bibfield  {journal} {\bibinfo
  {journal} {Science}\ }\textbf {\bibinfo {volume} {315}},\ \bibinfo {pages}
  {214} (\bibinfo {year} {2007})}\BibitemShut {NoStop}%
\bibitem [{\citenamefont {Chubukov}\ and\ \citenamefont
  {Maslov}(2009)}]{chubukov2009}%
  \BibitemOpen
  \bibfield  {author} {\bibinfo {author} {\bibfnamefont {A.~V.}\ \bibnamefont
  {Chubukov}}\ and\ \bibinfo {author} {\bibfnamefont {D.~L.}\ \bibnamefont
  {Maslov}},\ }\href {https://doi.org/10.1103/PhysRevLett.103.216401}
  {\bibfield  {journal} {\bibinfo  {journal} {Phys. Rev. Lett.}\ }\textbf
  {\bibinfo {volume} {103}},\ \bibinfo {pages} {216401} (\bibinfo {year}
  {2009})}\BibitemShut {NoStop}%
\bibitem [{\citenamefont {Hertz}(1976)}]{hertz1976}%
  \BibitemOpen
  \bibfield  {author} {\bibinfo {author} {\bibfnamefont {J.~A.}\ \bibnamefont
  {Hertz}},\ }\href {https://doi.org/10.1103/PhysRevB.14.1165} {\bibfield
  {journal} {\bibinfo  {journal} {Phys. Rev. B}\ }\textbf {\bibinfo {volume}
  {14}},\ \bibinfo {pages} {1165} (\bibinfo {year} {1976})}\BibitemShut
  {NoStop}%
\bibitem [{\citenamefont {Millis}(1993)}]{millis1993}%
  \BibitemOpen
  \bibfield  {author} {\bibinfo {author} {\bibfnamefont {A.~J.}\ \bibnamefont
  {Millis}},\ }\href {https://doi.org/10.1103/PhysRevB.48.7183} {\bibfield
  {journal} {\bibinfo  {journal} {Phys. Rev. B}\ }\textbf {\bibinfo {volume}
  {48}},\ \bibinfo {pages} {7183} (\bibinfo {year} {1993})}\BibitemShut
  {NoStop}%
\bibitem [{\citenamefont {Fu}(2015)}]{fu2015}%
  \BibitemOpen
  \bibfield  {author} {\bibinfo {author} {\bibfnamefont {L.}~\bibnamefont
  {Fu}},\ }\href {https://doi.org/10.1103/PhysRevLett.115.026401} {\bibfield
  {journal} {\bibinfo  {journal} {Phys. Rev. Lett.}\ }\textbf {\bibinfo
  {volume} {115}},\ \bibinfo {pages} {026401} (\bibinfo {year}
  {2015})}\BibitemShut {NoStop}%
\bibitem [{\citenamefont {Vollhardt}\ and\ \citenamefont
  {Wolfle}(1990)}]{vollhardt1990}%
  \BibitemOpen
  \bibfield  {author} {\bibinfo {author} {\bibfnamefont {D.}~\bibnamefont
  {Vollhardt}}\ and\ \bibinfo {author} {\bibfnamefont {P.}~\bibnamefont
  {Wolfle}},\ }\href@noop {} {\emph {\bibinfo {title} {The superfluid phases of
  helium 3}}}\ (\bibinfo  {publisher} {Taylor \& Francis},\ \bibinfo {address}
  {London},\ \bibinfo {year} {1990})\BibitemShut {NoStop}%
\bibitem [{\citenamefont {Gennes}\ and\ \citenamefont
  {Prost}(1995)}]{degennes}%
  \BibitemOpen
  \bibfield  {author} {\bibinfo {author} {\bibfnamefont {P.~G.~d.}\
  \bibnamefont {Gennes}}\ and\ \bibinfo {author} {\bibfnamefont
  {J.}~\bibnamefont {Prost}},\ }\href@noop {} {\emph {\bibinfo {title} {The
  Physics of Liquid Crystals}}}\ (\bibinfo  {publisher} {Clarendon Press; 2nd
  edition},\ \bibinfo {year} {1995})\BibitemShut {NoStop}%
\bibitem [{\citenamefont {Li}\ and\ \citenamefont {Wu}(2012)}]{liwu2012}%
  \BibitemOpen
  \bibfield  {author} {\bibinfo {author} {\bibfnamefont {Y.}~\bibnamefont
  {Li}}\ and\ \bibinfo {author} {\bibfnamefont {C.}~\bibnamefont {Wu}},\ }\href
  {https://doi.org/10.1103/PhysRevB.85.205126} {\bibfield  {journal} {\bibinfo
  {journal} {Phys. Rev. B}\ }\textbf {\bibinfo {volume} {85}},\ \bibinfo
  {pages} {205126} (\bibinfo {year} {2012})}\BibitemShut {NoStop}%
\bibitem [{\citenamefont {Lee}\ and\ \citenamefont {Wu}(2009)}]{leewu2009}%
  \BibitemOpen
  \bibfield  {author} {\bibinfo {author} {\bibfnamefont {W.-C.}\ \bibnamefont
  {Lee}}\ and\ \bibinfo {author} {\bibfnamefont {C.}~\bibnamefont {Wu}},\
  }\href {https://doi.org/10.1103/PhysRevB.80.104438} {\bibfield  {journal}
  {\bibinfo  {journal} {Phys. Rev. B}\ }\textbf {\bibinfo {volume} {80}},\
  \bibinfo {pages} {104438} (\bibinfo {year} {2009})}\BibitemShut {NoStop}%
\bibitem [{\citenamefont {Leggett}(1970)}]{leggett1970}%
  \BibitemOpen
  \bibfield  {author} {\bibinfo {author} {\bibfnamefont {A.~J.}\ \bibnamefont
  {Leggett}},\ }\href {http://stacks.iop.org/0022-3719/3/i=2/a=027} {\bibfield
  {journal} {\bibinfo  {journal} {Journal of Physics C: Solid State Physics}\
  }\textbf {\bibinfo {volume} {3}},\ \bibinfo {pages} {448} (\bibinfo {year}
  {1970})}\BibitemShut {NoStop}%
\bibitem [{\citenamefont {Corruccini}\ \emph {et~al.}(1971)\citenamefont
  {Corruccini}, \citenamefont {Osheroff}, \citenamefont {Lee},\ and\
  \citenamefont {Richardson}}]{corruccini1971}%
  \BibitemOpen
  \bibfield  {author} {\bibinfo {author} {\bibfnamefont {L.~R.}\ \bibnamefont
  {Corruccini}}, \bibinfo {author} {\bibfnamefont {D.~D.}\ \bibnamefont
  {Osheroff}}, \bibinfo {author} {\bibfnamefont {D.~M.}\ \bibnamefont {Lee}},\
  and\ \bibinfo {author} {\bibfnamefont {R.~C.}\ \bibnamefont {Richardson}},\
  }\href {https://doi.org/10.1103/PhysRevLett.27.650} {\bibfield  {journal}
  {\bibinfo  {journal} {Phys. Rev. Lett.}\ }\textbf {\bibinfo {volume} {27}},\
  \bibinfo {pages} {650} (\bibinfo {year} {1971})}\BibitemShut {NoStop}%
\bibitem [{\citenamefont {Osheroff}(1977)}]{osheroff1977}%
  \BibitemOpen
  \bibfield  {author} {\bibinfo {author} {\bibfnamefont {D.}~\bibnamefont
  {Osheroff}},\ }\href
  {https://doi.org/http://dx.doi.org/10.1016/0378-4363(77)90005-5} {\bibfield
  {journal} {\bibinfo  {journal} {Physica B+C}\ }\textbf {\bibinfo {volume}
  {90}},\ \bibinfo {pages} {20 } (\bibinfo {year} {1977})}\BibitemShut
  {NoStop}%
\bibitem [{\citenamefont {Greywall}(1983)}]{greywall1983}%
  \BibitemOpen
  \bibfield  {author} {\bibinfo {author} {\bibfnamefont {D.~S.}\ \bibnamefont
  {Greywall}},\ }\href {https://doi.org/10.1103/PhysRevB.27.2747} {\bibfield
  {journal} {\bibinfo  {journal} {Phys. Rev. B}\ }\textbf {\bibinfo {volume}
  {27}},\ \bibinfo {pages} {2747} (\bibinfo {year} {1983})}\BibitemShut
  {NoStop}%
\bibitem [{\citenamefont {Mydosh}\ \emph {et~al.}(2020)\citenamefont {Mydosh},
  \citenamefont {Oppeneer},\ and\ \citenamefont {Riseborough}}]{mydosh2020}%
  \BibitemOpen
  \bibfield  {author} {\bibinfo {author} {\bibfnamefont {J.~A.}\ \bibnamefont
  {Mydosh}}, \bibinfo {author} {\bibfnamefont {P.~M.}\ \bibnamefont
  {Oppeneer}},\ and\ \bibinfo {author} {\bibfnamefont {P.~S.}\ \bibnamefont
  {Riseborough}},\ }\href {https://doi.org/10.1088/1361-648x/ab5eba} {\bibfield
   {journal} {\bibinfo  {journal} {Journal of Physics: Condensed Matter}\
  }\textbf {\bibinfo {volume} {32}},\ \bibinfo {pages} {143002} (\bibinfo
  {year} {2020})}\BibitemShut {NoStop}%
\bibitem [{\citenamefont {Wolowiec}\ \emph {et~al.}(2021)\citenamefont
  {Wolowiec}, \citenamefont {Kanchanavatee}, \citenamefont {Huang},
  \citenamefont {Ran}, \citenamefont {Breindel}, \citenamefont {Pouse},
  \citenamefont {Sasmal}, \citenamefont {Baumbach}, \citenamefont {Chappell},
  \citenamefont {Riseborough},\ and\ \citenamefont {Maple}}]{wolowiece2021}%
  \BibitemOpen
  \bibfield  {author} {\bibinfo {author} {\bibfnamefont {C.~T.}\ \bibnamefont
  {Wolowiec}}, \bibinfo {author} {\bibfnamefont {N.}~\bibnamefont
  {Kanchanavatee}}, \bibinfo {author} {\bibfnamefont {K.}~\bibnamefont
  {Huang}}, \bibinfo {author} {\bibfnamefont {S.}~\bibnamefont {Ran}}, \bibinfo
  {author} {\bibfnamefont {A.~J.}\ \bibnamefont {Breindel}}, \bibinfo {author}
  {\bibfnamefont {N.}~\bibnamefont {Pouse}}, \bibinfo {author} {\bibfnamefont
  {K.}~\bibnamefont {Sasmal}}, \bibinfo {author} {\bibfnamefont {R.~E.}\
  \bibnamefont {Baumbach}}, \bibinfo {author} {\bibfnamefont {G.}~\bibnamefont
  {Chappell}}, \bibinfo {author} {\bibfnamefont {P.~S.}\ \bibnamefont
  {Riseborough}},\ and\ \bibinfo {author} {\bibfnamefont {M.~B.}\ \bibnamefont
  {Maple}},\ }\bibfield  {journal} {\bibinfo  {journal} {Proceedings of the
  National Academy of Sciences}\ }\textbf {\bibinfo {volume} {118}},\ \href
  {https://doi.org/10.1073/pnas.2026591118} {10.1073/pnas.2026591118} (\bibinfo
  {year} {2021}),\ \Eprint
  {https://arxiv.org/abs/https://www.pnas.org/content/118/20/e2026591118.full.pdf}
  {https://www.pnas.org/content/118/20/e2026591118.full.pdf} \BibitemShut
  {NoStop}%
\bibitem [{\citenamefont {Harter}\ \emph {et~al.}(2017)\citenamefont {Harter},
  \citenamefont {Zhao}, \citenamefont {Yan}, \citenamefont {Mandrus},\ and\
  \citenamefont {Hsieh}}]{harter2017}%
  \BibitemOpen
  \bibfield  {author} {\bibinfo {author} {\bibfnamefont {J.~W.}\ \bibnamefont
  {Harter}}, \bibinfo {author} {\bibfnamefont {Z.~Y.}\ \bibnamefont {Zhao}},
  \bibinfo {author} {\bibfnamefont {J.-Q.}\ \bibnamefont {Yan}}, \bibinfo
  {author} {\bibfnamefont {D.~G.}\ \bibnamefont {Mandrus}},\ and\ \bibinfo
  {author} {\bibfnamefont {D.}~\bibnamefont {Hsieh}},\ }\href
  {https://doi.org/10.1126/science.aad1188} {\bibfield  {journal} {\bibinfo
  {journal} {Science}\ }\textbf {\bibinfo {volume} {356}},\ \bibinfo {pages}
  {295每299} (\bibinfo {year} {2017})}\BibitemShut {NoStop}%
\bibitem [{\citenamefont {Harter}\ \emph {et~al.}(2018)\citenamefont {Harter},
  \citenamefont {Kennes}, \citenamefont {Chu}, \citenamefont {de~la Torre},
  \citenamefont {Zhao}, \citenamefont {Yan}, \citenamefont {Mandrus},
  \citenamefont {Millis},\ and\ \citenamefont {Hsieh}}]{harter2018}%
  \BibitemOpen
  \bibfield  {author} {\bibinfo {author} {\bibfnamefont {J.~W.}\ \bibnamefont
  {Harter}}, \bibinfo {author} {\bibfnamefont {D.~M.}\ \bibnamefont {Kennes}},
  \bibinfo {author} {\bibfnamefont {H.}~\bibnamefont {Chu}}, \bibinfo {author}
  {\bibfnamefont {A.}~\bibnamefont {de~la Torre}}, \bibinfo {author}
  {\bibfnamefont {Z.~Y.}\ \bibnamefont {Zhao}}, \bibinfo {author}
  {\bibfnamefont {J.-Q.}\ \bibnamefont {Yan}}, \bibinfo {author} {\bibfnamefont
  {D.~G.}\ \bibnamefont {Mandrus}}, \bibinfo {author} {\bibfnamefont {A.~J.}\
  \bibnamefont {Millis}},\ and\ \bibinfo {author} {\bibfnamefont
  {D.}~\bibnamefont {Hsieh}},\ }\href
  {https://doi.org/10.1103/PhysRevLett.120.047601} {\bibfield  {journal}
  {\bibinfo  {journal} {Phys. Rev. Lett.}\ }\textbf {\bibinfo {volume} {120}},\
  \bibinfo {pages} {047601} (\bibinfo {year} {2018})}\BibitemShut {NoStop}%
\bibitem [{\citenamefont {Norman}(2020)}]{norman2020}%
  \BibitemOpen
  \bibfield  {author} {\bibinfo {author} {\bibfnamefont {M.~R.}\ \bibnamefont
  {Norman}},\ }\href {https://doi.org/10.1103/PhysRevB.101.045117} {\bibfield
  {journal} {\bibinfo  {journal} {Phys. Rev. B}\ }\textbf {\bibinfo {volume}
  {101}},\ \bibinfo {pages} {045117} (\bibinfo {year} {2020})}\BibitemShut
  {NoStop}%
\end{thebibliography}%


\end{document}